\documentclass[11pt,twoside]{article}
\usepackage{cozumel2005}
\def\apjl{ApJ}%
\def\apjs{ApJS}%
\def\aap{A\&A}%
\usepackage{epsf}
\usepackage{psfig}
\usepackage{lscape}
\usepackage{natbib}
\usepackage{graphicx}
\usepackage{relsize}
\usepackage{amssymb}
\newcommand{\oversim}[2]{\protect{\mbox{\lower0.5ex\vbox{%
   \baselineskip=0pt\lineskip=0.2ex
   \ialign{$\mathsurround=0pt #1\hfil##\hfil$\crcr#2\crcr\sim\crcr}}}}} 
\newcommand{\simgreat}{\mbox{$\,\mathrel{\mathpalette\oversim>}\,$}} 
\newcommand{\simless} {\mbox{$\,\mathrel{\mathpalette\oversim<}\,$}} 
\begin{document}
\title{The initial mass function of simple and composite stellar
populations} \author{Pavel Kroupa$^{1,2}$} \affil{$^1$Argelander-Institut f\"ur
Astronomie, Universit\"at Bonn, Auf dem H\"ugel 71, D-53121~Bonn, Germany}
\affil{$^2$The Rhine Stellar Dynamical Network}

\begin{abstract} 
The distribution of stellar masses that form together, the initial
mass function (IMF), is one of the most important astrophysical
distribution functions. The determination of the IMF is a very
difficult problem because stellar masses cannot be measured directly
and because observations usually cannot assess all stars in a
population requiring elaborate bias corrections. Nevertheless,
impressive advances have been achieved during the last decade, such
that the shape of the IMF is reasonably well understood from low-mass
brown dwarfs (BDs) to very massive stars. The case can be made for a
rather universal form that can be well approximated by a two-part
power-law function in the stellar regime. However, there exists a
possible hint for a systematic variation with metallicity.  From very
elaborate observational surveys a picture is emerging according to
which the binary properties of very-low-mass stars (VLMSs) and BDs may
be fundamentally different from those of late-type stars implying the
probable existence of a discontinuity in the IMF, but the surveys also
appear to suggest the number of BDs per star to be independent of the
physical conditions of current Galactic star formation.  Star-burst
clusters and thus globular cluster may, however, have a much larger
abundance of BDs.  Very recent advances have allowed the measurement
of the physical upper stellar mass limit, which also appears to be
disconcertingly robust to variations in metallicity.  Furthermore, it
now appears that star clusters are formed in a rather organised
fashion from low- to high stellar masses, such that the most-massive
stars just forming terminate further star-formation within the
particular cluster. Populations formed from many star clusters, {\it
composite populations}, would then have steeper IMFs (fewer massive
stars per low-mass star) than the simple populations in the
constituent clusters. A near invariant star-cluster mass function
implies the maximal cluster mass to correlate with the galaxy-wide
star-formation rate. This then leads to the result that the
composite-stellar IMFs vary in dependence of galaxy type, with
potentially dramatic implications for theories of galaxy formation and
evolution.

\end{abstract}


\section{Introduction}
\label{sec:intro}

The stellar initial mass function (IMF) is perhaps the most important
macroscopic distribution function in astrophysics, because it defines
the mass-spectrum of stars born together.  The other fundamental
function is the star-formation history of a stellar system. They are
connected through complex microscopic physical processes that defy
detailed treatment on galactic scales.  Together they contain the
essential information on the transformation of dark gas to shining
stars and the spectral energy distribution thereof. They also contain
the essential information on the cycle of matter, which fraction of it
is locked up for ``ever'' in feeble stars and sub-stellar objects, and
how much of it is returned enriched with higher chemical elements to
the interstellar medium or atmosphere of a galaxy.

Given the importance of the IMF a huge research effort has been
invested to distill its shape and variability. The seminal
contribution by \cite{S55} whilst staying in Canberra first described
the IMF as a power-law, $dN=\xi(m)\,dm=k\,m^{-\alpha} = dN$, where
$dN$ is the number of stars in the mass interval $m,m+dm$ and $k$ is
the normalisation constant.  By modelling the spatial distribution of
the then observed stars with assumptions on the star-formation rate,
Galactic-disk structure and stellar evolution time-scales, Salpeter
arrived at the power-law index $\alpha=2.35$ for $0.4 \simless
m/M_\odot \simless 10$, which today is known as the ``Salpeter IMF''.

This IMF form implies a diverging mass density for $m\rightarrow 0$,
which was interesting since dark matter was speculated, until the
early 1990's, to possibly be made-up of faint stars or sub-stellar
objects.  Studies of the stellar velocities in the solar-neighbourhood
also implied a large amount of missing, or dark, mass in the disk of
the Milky Way (MW) \citep{B84}. Careful compilation in Heidelberg of
the Gliese {\it Catalogue of Nearby Stars} beginning in the 1960's
\citep{JW97}\footnote{The latest version of the catalogue can be found
at http://www.ari.uni-heidelberg.de/aricns/, while
http://www.nstars.nau.edu/ contains the Nearby Stars (NStars)
database.}, and the application at the beginning of the 1980's of an
innovative photographic pencil-beam survey-technique reaching deep
into the Galactic field in Edinburgh by \citet{RG82} significantly
improved knowledge of the space density of VLMSs ($0.072 \simless
m/M_\odot \simless 0.14$).

Major studies extending Salpeter's work to lower and larger masses
followed, showing that the mass function (MF) of Galactic-field stars
turns over below one solar mass thus avoiding the divergence. Since
stars with masses $m\simless 0.8\,M_\odot$ do not evolve significantly
over the age of the Galactic disk, the MF equals the IMF for these.
While the work of \cite{MS79} relied on using the nearby stellar
sample to define the IMF for $m<1\,M_\odot$, \cite{Sc86} relied mostly
on a more recent deep pencil-beam star-count survey.  \cite{Sc86}
stands out as the most thorough and comprehensive analysis of the IMF
in existence, laying down notation and ideas still in use today.  The
form of the IMF for low-mass stars was revised in the early 1990's in
Cambridge, especially through the quantification of significant
non-linearities in the stellar mass--luminosity relation and
evaluation of the bias due to unresolved binary systems
\citep{KTG90,KTG91,KTG93}. On the one hand this work lead to a
detailed understanding of the shape of the stellar luminosity function
(LF) in terms of stellar physics, and on the other the difference
between the results obtained by \cite{MS79} and \cite{Sc86} for
low-mass stars was resolved by this work through rigorous modelling of
all biases affecting local trigonometric-based and distant
photometric-parallax-based surveys, such as come from an intrinsic
metallicity scatter, evolution along the main sequence and contraction
to the main sequence.  In doing so this work also included an updated
local stellar sample {\it and} the then best-available deep
pencil-beam survey by \cite{SIP}. As such it stands unique today as
being the only rigorous analysis of the late-type-star MF using
simultaneously both the {\it nearby trigonometric parallax} and the
{\it far, pencil-beam} star-count data.  This study was further
extended to an analysis of other ground-based pencil-beam surveys
\citep{K95a}.  \cite{Zheng01} employed the HST to measure the LF
through the entire thickness of the MW disk finding excellent
agreement with the pencil-beam (photometric-parallax) LFs
(Fig.~\ref{fig:MWlf} below).  The results of \citet{KTG93}
($\alpha\approx 1.3, 0.08-0.5\,M_\odot$), were confirmed by
\citet{RGH02} using updated local star-counts that included Hipparcos
parallax data.  Indeed, the continued long-term observational effort
on nearby stars by Neill Reid, John Gizis and collaborators forms one
of the very major pillars of modern IMF work; continued discussion of
controversial interpretations have much improved and sharpened our
general understanding of the issues.  A re-analysis of the nearby mass
function of stars in terms of a log-normal form in the mass range
$0.07-1\,M_\odot$ was provided by Chabrier, finding agreement to the
deep HST star-count data once unresolved multiple stars and a
metal-deficient colour-magnitude relation for thick-disk M~dwarfs are
accounted for \citep{Ch03}.

The Galactic-field IMF for $0.08<m/M_\odot<1$ can thus be regarded as
being reasonably well-constrained, but some unresolved issues
nevertheless remain.  The exact form of the IMF is still under
dispute. The mathematical necessity for the IMF to turn-over at some
small mass and the corresponding empirical result by \cite{MS79} lead
to the development of a theory of the IMF based on gravitational
instabilities in the turbulent inter-stellar medium \citep{Fleck82,
Ferrini_etal83} and random hierarchical fragmentation \citep{Zinn84},
thereby accounting for the approximately log-normal shape of the IMF.
While a log-normal form has often been adopted given the turnover near
a few tenths of a solar mass \citep{MS79, Ch01} and some theoretical
ideas (e.g. \citealt{AF96, Chrev03}), it has never been demonstrated that
the log-normal form is consistent with the sharp peak in the stellar
luminosity function (LF) near absolute visual and I-band magnitudes
$M_{\rm V}\approx12, M_{\rm I}\approx8.5$, respectively. A power-law
description together with a semi-empirical mass--luminosity relation
has, in contrast, been found to fit the nearby and deep LFs well
\citep{KTG93, K95a}.  Furthermore, most approaches have relied on
using parametrised formulae, such as the multi-power-law form or the
log-normal form, and so it would be important to develop a
non-parametric method that allows inclusion of the biases through
unresolved multiple systems.  Such a method would be very useful for
studying the shape of the IMF in star clusters.

In contrast to the Galactic-field sample, where stars of many ages and
metallicities are mixed, star clusters offer the advantage that the
stars have the same age and metallicity and distance. And so a very
large effort has been invested to try to extract the IMF from open
\citep[e.g.]{SG01, Piskunov_etal04, MKB04} and embedded
\citep[e.g.]{Hill97, Muench_etal02, Luhman04} clusters, as well as
associations \citep{Goulier_etal05}. Here the continued
methodologically consistent observational work on a number of
different very young populations by Kevin Luhman and collaborators has
had a major impact on our understanding of the IMF at low masses. On
the theoretical side, the ever-improving modelling of stellar and BD
atmospheres being pushed forward with excellent results notably by the
Lyon group (Isabelle Baraffe and Gilles Chabrier), has allowed
consistently better constraints on the faint-star MF by a wide variety
of observational surveys \footnote{Here it should be emphasised and
acknowledged that the intensive and highly fruitful discourse between
Guenther Wuchterl and the Lyon group has led to the important
understanding that the classical evolution tracks computed in Lyon and
by others are unreliable for ages less than a few~Myr
\citep{WT03}.}. Furthermore, the development of high-precision
$N$-body codes that rely on complex mathematical and algorithmic
procedures through the work of Sverre Aarseth and Seppo Mikkola
\citep{Aarseth99} and others has led to important progress on
understanding the variation of the dynamical properties of stellar
populations in individual clusters and the Galactic field
\citep{K95c,Fuente97,K01a,KAH,Port_etal01,Port_etal02,BaumMakin03,MKB04,
Hurley_etal05}.  In general, the MF found for clusters is consistent
with the Galactic-field IMF for $m<1\,M_\odot$, but some problems
remain.  While a huge progress has been achieved in modelling cluster
populations over time, it is for example still unclear as to why open
clusters have a significant deficit of white dwarfs
\citep{Fellhauer_etal03}. A further problem is posed by the rapid and
violent early dynamical evolution of clusters
\citep{Lada_etal84,Goodwin97, KAH, GeyerBurkert01,MKB04, Kr_paris05}
and the associated loss of a large fraction of the cluster population,
and due to the density-dependent disruption of primordial binary
systems. Young clusters have thus undergone a highly complex dynamical
evolution which continues into old age \citep{BaumMakin03} and are
therefore subject to biases that can only be studied effectively with
full-scale $N$-body methods, thus imposing a complexity of analysis
that surpasses that for the Galactic-field sample.

For massive stars, the \citet{Sc86} IMF is based on a combination of
Galactic-field star-counts and OB association data and has a slope
$\alpha\approx 2.7$ for $m \simgreat 2\,M_\odot$ with much uncertainty
for $m\simgreat 10\,M_\odot$.  The previous determination by
\cite{MS79} also argued for a relatively steep field-IMF with $\alpha
= 2.5, 1 \le m/M_\odot \le 10$ and $\alpha=3.3, m>10\,M_\odot$.
However, Massey's work at Tucson demonstrated, through extensive
spectroscopic classification, that $\alpha =2.35\pm 0.1,
m\simgreat10\,M_\odot$ \citep{M98}, for a large variety of physical
environments, namely OB associations and dense clusters for
populations with metallicity ranging from near-solar-abundance to
about $1/10$th metal abundance
(Fig.~\ref{fig:kroupa_figmassey}). Consequently, above about
$0.5\,M_\odot$ the empirical IMF can be described well by a single
power-law form with the {\it Salpeter/Massey index},
$\alpha=2.35$,therewith being remarkably invariant.\footnote{Note that
\cite{Sc98} emphasises that the IMF remains poorly constrained owing
to the small number of massive stars in any one sample. This is a true
albeit conservative stand-point, and the present author prefers to
accept Massey's result as a working hypothesis.  This hypothesis that
there exists a universal parent distribution function is tested in
individual clusters and OB associations for possible significant
deviations (cf. \citealt{P-AK06a})\label{pk_fn1}.}

While the case is often made that star-bursts may have an IMF that is
over-abundant in massive stars (e.g. \citealt{Eisenh01}),
time-dependent mass segregation mimics just such an effect
\citep{Boily_etal05}.  Local well-resolved star-burst clusters also do
not support top-heavy IMFs. Notable well-studied cases with masses in
the range $10^{4-5}M_\odot$ and central densities near
$10^5$~stars/pc$^3$ are the 30~Dor cluster (R136) in the LMC, NGC~3603
in the MW, and the Arches cluster near the Galactic centre. The 30~Dor
star-burst cluster (NGC~2070) has been found by \citet{MH98} to have a
Salpeter MF for $2.8 \simless m/M_\odot \simless 120$, which is
confirmed by \citet{Setal99} who apply corrections for variable
reddening. But they also note that the core of NGC~2070 is clearly
mass-segregated with a flatter MF. For NGC~3603, \cite{SB04} find a
highly mass-segregated system with a flat MF in the core, but the MF
beyond the core has $\alpha=1.9\pm0.1, m\simgreat 1.6\,M_\odot$, thus
again being consistent with a Salpeter value. The Arches cluster is
situated at a projected distance of 25~pc from the Galactic centre and
is therefore in a very exotic environment being strongly influenced by
tidal forces. \citet{Fetal99} and \citet{Stolteetal02} find the
innermost region of the cluster to have $\alpha\approx 1$, but at
larger radii $\alpha\approx 2.7$. \citet{Kimetal2000} and
\cite{Port_etal02} study the evolution of the Arches using $N$-body
modelling and find that the cluster evolves rapidly, loosing memory of
its birth configuration within about 1~Myr.  In particular, they find
that the IMF in the Arches may have been quite normal. 

This short discussion thus indicates that the evidence for a top-heavy
IMF is not strong in well-resolved star-burst clusters, and that
dynamical evolution of the clusters needs to be modelled in detail to
understand possibly deviant {\it observed} IMF's.  A highly
interesting finding reported by \citet{M98, Massey03} is that the OB
stellar population found in the field of the LMC has a very steep MF,
$\alpha \approx 4.5$, which can be interpreted to be due to the
preferred formation of small groups or even isolated O and B-type
stars. Another interpetation that would not need to resort to exotic
star-fomration events is the dynamical ejection of OB stars from
dynamically unstable cores of young clusters. This may lead to such a
steep IMF because simple estimates show that preferentlially the
less-massive members of a core are ejected \citep{ClarkePringle92}.
This process needs to be studied using fully-consistent and
high-precission $N-$body modelling of youing clusters, to see if the
observed distribution of field-OB stars can be accounted for with this
process alone, or if indeed an exotic star-formation mode needs to be
invoked to explain some of the observations.  

{\it The ``secret'' ingredient for uncovering the true nature of an
IMF is thus the high-precision $N$-body experiment, which therewith
achieves an importance well beyond merely the methodological
\citep[e.g.]{Kimetal06,P-AK06a}.}

Although we have essentially no knowledge about the low-mass end of
the IMF in star-burst clusters, due to their rarity and thus generally
large distances, today we do know that the MF of low-mass stars in
globular clusters, that presumably formed as star-burst clusters, is
quite similar to that of the solar-neighbourhood and in young open
clusters. This suggests a remarkable invariance of the low-mass IMF
\citep[e.g.]{deMP95a,deMP95b,PdeMR95}.  The case has been made for a
systematic variation with metallicity in the sense that the
metal-poorer and older populations may have flatter MFs as expected
from simple Jeans-mass arguments (\citealt{K01a},
eq.~\ref{eq:systemvar} below), but this suggestion remains speculative
rather than conclusive because stellar-dynamical processes skew
observed MFs \citep{K01a,BaumMakin03}.

The MF for BDs is even shallower, as two recent studies of the
available solar-neighbourhood star-count data demonstrate, these
constraints being consistent with those from young star clusters
\citep{Ch02, Allen_etal04}.  From their non-stellar binary properties
\citep{Martin_etal03, Bouy_etal03, Close_etal03} it is emerging,
however, that BDs may not be a continuation of the stellar
population. Although their formation is linked to that of their
sibling stars, the IMF is likely to be discontinuous near the
hydrogen-burning mass limit, and the overall result appears to be that
about one~BD is born per five~stars (\citealt{Kr_Bouv03b},
\S~\ref{sec:BDnr} below).  We have thus had to realise that BDs
contribute a negligible dynamical mass to any stellar population.
Faint main-sequence stars account for most of the mass in the MW disk,
and the need for dark matter in the MW disk also disappeared as
improved kinematical data and analysis thereof became available
\citep{Kuijken,FF94}. Furthermore, the suggestion has been made by two
independent teams, based on an extremely high-quality spectral
analysis of solar-neighbourhood stars and by accounting for the
relative death rate of stars, that the Galactic thick disk may be
substantially more massive than is currently thought
\citep{Soubiran_etal03,Fuhrmann04}. This would further reduce the mass
discrepancy evident in the solar motion about the Galaxy.

Five very recent further developments have enriched the study of the
mass-distribution of stars appreciably. \cite{WK04} demonstrated for
the first time that the absence of stars more massive than about
$150\,M_\odot$ in R136 in the LMC must imply a physical mass limit to
stars near $150\,M_\odot$, with high statistical significance, unless
$\alpha\simgreat 2.8$ for $m\simgreat 1\,M_\odot$. Such a steep IMF
may be the true mass-distribution if unresolved multiple systems are
corrected for, but an affirmation of this and thus the negation of the
physical stellar mass limit awaits a detailed investigation of this
issue. \cite{Fi05} followed by performing the same analysis with his
HST data on the Arches cluster confirming the same physical mass
limit, and \cite{OC05} published a statistical analysis of a number of
OB associations and clusters again confirming that a physical mass
limit exists in the mass range $120-200\,M_\odot$. Thus, stellar
masses are limited near $0.072\,M_\odot$ \citep{Chrev03} {\it and}
near $150\,M_\odot$.  The other notion recently re-affirmed is that
star clusters appear to limit the masses of the most massive stars in
them in such a way that more massive clusters permit higher masses of
their most massive stars \citep{WK05a}. The third very recent insight,
resulting from the above two, is that the composite IMFs of whole
galaxies must be steeper than the stellar IMF for $m\simgreat
1\,M_\odot$, because composite populations result from the addition of
the stellar IMFs in the star clusters which themselves are distributed
according to a power-law cluster mass function
\citep{KW03,WK05b}. This notion had been raised already by
\cite{Vanbev82} and a similar ansatz has been followed by \cite{OC98}
for the construction of HII LFs in galaxies, and the recent advances
on this issue thus suggest that the discrepancy between Scalo's
field-star IMF and Massey's stellar IMF in star clusters may have
found a natural resolution \citep{KW03}.  The fourth recent insight
\citep{WKL04} is that the star-formation rate of a galaxy dictates the
mass of the most massive star cluster forming within it.  This then
implies the fifth recent result \citep{WK05a} that galaxies have
different integrated (composite) IMFs. According to this picture,
galaxies with a small stellar mass ($\simless 10^7\,M_\odot$) may,
over their life-times, be significantly deficient in stars more
massive than a few $M_\odot$.

Given these results, it thus becomes necessary to distinguish between
{\it simple stellar populations} and {\it composite populations}.  A
simple population is found in a star cluster and consists of stars of
equal age and metallicity. A composite population consists of more
than one cluster.  The {\it stellar IMF} refers to the IMF of stars in
a simple population, while the {\it composite IMF}, or the {\it
integrated galactic IMF} (IGIMF) is the IMF of a composite population,
i.e. a population composed of many star clusters, most of which may be
dissolved.

This treatise attempts to provide an overview of the general methods
used to derive the IMF with special attention on the pitfalls that are
typically encountered. The somewhat uncomfortable result until now is
that there is no reliable, or confirmed, evidence for stellar-IMF
differences in simple populations with different physical
properties. While this must be recognised as being a problem, because
elementary arguments would imply variations with different conditions,
it does ease modelling of simple stellar systems. While the IMF for
$0.08\simless m/M_\odot \simless 1\,M_\odot$ is by now quite well
constrained, for massive stars the effects of unresolved multiple
systems remains unknown\footnote{see also
footnote~\ref{pk_fn1}}. Corrections for these may significantly affect
the shape of the IMF with rather major implications for much of
extragalactic and Galactic astrophysics. Understanding these
corrections is thus of major importance.  The very recent realisation
that star-clusters limit the mass spectrum of their stars has
interesting implications for the formation of stars in a cluster and
leads to the insight that composite populations must show IMFs that
differ from the stellar IMF in each cluster. A detailed analysis
indicates that the composite IMF, i.e. the IGIMF, should vary between
galaxies.  With this finale, this treatise reaches the cosmological
arena.

Other relatively recent reviews of the IMF are by \cite{Sc98},
\cite{K02}, \cite{Chrev03}. The proceedings of the ``38th Herstmonceux
Conference on the Stellar Initial Mass Function'' \citep{GilHow98} and
the proceedings of the ``IMF\@50'' meeting in celebration of Ed
Salpeter's 80th birthday \citep{IMF50} contain a wealth of important
contributions to the field.

\section{Some essentials}
Assuming all binary and higher-order stellar systems can be resolved
into individual stars in some population such as the solar
neighbourhood and that only main-sequence stars are selected for, then
the number of single stars per pc$^3$ in the mass interval $m$ to
$m+dm$ is $dN=\Xi(m)\,dm$, where $\Xi(m)$ is the present-day mass
function (PDMF). The number of single stars per pc$^3$ in the absolute
V-band magnitude interval $M_P$ to $M_P+dM_P$ is
$dN=-\Psi(M_P)\,dM_P$, where $\Psi(M_P)$ is the stellar luminosity
function (LF) which is constructed by counting the number of stars in
the survey volume per magnitude interval, and $P$ signifies an
observational photometric pass-band such as the $V$- or $I$-band.
Thus
\begin{equation}
\Xi(m) = -\Psi(M_P)\,({dm \over dM_P})^{-1}.
\label{eq:mf_lf}
\end{equation}
Note that the the minus sign comes-in because increasing mass leads to
decreasing magnitudes, and that the LF constructed in one photometric
pass band $P$ can be transformed into another band $P'$ by
\begin{equation}
\Psi(M_P) = {dN \over dM_{P'}} {dM_{P'} \over dM_P} = 
\Psi(M_{P'}) {dM_{P'} \over dM_P}
\label{eq:passband}
\end{equation}
if the function $M_{P'} = {\rm fn}(M_P)$ is known.  

Since the derivative of the stellar mass--luminosity relation (MLR),
$m(M_P)=m(M_P,Z,\tau,\mathbf{s})$, enters the calculation of the MF,
any uncertainties in stellar structure and evolution theory on the one
hand (if a theoretical MLR is relied upon) or in observational ML-data
on the other hand, will be magnified accordingly. This problem cannot
be avoided if the mass function is constructed by converting the
observed stellar luminosities one-by-one to stellar masses using the
MLR and then binning the masses, because the derivative of the MLR
nevertheless creeps-in through the binning process, because {\it equal
luminosity intervals are not mapped into equal mass intervals}.  The
dependence of the MLR on the star's chemical composition, $Z$, it's
age, $\tau$, and spin vector $\mathbf{s}$, is explicitly stated here,
since stars with fewer metals than the Sun are brighter (lower
opacity), main-sequence stars brighten with time and loose mass, and
rotating stars are dimmer because of the reduced internal
pressure. Mass loss and rotation are significant factors for
intermediate and especially high-mass stars \citep{MPV01}.

The IMF follows by correcting the observed number of main sequence
stars for the number of stars that have evolved off the main sequence.
Defining $t=0$ to be the time when the Galaxy that now has an age
$t=\tau_{\rm G}$ began forming, the number of stars per pc$^3$ in the
mass interval $m,m+dm$ that form in the time interval $t,t+dt$ is
$dN=\xi(m,t)\,dm\times b(t)\,dt$, where the expected time-dependence
of the IMF is explicitly stated (\S~\ref{sec:comppop}), and where
$b(t)$ is the normalised SFH, $(1/\tau_{\rm G}) \int_0^{\tau_{\rm
G}}b(t)\,dt = 1$.  Stars that have main-sequence life-times $\tau(m) <
\tau_{\rm G}$ leave the stellar population unless they were born
during the most recent time interval $\tau(m)$. The number density of
such stars with masses in the range $m,m+dm$ still on the main
sequence and the total number density of stars with $\tau(m) \ge
\tau_{\rm G}$, are, respectively,
\begin{equation}
\Xi(m) = 
   \xi(m){1\over \tau_{\rm G}} \times
   \left\{ 
   \begin{array}{l@{\quad\quad,\quad}l}
   \int_{\tau_{\rm G}-\tau(m)}^{\tau_{\rm G}} b(t)dt &
   \tau(m) < \tau_{\rm G},\\
   \int_0^{\tau_{\rm G}} b(t)\,dt & \tau(m) \ge \tau_{\rm G},
   \end{array}\right.
\label{eq:imf_pdmf}
\end{equation}
where the time-averaged IMF, $\xi(m)$, has now been defined. Thus, for
low-mass stars $\Xi=\xi$, while for a sub-population of massive stars
that has an age $\Delta t \ll \tau_{\rm G}$, $\Xi=\xi\,(\Delta
t/\tau_{\rm G})$ for those stars of mass $m$ for which the
main-sequence life-time $\tau(m)>\Delta t$, indicating how an observed
high-mass IMF in an OB association, for example, has to be scaled to
the Galactic-field IMF for low-mass stars.\footnote{assuming
continuity of the IMF} In this case the different spatial distribution
via different disk-scale heights of old and young stars also needs to
be taken into account, which is done globally by calculating the
stellar surface density in the MW disk \citep{MS79,Sc86}.  Thus we can
see that joining the cumulative low-mass star counts to the snap-shot
view of the massive-star IMF is non-trivial and affects the shape of
the IMF in the notorious mass range $\approx 0.8-3\,M_\odot$, where
the main-sequence life-times are comparable to the age of the MW disk.
In a star cluster or association with an age $\tau_{\rm
cl}\ll\tau_{\rm G}$, $\tau_{\rm cl}$ replaces $\tau_{\rm G}$ in
eq.~\ref{eq:imf_pdmf}.  Examples of the time-modulation of the IMF are
$b(t)=1$ (constant star-formation rate) or a Dirac-delta function,
$b(t)=\tau_{\rm cl} \times \delta(t-t_0)$ (all stars formed at the
same time $t_0$).

Often used is the ``logarithmic mass function'' (Table~\ref{tab:imfs} below),
\begin{equation}
\xi_{\rm L}(m) = \left( m\,{\rm ln}10\right)\, \xi(m),
\end{equation}
where $dN=\xi_{\rm L}(m)\,dlm$ is the number of stars with mass in the
interval $lm,lm+dlm$ ($lm \equiv {\rm log}_{10}m$)\footnote{Note that
\citet{Sc86} calls $\xi_{\rm L}(m)$ the {\it mass function} and
$\xi(m)$ the {\it mass spectrum}. }.

\section{The massive stars}
\label{sec:massst}

Studying the distribution of massive stars is complicated because they
radiate most of their energy at far-UV wavelengths that are not
accessible from Earth, and through their short main-sequence
life-times, $\tau$, that remove them from star-count surveys.  For
example, a $85\,M_\odot$ star cannot be distinguished from a
$40\,M_\odot$ star on the basis of $M_V$ alone \citep{M98,Massey03}.
Constructing $\Psi(M_V)$ in order to arrive at $\Xi(m)$ for a
mixed-age population does not lead to success if optical or even
UV-bands are used.  Instead, spectral classification and broad-band
photometry for estimation of the reddening on a star-by-star basis has
to be performed to measure the effective temperature, $T_{\rm eff}$,
and the bolometric magnitude, $M_{\rm bol}$, from which $m$ is
obtained allowing the construction of $\Xi(m)$ directly (whereby
$\Psi(M_{\rm bol})$ and $\Xi(m)$ are related by eq.~\ref{eq:mf_lf}).
Having obtained $\Xi(m)$ for a population under study, the IMF follows
by applying eq.~\ref{eq:imf_pdmf}.  A more straight-forward method
often used (e.g. \citealt{Hill97, M98, Massey03}) is to evolve each
measured stellar mass to its initial value using theoretical stellar
evolution tracks, and to construct $\xi(m)$ from this set of masses.

\begin{figure}
\begin{center}
\rotatebox{0}{\resizebox{0.75 \textwidth}{!}{\includegraphics{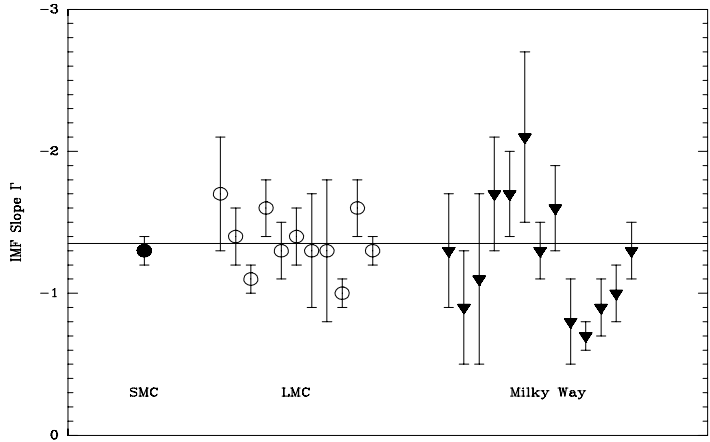}}}
\vskip -4mm
\caption{\small{ The IMF slope $\Gamma=1-\alpha$ determined in a
homogeneous manner by \citet{Massey03} for OB associations and
clusters in the MW, LMC and SMC.  The Small Magellanic Cloud (SMC) has
a metallicity $Z=0.005$, the Large Magellanic Cloud (LMC) has
$Z=0.008$ and the Milky Way (MW) has $Z=0.018$ within a distance of
3~kpc around the Sun.  With kind permission from \cite{Massey03}.  }}
\label{fig:kroupa_figmassey}
\end{center}
\end{figure}

\cite{Massey03} stresses that studies that only rely on broad-band
optical photometry consistently arrive at IMFs that are significantly
steeper with $\alpha_3\approx3$ (see eq.~\ref{eq:imf} below), rather
than $\alpha_3=2.2\pm0.1$ consistently found for a wide range of
stellar populations.  Indeed, the application of the same methodology
by Massey on a number of young populations of different metallicity
and density shows a remarkable uniformity of the IMF above about
$10\,M_\odot$ (Fig.~\ref{fig:kroupa_figmassey}).

The available IMF measurements do not take into account the bias
through unresolved systems which may be substantial since the
proportion of multiple stars is higher for massive stars than for
low-mass Galactic-field stars (e.g. \citealt{Duchene01}).  For
example, in the Orion Nebula Cluster (ONC) each massive star has, on
average, 1.5 companions \citep{Preibisch99}, while in the cluster
NGC~6231 \cite{GM01} find that 80~\% of all O~stars are
radial-velocity binaries.  \cite{SR91} estimate the binary-star bias
on the IMF for stars in the mass range 2~to $14\,M_\odot$ assuming
each star has one companion.  The IMF would steepen to the Scalo value
$\alpha\approx2.7$ for a measured $\alpha=2.3$. This correction,
however, depends on the distribution of companions that is not yet
known very well \citep{Preibisch99,Duchene01}. A larger value,
$\alpha\approx3\pm0.1$, is also suggested from a completely
independent but indirect approach relying on the distribution of
ultra-compact HII regions in the Galaxy \citep{Cass00,OC98}, but this
may be a result of the composite nature of galaxy-wide populations
(\S~\ref{sec:comppop}).  Clearly, the effect multiplicity has on the
massive-star IMF needs further exploration.

\begin{figure}
\begin{center}
\rotatebox{0}{\resizebox{0.55 \textwidth}{!}{\includegraphics{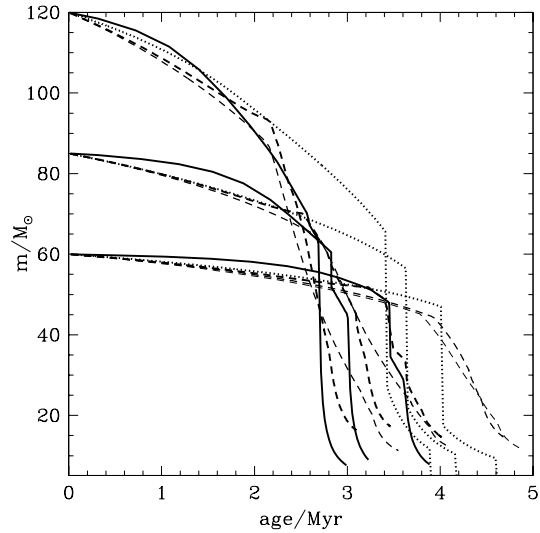}}}
\vskip -20mm
\caption{\small{The mass-evolution of massive stars according to
theoretical work: {\it Solid lines}: Geneva models
\citep{Schaller_etal92}, {\it Dashed lines}: Geneva with rotation
\cite[(thick: none, medium: 300 km/s, thin: 500 km/s;]{MM03} {\it
Dotted lines}: stellar evolution package from Hurley based on the Pols
non-rotating models \citep{HPT00}.  From \cite{WK05a}.  }}
\label{fig:kroupa_massevol}
\end{center}
\end{figure}

\begin{figure}
\begin{center}
\rotatebox{0}{\resizebox{0.55
\textwidth}{!}{\includegraphics{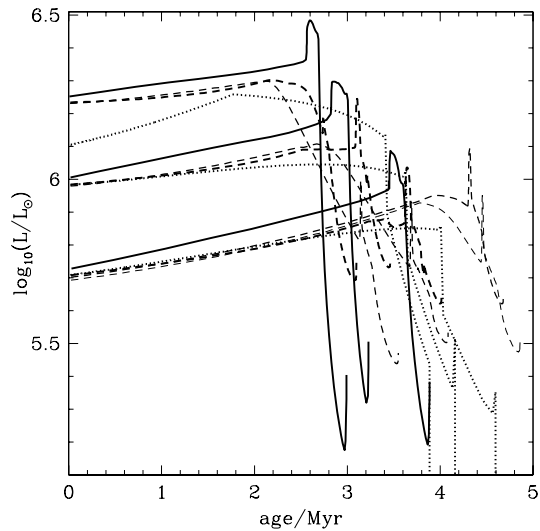}}}
\vskip -20mm
\caption{\small{ The luminosity-evolution of massive stars according
to theoretical work. Line-types and initial masses are as in
Fig.~\ref{fig:kroupa_massevol}.  From \citet{WK05a}}}
\label{fig:kroupa_lumevol}
\end{center}
\end{figure}

Massive main-sequence stars have substantial winds flowing outwards
with velocities of a few~100 to a few~1000~km/s \citep{KP00}.  For
example, $10^{-6.5}<\dot{M}<10^{-6}\,M_\odot$/yr for $m=35\,M_\odot$
with $\tau=4.5$~Myr \citep{Garcia96a}, and
$10^{-5.6}<\dot{M}<10^{-5.8}\,M_\odot$/yr for $m=60\,M_\odot$ with
$\tau=3.45$~Myr \citep{Garcia96b}.  More problematical is that stars
form rapidly rotating and are sub-luminous as a result of reduced
internal pressure. But they decelerate significantly during their
main-sequence life-time owing to the angular-momentum loss through
their winds and become more luminous more rapidly than non-rotating
stars \citep{MM00}.  A comparison of such models is available in
Figs.~\ref{fig:kroupa_massevol} and~\ref{fig:kroupa_lumevol}.
Evidently, for ages less than 2.5~Myr the models deviate only by
5--13~\% from each other in mass, luminosity or temperature
\citep{WK05a}. Large deviations are evident for advanced stages of
evolution though.

The mass--luminosity relation for a population of stars that have a
range of ages is therefore broadened making mass estimates from
$M_{\rm bol}$ uncertain by up to 50~per cent \citep{MPV01}, a bias
that probably needs to be taken into account more thoroughly in the
derivations of the IMF. Another problem is that
$m\simgreat40\,M_\odot$ stars may finish their assembly after burning
a significant proportion of their central~H so that a zero-age-main
sequence may not exist for massive stars \citep{MB01}. However, the
agreement between slowly-rotating tidally-locked massive O-type
binaries with standard non-rotating theoretical stellar models is very
good \citep{MPV01}.

\subsection{The maximum stellar mass and cluster formation}
\label{sec:maxlim}
\subsubsection{A brief history:}

The empirically determined range of stellar masses poses important
constraints on the physics of stellar formation, structure and stellar
evolution, as well as on the feedback energy injected into a galaxy's
atmosphere by a population of brand-new stars. The physical limit at
low masses is now well established (\S~\ref{sec:bds}), and an
upper mass limit appears to have been found recently.

A theoretical physical limitation to stellar masses has been known
since many decades.  \citet{Edd26} calculated the limit which is
required to balance radiation pressure and gravity, the {\it Eddington
limit}: $L_{\rm Edd}/L_{\odot}\,\approx\,3.5\,\times\,10^{4}\,m /
M_{\odot}$. Hydrostatic equilibrium will fail if a star of a certain
mass $m$ has a theoretical luminosity that exceeds this limit, which
is the case for $m\simgreat 60\,M_{\odot}$. It is not clear if stars
above this limit cannot exist, as massive stars are not fully
radiative but have convective cores. But more massive stars will loose
material rapidly due to strong stellar winds.  \citet{SH59} inferred a
limit of $\approx\,60\,M_{\odot}$ beyond which stars should be
destroyed due to pulsations. But later studies suggested that these
may be damped \citep{BM94}.  \cite{St92} showed that the limit
increases to $m_{\rm max*}\approx 120-150\,M_\odot$ for more recent
Rogers-Iglesia opacities and for metallicities [Fe/H]$\approx0$. For
[Fe/H]$\approx -1$, $m_{\rm max*}\approx 90\,M_\odot$. A larger
physical mass limit at higher metallicity comes about because the
stellar core is more compact, the pulsations driven by the core having
a smaller amplitude, and because the opacities near the stellar
boundary can change by larger factors than for more metal-poor stars
during the heating and cooling phases of the pulsations thus damping
the oscillations. Larger physical mass limits are thus allowed to
reach pulsational instability.

Related to the pulsational instability limit is the problem that
radiation pressure also opposes accretion for proto-stars that are
shining above the Eddington luminosity.  Therefore the question
remains how stars more massive than 60 $M_{\odot}$ may be formed.
Stellar formation models lead to a mass limit near $40-100\,M_\odot$
imposed by feedback on a spherical accretion envelope
(\citealt{Kahn74}; \citealt{Wolf86}; \citealt{Wolf87}). Some
observations suggest that stars may be accreting material in discs and
not in spheres (e.g. \citealt{CHK04}). The higher density of the
disc-material may be able to overcome the radiation at the equator of
the proto-star. But it is unclear if the accretion-rate can be boosted
above the mass-loss rate from stellar winds by this mechanism.
Theoretical work on the formation of massive stars through
disk-accretion with high accretion rates thereby allowing thermal
radiation to escape pole-wards (e.g. \citealt{Nakano89};
\citealt{JA96}) indeed lessen the problem and allow stars with larger
masses to form.

Another solution proposed is the merging scenario. In this case
massive stars form through the merging of intermediate-mass
proto-stars in the cores of dense stellar clusters driven by
core-contraction due to very rapid accretion of gas with low specific
angular momentum, thus again avoiding the theoretical feedback-induced
mass limit (\citealt{Bonn98}; \citealt{SPH00}).  It is unclear though
if the very large central densities required for this process to act
are achieved in reality, but it should be kept in mind that an
observable cluster is, by necessity, exposed from its natal cloud and
is therefore likely to be always observed in an expanding phase
\citep{Kr_paris05}.

The search for a possible maximal stellar mass can only be performed
in massive, star-burst clusters that contain sufficiently many stars
to sample the stellar initial mass function beyond $100\,M_\odot$.
Observationally, the existence of a finite physical stellar mass limit
was not evident until very recently. Indeed, observations in the
1980's of R136 in the Large Magellanic Cloud (LMC) suggested this
object to be one single star with a mass of about
$2000-3000\,M_\odot$.  \citet{WB85} for the first time resolved the
object into eight components using digital speckle interferometry,
therewith proving that R136 is a massive star cluster rather than one
single super-massive star.  The evidence for any physical upper mass
limit became very uncertain, and \citet{Elm97} stated that
``observational data on an upper mass cutoff are scarce, and it is not
included in our models [of the IMF from random sampling in a turbulent
fractal cloud]''. Although \citet{MH98} found stars in R136 with
masses ranging up to $140-155\,M_\odot$, \citet{Massey03} explains that
the observed limitation is statistical rather than physical.  We refer
to this as the {\it Massey assertion}, i.e. that $m_{\rm
max*}=\infty$. Meanwhile, \citet{Setal99} found, from their
observations, a probable upper mass limit in the LMC near about
$130\,M_\odot$, but they did not evaluate the statistical significance
of this suggestion. \citet{Figer02} discussed the apparent cut-off of the
stellar mass-spectrum near $150\,M_\odot$ in the Arches cluster near
the Galactic centre, but again did not attach a statistical analysis
of the significance of this observation.  \citet{Elm00} also noted
that random sampling from an unlimited IMF for all star-forming
regions in the Milky Way (MW) would lead to the prediction of stars
with masses $\simgreat 1000\,M_\odot$, unless there is a rapid
turn-down {\it in the IMF} beyond several hundred~$M_\odot$. However,
he also stated that no upper mass limit to star formation has ever
been observed, a view also emphasised by \citet{Larson03}.

Thus, while theory clearly expected a physical stellar upper mass
limit, the observational evidence in support of this was very unclear.
This, however, changed in~2004.

\subsubsection{Empirical results}

Given the observed rather sharp drop-off of the IMF in R136 near
$150\,M_\odot$, \citet{WK04} studied the {\it Massey assertion} in
some detail.

R136 has an age $\le$~2.5~Myr \citep{MH98} which is young enough such
that stellar evolution will not have removed stars through supernova
explosions. It has a metallicity of [Fe/H]$\approx -0.5$~dex
\citep{deBoer_etal85}.

From the radial surface density profile \citet{Setal99} estimated
there to be 1350~stars with masses between 10~and $40\,M_\odot$ within
20~pc of the 30~Doradus region, within the centre of which lies R136.
\citet{MH98} and \citet{Setal99} found that the IMF can be
well-approximated by a Salpeter power-law with exponent $\alpha=2.35$
for stars in the mass range 3~to $120~M_\odot$. This corresponds to
8000 stars with a total mass of $0.68\times 10^5\,M_\odot$.
Extrapolating down to $0.1\,M_\odot$ the cluster would contain
$8\times 10^5$~stars with a total mass of
$2.8\times10^5\,M_\odot$. Using a standard IMF with a slope of
$\alpha=1.3$ (instead of the Salpeter value of 2.35) between 0.1 and
$0.5\,M_\odot$ this would change to $3.4\times 10^5$~stars with a
combined mass of $2\times 10^5\,M_\odot$, for an average mass of
$0.61\,M_\odot$ over the mass range $0.1-120\,M_\odot$.  Based on the
observations by \citet{Setal99}, \citet{WK04} assumed that R136 has a
mass in the range $5 \times 10^{4} \le M_{\rm R136}/M_{\odot} \le 2.5
\times 10^{5}$. This mass range can be used to investigate the
expected number of stars above mass $m$,
\begin{equation}
N(>m) = \int_{m}^{m_{\rm max*}} \xi(m')\,dm',
\label{eq:Nm}
\end{equation}
with the mass in stars of the whole (originally embedded) cluster
being
\begin{equation}
M_{\rm ecl} = \int_{m_{\rm low}}^{m_{\rm max*}} m'\,\xi(m')\,dm',
\label{eq:Mecl1}
\end{equation}
where $m_{\rm low}\,=\,0.01\,M_{\odot}$ and $m_{\rm max*}=\infty$ (the
Massey assertion). Here the assumption is that the cluster is still
compact despite having-blown out its residual gas. There are two
unknowns ($N(>m)$ and $k$) that can be solved for using the two
equations above.

Using the {\it standard stellar IMF} (eq.~\ref{eq:imf} below), $N(>m)$
is plotted in Fig.~\ref{fig:pk_numbs} for the two mass estimates of
the cluster. The solid vertical line indicates $150\,M_{\odot}$, the
approximate maximum mass observed in R136 \citep{MH98}.  \citet{WK04}
found that $N(>150\,M_\odot)=40$~stars are missing if $M_{\rm
ecl}=2.5\times 10^5\,M_\odot$, while $N(>150\,M_\odot)= 10$~stars are
missing if $M_{\rm ecl}=5\times 10^4\,M_\odot$. The probability that
no stars are observed although 10 are expected, assuming $m_{\rm
max*}=\infty$, is $P=4.5\times 10^{-5}$. \citet{WK04} concluded that
the observations of the massive stellar content of R136 suggest a
physical stellar mass limit near $m_{\rm max*}=150\,M_\odot$.

\begin{figure}
\begin{center}
\rotatebox{0}{\resizebox{0.5 \textwidth}{!} {\includegraphics[width=8cm]{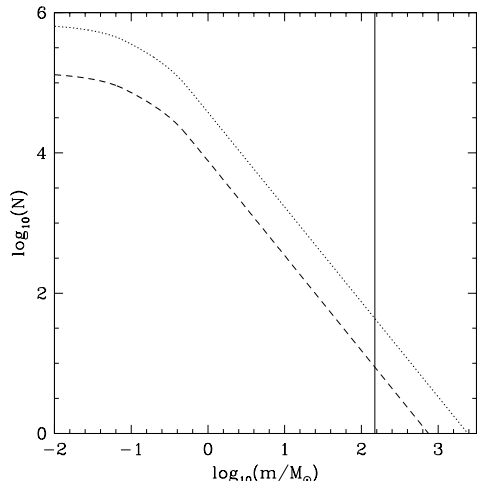}}}
\vspace*{-2.0cm}
\caption{Number of stars above mass $m$ for R136 with different mass
  estimates (dotted line: $M_{\rm R136} = 2.5 \times 10^5\, M_\odot$,
  dashed line: $M_{\rm R136} = 5 \times 10^4\, M_\odot$,
  \citep{Setal99}. The vertical solid line marks
  $m=150\,M_\odot$. Taken from \citet{WK04}.}
\label{fig:pk_numbs}
\end{center}
\end{figure}

Furthermore, \citet{WK04} deduced that the {\it Massey assertion} would
be correct for both cluster masses if the IMF had a slope $\alpha_{3}
\simgreat 2.8$. Such a steep slope would make the observed limit
consistent with random selection from the IMF, and it may be the true
power-law index if unresolved multiple systems among O~stars are
corrected for, but this awaits a detailed study.  A further caveat
comes from unresolved multiple systems which would allow an $m_{\rm
max*, true}$ as small as $\approx m_{\rm max*}/2$ if $\alpha_3\approx
2.35$.

Similar results were obtained by Don Figer for the Arches cluster.
The Arches is a star-burst cluster within 30~pc in projected distance
from the Galactic centre. It has a mass $M\approx 1\times
10^5\,M_\odot$ \citep{Bosch_etal01}, age $2-2.5$~Myr
\citep{Najarro_etal04} and [Fe/H]$\approx 0$
\citep{Najarro_etal04}. It is thus a counterpart to R136 in that the
Arches is metal rich and was born in a very different tidal
environment to R136.

Using his HST observations of the Arches, \citet{Fi05} performed the
same analysis as \citet{WK04} did for R136. The Arches appears to be
dynamically evolved, with substantial mass loss through the strong
tidal forces \citep{Port_etal02} and the stellar mass
function with $\alpha=1.9$ is thus flatter than the Salpeter
IMF. Using his updated IMF measurement, Figer calculated the expected
number of stars above $150\,M_\odot$ to be 33, while a Salpeter IMF
would predict there to be 18 stars. Observing no stars but expecting
to see~18 has a probability of $P=10^{-8}$, again strongly suggesting
$m_{\rm max*}\approx 150\,M_\odot$.

Given the importance of knowing if a finite physical upper mass limit
exists and how it varies with metallicity, \citet{OC05} studied the
massive-star content in 9~clusters and OB associations in the MW, the
LMC and the SMC.  They predicted the expected masses of the most
massive stars in these clusters for different upper mass limits ($120,
150, 200, 1000\,{\rm and}\,10000\,M_{\odot}$). For all populations
they found that the observed number of massive stars supports with
high statistical significance the existence of a general upper mass
cutoff in the range $m_{\rm max*}\in
(120,\,200\,M_{\odot})$.\footnote{More recent work on the physical
upper mass limit can be found in \cite{Koen06,Jesus07}.}

The general indication thus is that a physical stellar mass limit near
$150\,M_\odot$ seems to exist. While biases due to unresolved
multiples that may steepen the IMF and/or reduce the true maximal mass
need to be studied further, the absence of variations of $m_{\rm
max*}$ with metallicity poses a problem.  A constant $m_{\rm max*}$
would only be apparent for a true variation as proposed by the
theoretical models, {\it if metal-poor environments have a larger
stellar multiplicity}, the effects of which would have to compensate
the true increase of $m_{\rm max*}$ with metallicity.

\subsubsection{Maximal stellar mass in clusters}
\begin{figure}
\begin{center}
\rotatebox{0}{\resizebox{0.55 \textwidth}{!}{\includegraphics[width=8cm]{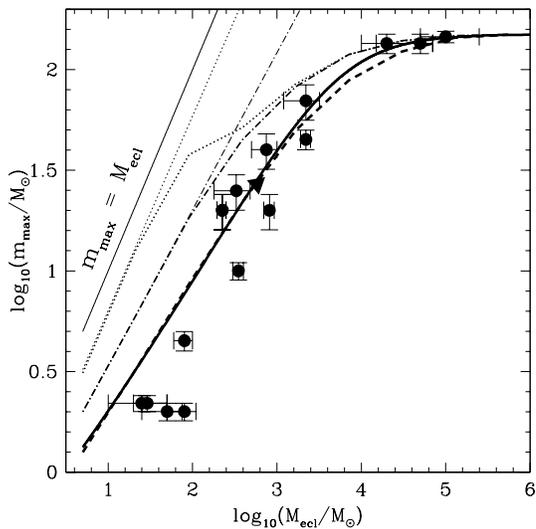}}}
\vspace*{-20mm}
\caption{The {\it thick solid line} shows the dependence of the mass
of the most-massive star in a cluster on the cluster mass according to
the semi-analytical model.  The {\it thick dashed line} shows the mean
maximum stellar mass for sorted sampling. The {\it dot-dashed lines}
are mass-constrained random-sampling results with a physical upper
mass limit of $m_{\rm max*}=150\,M_{\odot}$ ({\it thick line}) and
$10^{6}\,M_{\odot}$ ({\it thin line}). Pure random sampling models are
plotted as {\it dotted lines}. The {\it thick} one is sampled to
$m_{\rm max*}\,=\,150\,M_{\odot}$ while the {\it thin} one up to
$10^{6}\,M_{\odot}$. The {\it thin solid line} shows the identity
relation, where a ``cluster'' consists only of one star. The {\it
dots} with error bars are observed clusters, while the {\it triangle}
is a result from a star-formation simulation with an SPH code
\citep{BBV03}. Taken from \citet{WK05a}.}
\label{fig:pk_mmaxf}
\end{center}
\end{figure}

Above we have seen that there seems to exist a universal physical
stellar mass limit. However, an elementary argument suggests that
star-clusters must also limit the masses of their constituent stars: A
pre-star-cluster gas core with a mass $M_{\rm core}$ can, obviously,
not form stars with masses $m>\epsilon\, M_{\rm core}$, where
$\epsilon\approx 0.33$ is the star-formation efficiency
\citep{Lada_Lada03}. Thus, given a freshly hatched cluster with
stellar mass $M_{\rm ecl}$, stars in that cluster cannot surpass
masses $m_{\rm max}=M_{\rm ecl}$, which is the identity relation
corresponding to a ``cluster'' consisting of one massive
star. Assuming the stellar IMF is a continuous density distribution
function and that clusters are filled with stars distributed according
to the stellar IMF, this can be generalised by stating that each
cluster can have only one most massive star,
\begin{equation}
1 = \int_{m_{\rm max}}^{m_{\rm max*}} \xi(m')\,dm',
\label{eq:mm}
\end{equation}
with
\begin{equation}
M_{\rm ecl}(m_{\rm max}) = \int_{m_{\rm low}}^{m_{\rm max}}
m'\,\xi(m')\,dm'
\label{eq:Mecl}
\end{equation}
as a further condition, as above. These two equations need to be
solved numerically and give the semi-analytical relation $m_{\rm
max}\,=\,{\cal F}(M_{\rm ecl})$ \citep{WK04}.  It is plotted in
Fig.~\ref{fig:pk_mmaxf} as the thick-solid curve.

A compilation of clusters from the literature for which the cluster
mass and the initial mass of the heaviest star can be estimated
\citep{WK05a} shows that the cluster mass indeed appears to have a
limiting influence on the stellar mass within it. The observational
data are plotted in Fig.~\ref{fig:pk_mmaxf}, finding rather excellent
agreement with the semi-analytical description above.

However, it would be undisputed that a stochastic sampling effect from
the IMF must be present when stars form. This can be mimicked in the
computer by performing various Monte-Carlo experiments \citep{WK05a}.  The
Monte-Carlo experiments are conducted in three different ways,
\begin{itemize}
\itemsep=-1mm
\item[-] pure random sampling ({\it random sampling})
\item[-] mass constrained random sampling ({\it constrained sampling})
\item[-] mass constrained random sampling with sorted adding ({\it sorted sampling})
\end{itemize}

\noindent For the {\it random sampling model}, 10~million clusters are randomly
taken from a cluster distribution with a power-law index of $\beta_N =
2.35$ between 12~and $2.5\times10^7$~stars. The relevant distribution
function is the embedded-cluster star-number function (ECSNF),
\begin{equation}
dN_{\rm ecl} \propto N^{-\beta_N},
\end{equation}
which is the number of clusters containing $N \in [N',N'+dN')$~stars.
Each cluster is then filled with $N$ stars randomly from the standard
IMF (eq.~\ref{eq:imf} without a mass limit, or by
imposing the physical stellar mass limit,
$m\,\le\,150\,M_{\odot}$. The stellar masses are added to get the
cluster mass, $M_{\rm ecl}$. For each cluster the maximal stellar mass
is searched for. For each cluster in a mass bin $M_{\rm ecl} - \Delta
M_{\rm ecl}/2, M_{\rm ecl} + \Delta M_{\rm ecl}/2$ the average $m_{\rm
max}$ is calculated, and the set of average $m_{\rm max}$ values
define the relation
\begin{equation}
m_{\rm max} = m_{\rm max}^{\rm ran}(M_{\rm ecl}).
\end{equation}

For the {\it constrained sampling model}, $5\times 10^7$~clusters are
randomly taken from the embedded-cluster mass function (ECMF),
\begin{equation}
\xi_{\rm ecl}(M_{\rm ecl}) \propto M_{\rm ecl}^{-\beta}, 
\label{eq:ECMF}
\end{equation} 
 between $5\,M_{\odot}$ (the minimal, Taurus-Auriga-type, star-forming
 ``cluster'' counting $\approx$ 15 stars) and $10^{7}\,M_{\odot}$ (an
 approximate maximum mass for a single stellar population that
 consists of one metallicity and age, \citealt{WKL04}) and again with
 $\beta=2.35$. Note that $\beta_N\approx\beta$ because the ECSNF and
 the ECMF only differ by a nearly-constant average stellar mass. Then
 stars are taken randomly from the standard IMF and added until they
 reach or surpass the respective cluster mass, $M_{\rm
 ecl}$. Afterwards the clusters are searched for their maximum stellar
 mass.  For each cluster in a mass bin $M_{\rm ecl} - \Delta M_{\rm
 ecl}/2, M_{\rm ecl} + \Delta M_{\rm ecl}/2$ the average $m_{\rm max}$
 is calculated, and the set of average $m_{\rm max}$ values define the
 relation
\begin{equation}
m_{\rm max} = m_{\rm max}^{\rm con}(M_{\rm ecl}).
\end{equation}

For the {\it sorted sampling model} again $5\times10^7$~clusters are
randomly sampled from the ECMF (eq.~\ref{eq:ECMF}) between 5
$M_{\odot}$ and $10^{7}\,M_{\odot}$ and with $\beta=2.35$. However,
this time the number $N$ of stars which are to populate the cluster is
estimated from $N=M_{\rm ecl}/m_{\rm av}$, where $m_{\rm
av}\,=\,0.36\,M_{\odot}$ is the average stellar mass for the standard
IMF (eq.~\ref{eq:imf}) between 0.01 $M_{\odot}$ and 150
$M_{\odot}$. These stars are added to give $M_{\rm ecl, ran}$,
\[M_{\rm ecl, ran} = \sum_{\rm N} m_{i},
\]
such that $m_{i} \le m_{i+1}$. If $M_{\rm ecl, ran} < M_{\rm ecl}$ in
this first step, an additional number of stars, $\Delta N$, is picked
randomly from the IMF, where $\Delta N=(M_{\rm ecl} - M_{\rm
ran})/m_{\rm av}$ (we assume $m_{\rm av}=\;$constant). Again these
stars are added to obtain an improved estimate of the desired cluster
mass,
\[^{2}M_{\rm ecl, ran} = \sum_{\rm N + \Delta N} m_{i}, \quad m_i \le m_{i+1}.
\]
This is done such that $^{2}M_{\rm ecl, ran} \approx M_{\rm ecl}$ (for
details of the method see \citealt{WK05a}).  The procedure is repeated
until all clusters from the ECMF are 'filled'. They are then also
searched for the most massive star in each cluster, as above.  For
each cluster in a mass bin $M_{\rm ecl}- \Delta M_{\rm ecl}/2, M_{\rm
ecl} + \Delta M_{\rm ecl}/2$ the average $m_{\rm max}$ is calculated,
and the set of average $m_{\rm max}$ values define the relation
\begin{equation}
m_{\rm max} = m_{\rm max}^{\rm sort} (M_{\rm ecl}).
\end{equation}

All three relations are plotted in Fig.~\ref{fig:pk_mmaxf}. We noted
already that the observations follow the semi-analytic relation
remarkably well. Furthermore, Fig.~\ref{fig:pk_mmaxf} also suggests that
the different Monte-Carlo schemes can be selected for. Thus, the
sorted-sampling algorithm leads to virtually the same results as the
semi-analytical relation, and it fits the data very well indeed. The
correspondence of the sorted-sampling algorithm to the semi-analytical
result is not really surprising, because the algorithm is Monte-Carlo
integration of the same problem.  The constrained-sampling and
random-sampling algorithms, on the other hand, can be excluded with
very high confidence by performing statistical tests on the
observational data that are reported in detail in \citet{WK05a}.
 
On a historical note, \citet{Larson82} had pointed out that more
massive and dense clouds correlate with the mass of the most massive
stars within them and he estimated that $m_{\rm max}=0.33\,M_{\rm
cloud}^{0.43}$ (masses are in $M_\odot$). An updated relation was
derived by \citet{Larson03} by comparing $m_{\rm max}$ with the
stellar mass in a few clusters, $m_{\rm max}\approx 1.2\,M_{\rm
cluster}^{0.45}$. Both are flatter than the semi-analytical relation,
and therefore do not fit the data in Fig.~\ref{fig:pk_mmaxf} as well
\citep{WK05a}. \citet{Elm83} constructed a relation between cluster
mass and its most massive star based on an assumed equivalence between
the luminosity of the cluster population and its binding energy, for a
Miller-Scalo IMF. This function is even shallower than the one
estimated by \citet{Larson03} relation. Assuming $m_{\rm
max*}=\infty$, \citet{Elm00} solved eqs~\ref{eq:mm} and~\ref{eq:Mecl}
above for a single Salpeter power-law stellar IMF finding a $m_{\rm
max}(M_{\rm ecl})$ relation quite consistent with the data in
Fig.~\ref{fig:pk_mmaxf} \citep{WK05a}.

\subsubsection{Implications: stellar astrophysics and the formation of star clusters}
\label{sec:stmass_clmass}

We are now in the happier situation that a physical stellar mass limit
seems to have been found. But the absence of clear variation of this
limit with metallicity poses a potential problem, although it may be
too early to make definite statements. Further observational work on
many more very young and massive clusters is needed to ascertain the
findings reported here, and to quantify the multiplicity properties of
massive stars, as noted above.

That the sorted-sampling algorithm for making star clusters fits the
observational maximal-stellar-mass--star-cluster-mass data so well
would appear to imply that {\it clusters form in an organised
fashion}. The physical interpretation of the algorithm (i.e. of the
Monte-Carlo integration) is that as a pre-cluster core contracts under
self gravity the gas densities increase and local density fluctuations
in the turbulent medium lead to low-mass star formation, perhaps
similar to what is seen in Taurus-Aurigae. As the contraction proceeds
and before feedback from young stars begins to disrupt the cloud,
star-formation activity increases in further density fluctuations with
larger amplitudes thereby forming more massive stars. The process
stops when the most massive stars that have just formed supply
sufficient feedback energy to disrupt the cloud \citep{Elm83}. Thus,
less-massive pre-cluster cloud-cores would die at a lower maximum
stellar mass than more massive cores. But in all cases stellar masses
are limited, $m\le m_{\rm max}(M_{\rm ecl}) \le m_{\rm max*}$.

This scenario is nicely consistent with the hydrodynamic cluster
formation calculations presented by \citet{BBV03} and \citet{BVB04},
as is reported in more detail in \cite{WK05a}.  We note here that
\citet{BVB04} found their theoretical clusters to form hierarchically
from smaller sub-clusters, and together with continued competitive
accretion this leads to the relation $m_{\rm max}\propto M_{\rm
ecl}^{2/3}$ in excellent agreement with the compilation of
observational data.  While this agreement is stunning, the detailed
outcome of the currently available SPH modelling in terms of stellar
multiplicities is not right \citep{Goodwin_etal04,GK05}, and feedback
that ultimately dominates the process of star-formation, given the
generally low star-formation efficiencies observed in cluster-forming
volumes, is not yet incorporated in the modelling. This poses a severe
computational hurdle because radiation transfer requires the correct
treatment of atomic processes in remote regions of space that are
exchanging radiation (eg. \citealt{Baes_etal05}).

\subsubsection{Caveats}
Unanswered questions regarding the formation and evolution of massive
stars remain. There may be stars with $m\,\ge\,m_{\rm max *}$ which
implode ``invisibly'' after 1 or 2 Myr.  The explosion mechanism
sensitively depends on the presently still rather uncertain mechanism
for shock revival after core collapse (e.g. \citealt{Janka01}).  Since
such stars would not be apparent in massive clusters older than 2~Myr
they would not affect the empirical maximal stellar mass, and $m_{\rm
max*, true}$ would be unknown at present.

Furthermore, and as stated already above, stars are often in multiple
systems. Especially massive stars seem to have a binary fraction of
80\% or even larger \citep{GM01} and apparently tend to be in binary
systems with a preferred mass-ratio near unity. Thus, if all O~stars
would be in equal-mass binaries, then $m_{\rm max *\,true} \approx
m_{\rm max *}/2$.

Finally, it is disconcerting that $m_{\rm max
  *}\,\approx\,150\,M_{\odot}$ appears to be the same for
low-metallicity environments ([Fe/H]$ = -0.5$, R136) and metal-rich
environments ([Fe/H]$ = $0, Arches), in apparent contradiction to the
theoretical values \citep{St92}. Clearly, this issue needs further
study.

\section{Low-mass stars}
\label{sec:lmst}

There are three well-tried approaches to determine $\Psi(M_V)$ in
eq.~\ref{eq:mf_lf} \citep{K01b}.  The first two are applied to
Galactic-field stars, and the third to star clusters. The sample of
Galactic-field stars close to the Sun is especially important because
it is the most complete and well-studied stellar sample at our
disposal.

\subsection{Galactic-field stars} 

Galactic-field stars have an average age of about 5~Gyr and represent
a mixture of many star-formation events. The IMF deduced for these is
therefore a time-averaged composite IMF. For $m\simless 1.3\,M_\odot$
the composite IMF equals the simple-stellar IMF
(\S~\ref{sec:comppop}), and so it is an interesting quantity for at
least two reasons: For the mass-budget of the Milky-Way disk, and as a
bench-mark against which the IMFs measured in presently occurring
star-formation events can be compared to distill possible variations
about the mean.

The first and most straightforward method consists of creating a local
volume-limited catalogue of stars. Completeness of the modern {\it
Jahreiss--Gliese Catalogue of Nearby Stars} extends to about 25~pc for
$m\simgreat0.6\,M_\odot$, trigonometric distances having been measured
using the Hipparcos satellite, and only to about 5~pc for less massive
stars for which we still rely on ground-based trigonometric parallax
measurements\footnote{\label{fn:compl} Owing to the poor statistical
definition of $\Psi_{\rm near}(M_V)$ for $M_V\simgreat 10, m\simless
0.5$, it is important to increase the sample of nearby stars, but
controversy exists as to the maximum distance to which the VLMS census
is complete. Using spectroscopic parallax it has been suggested that
the local census of VLMSs is complete to within about 15~\% to
distances of 8~pc and beyond \citep{RG97}.  However, Malmquist bias
allows stars and unresolved binaries to enter such a flux-limited
sample from much larger distances \citep{K01b}.  The increase of the
number of stars with distance using trigonometric distance
measurements shows that the nearby sample becomes incomplete for
distances larger than 5~pc and for $M_V>12$ \citep{J94,Henry97}. The
incompleteness in the northern stellar census beyond 5~pc and within
10~pc amounts to about 35~\% \citep{Hetal03}, and recently discovered
companions (e.g. \citealt{Detal99,Beuzit01}) to known primaries in the
distance range $5<d<12$~pc indeed suggest that the extended sample may
not yet be complete.  Based on the work by \cite{Retal03a} and
\cite{Retal03b}, \cite{Luhman04} however argues that the incompleteness
is only about 15~\%.}.  The advantage of the LF, $\Psi_{\rm
near}(M_V)$, created using this catalogue is that virtually all
companion stars are known, that it is truly distance limited and that
the distance measurements are direct.

The second method is to make deep pencil-beam surveys using
photographic plates or CCD cameras to extract a few hundred low-mass
stars from a hundred-thousand stellar and galactic images.  This
approach leads to larger stellar samples, especially so since many
lines-of-sight into the Galactic field ranging to distances of a
few~100~pc to a~few~kpc are possible. The disadvantage of the LF,
$\Psi_{\rm phot}(M_V)$, created using this technique is that the
distance measurements are indirect relying on photometric
parallax. Such surveys are flux limited rather than volume limited and
pencil-beam surveys which do not pass through virtually the entire
stellar disk are prone to Malmquist bias \citep{SIP}. This bias results
from a spread of luminosities of stars that have the same colour
because of their dispersion of metallicities and ages, leading to
intrinsically more luminous stars entering the flux-limited sample and
thus biasing the inferred absolute luminosities and the inferred
stellar space densities.  Furthermore, binary systems are not resolved
in the deep surveys.  

\begin{figure}
\begin{center}
\rotatebox{0}{\resizebox{0.8
\textwidth}{!}{\includegraphics{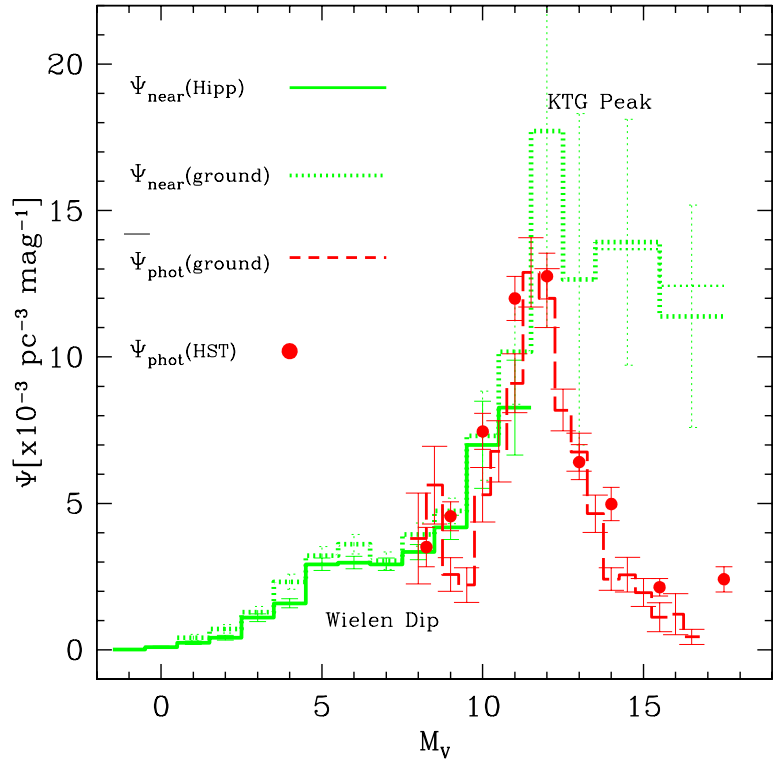}}}
\vskip -28mm
\caption{\small{Stellar luminosity functions (LFs) for
solar-neighbourhood stars. The photometric LF corrected for Malmquist
bias and at the mid-plane of the Galactic disk ($\Psi_{\rm phot}$) is
compared with the nearby LF ($\Psi_{\rm near}$). The average,
ground-based $\overline{\Psi}_{\rm phot}$ (dashed histogram, data
pre-dating 1995, \citealt{K95a}) is confirmed by Hubble-Space-Telescope
(HST) star-count data which pass through the entire Galactic disk and
are thus less prone to Malmquist bias (solid dots, \citealt{Zheng01}).
The ground-based volume-limited trigonometric-parallax sample (dotted
histogram) systematically overestimates $\Psi_{\rm near}$ due to the
Lutz-Kelker bias, thus lying above the improved estimate provided by
the Hipparcos-satellite data (solid histogram,
\citealt{JW97,K01b}). The depression/plateau near $M_V=7$ is the {\it
Wielen dip}.  The maximum near $M_V\approx 12, M_I\approx 9$ is the
{\it KTG peak}. The thin dotted histogram at the faint end indicates
the level of refinement provided by recent stellar additions
\citep{K01b} demonstrating that even the immediate neighbourhood
within 5.2~pc of the Sun probably remains incomplete at the faintest
stellar luminosities. }}
\label{fig:MWlf}
\end{center}
\end{figure}

\begin{figure}
\begin{center}
\rotatebox{0}{\resizebox{0.8
\textwidth}{!}{\includegraphics{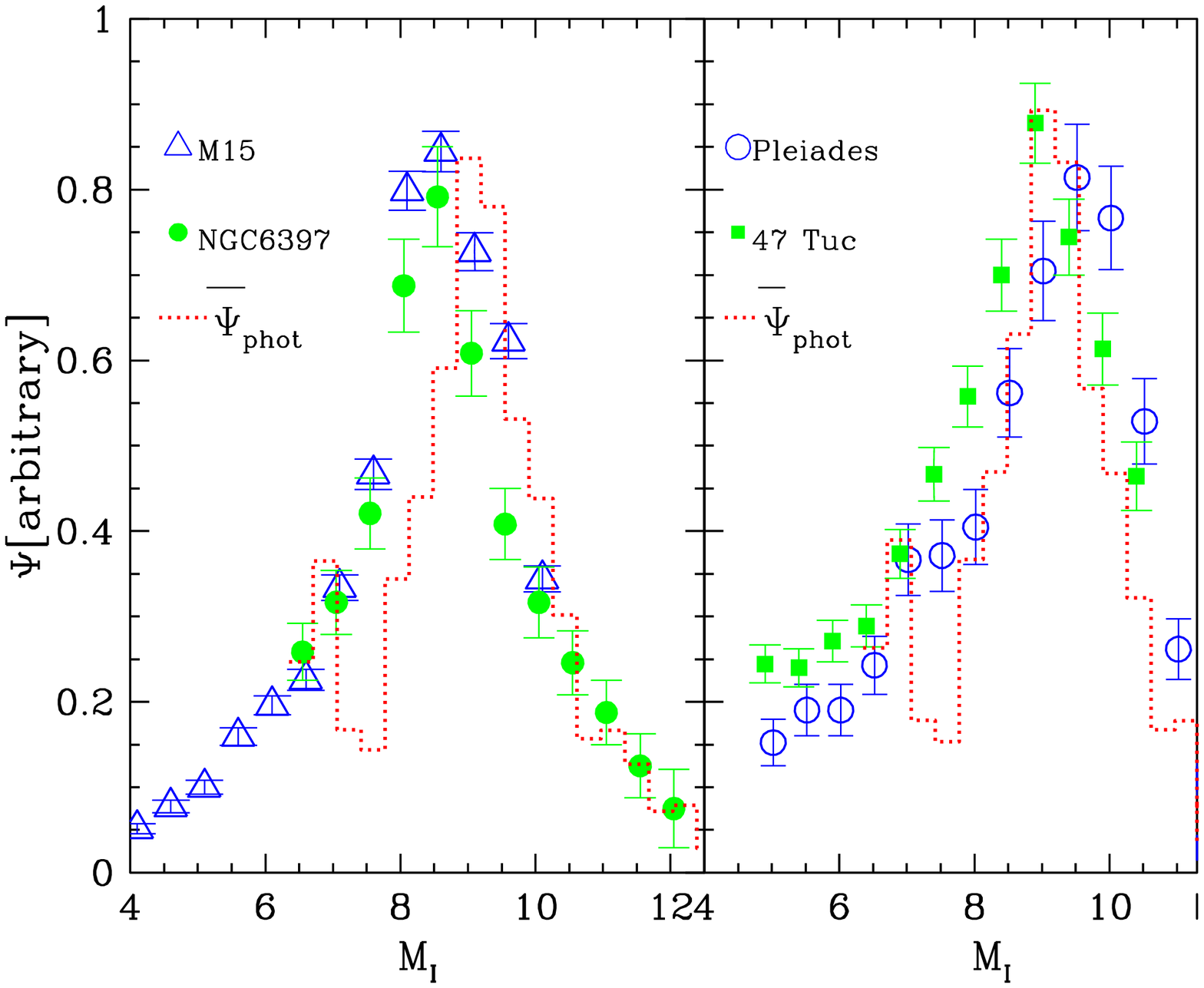}}}
\vskip -28mm
\caption{\small{$I$-band LFs of stellar {\it systems} in four star
clusters: globular cluster (GC) {\it M15} (\citealt{deMP95a}, distance
modulus $\Delta m=m-M=15.25$~mag); GC {\it NGC~6397}
(\citealt{PdeMR95}, $\Delta m=12.2$); young Galactic cluster {\it
Pleiades} (\citealt{HJH91}, $\Delta m=5.48$); GC {\it 47~Tuc}
(\citealt{deMP95b}, $\Delta m=13.35$). The dotted histogram is
$\overline{\Psi}_{\rm phot}(M_I)$ from the upper panel, transformed to
the $I$-band using the linear colour--magnitude relation
$M_V=2.9+3.4\,(V-I)$ \citep{KTG93} and $\Psi_{\rm phot}(M_I) =
(dM_V/dM_I) \, \Psi_{\rm phot}(M_V)$ (eq.~\ref{eq:passband}).  }}
\label{fig:cllf}
\end{center}
\end{figure}

The local, {\it nearby LF} and the Malmquist-corrected deep {\it
photometric LF} are displayed in Fig.~\ref{fig:MWlf}. They differ
significantly for stars fainter than $M_V\approx11.5$ which caused
significant controversy in the past \citep{K95a}. That the local
sample has a spurious but significant over-abundance of VLMSs can be
ruled out by virtue of the large velocity dispersion in the disk,
$\approx30$~pc/Myr. Any significant overabundance of stars within a
sphere with a radius of 30~pc would disappear within one~Myr, and
cannot be created by any physically plausible mechanism from a
population of stars with stellar ages spanning the age of the Galactic
disk. The shape of $\Psi_{\rm phot}(M_V)$ for $M_V\simgreat 12$ is
confirmed by many independent photometric surveys. That all of these
could be making similar mistakes, such as in colour transformations,
becomes unlikely on consideration of the LFs constructed for
completely independent stellar samples, namely star clusters
(Fig.~\ref{fig:cllf}).

\begin{figure}
\begin{center}
\rotatebox{0}{\resizebox{0.8 \textwidth}{!}{\includegraphics{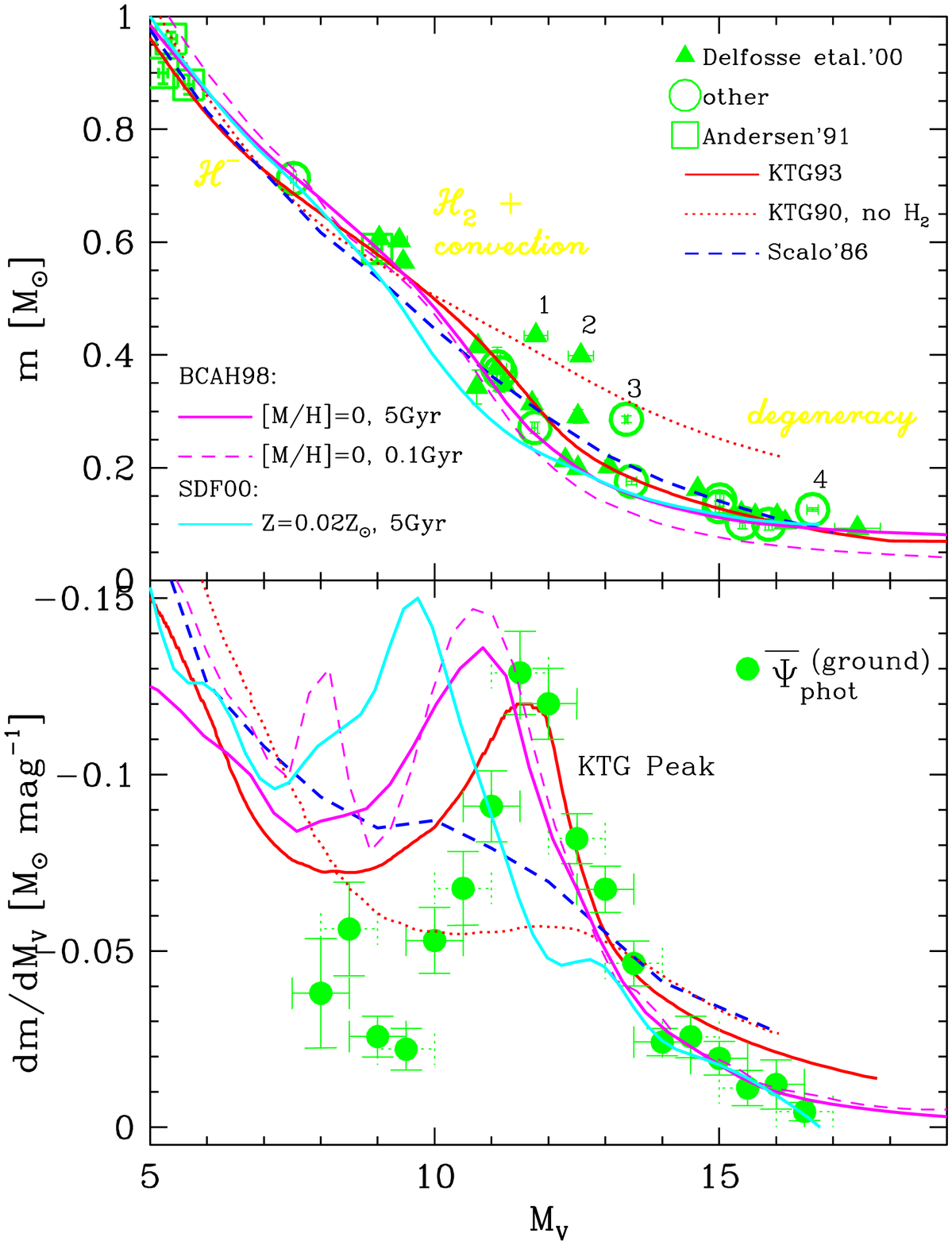}}}
\vskip -8mm
\caption {\small{The mass-luminosity relation (MLR, upper panel) and
its derivative (lower panel) for late-type stars. {\it Upper panel:}
The most recent observational data (solid triangles and open circles,
\citealt{Detal00}; open squares, \citealt{A91}) are compared with the
empirical MLR of \citet{Sc86} and the semi-empirical KTG93-MLR
\citep{KTG93}. The under-luminous data points~1,2 are GJ2069Aa,b and
~3,4 are Gl791.2A,B. All are probably metal-rich by $\sim$0.5~dex
\citep{Detal00}.  Recent theoretical MLRs from \citet{BCAH98} (BCAH98)
and \citet{SDF00} (SDF00) are also shown.  The observational data
\citep{A91} show that log$_{10}[m(M_V)]$ is approximately linear for
$m>2\,M_\odot$. {\it Lower panel:} The derivatives of the same
relations plotted in the upper panel are compared with
$\overline{\Psi}_{\rm phot}$ from Fig.~\ref{fig:MWlf}.  Note the good
agreement between the location, amplitude and width of the KTG peak in
the LF and the extremum in $dm/dM_V$.  }}
\label{fig:mlr}
\end{center}
\end{figure}

\begin{figure}
\begin{center}
\rotatebox{0}{\resizebox{0.8
\textwidth}{!}{\includegraphics{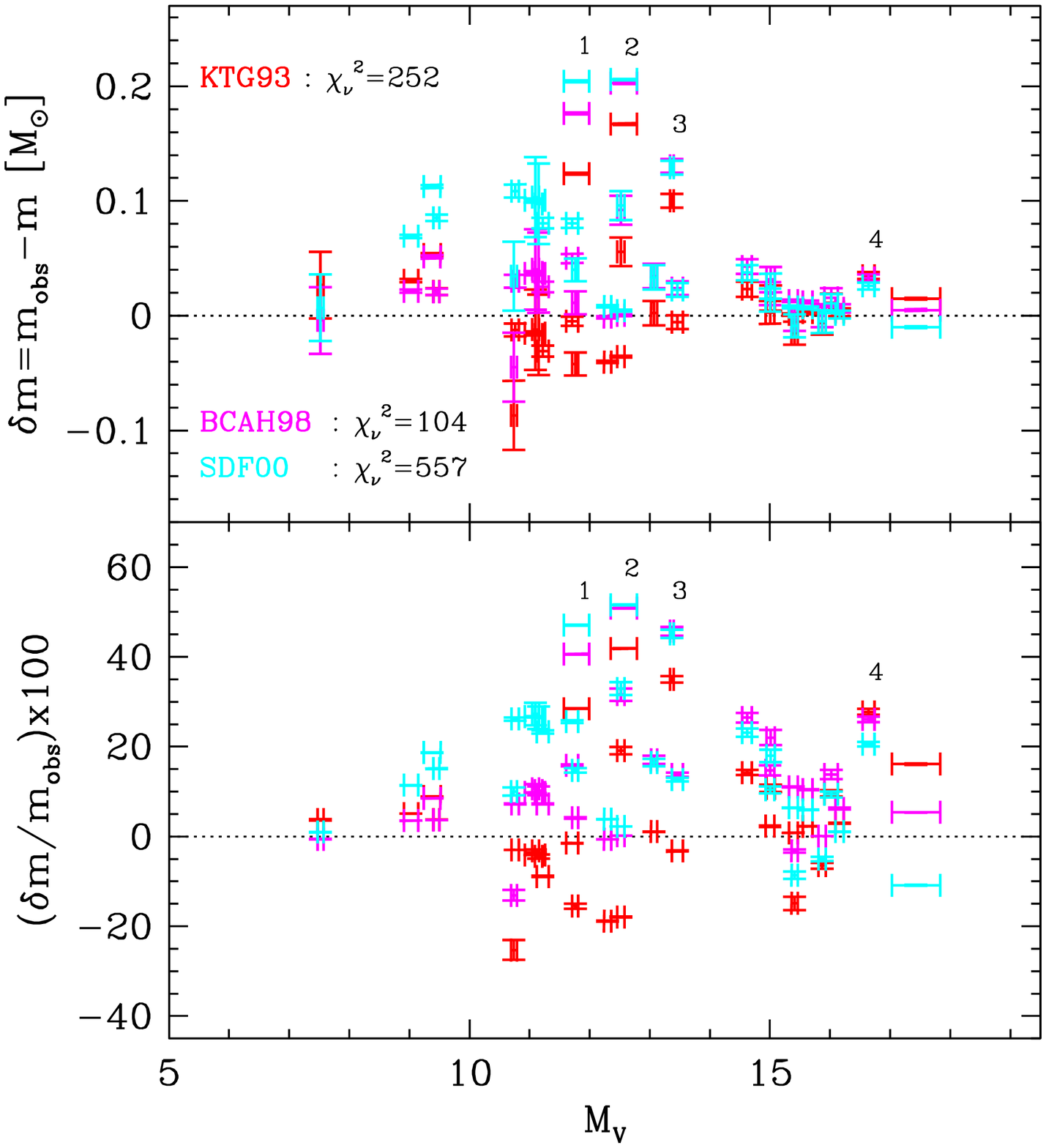}}}
\vskip -8mm
\caption {\small{Deviations of the MLRs ($\delta m=m_{\rm
obs}-m(M_V)$) from the empirical data with errors $\epsilon_m$ shown
in Fig.~\ref{fig:mlr} in $M_\odot$ (upper panel) and in percent
(lower panel with uncertainties $-(m(M_V)/m_{\rm
obs}^2)\times\epsilon_m$). Reduced $\chi_\nu^2$ ($\nu=26$ for 31 data
points, ignoring the four outliers) values indicate the formal
goodness-of-fit. Formally, none of the MLRs available is an acceptable
model for the data. This is not alarming though, because the models
are for a single-metallicity, single-age population while the data
span a range of metallicities and ages typical for the
solar-neighbourhood stellar population, as signified by $\delta m \gg
\epsilon_m$ in most cases.  The $\chi_\nu^2$ values confirm that the
BCAH98 models \citep{BCAH98} and the semi-empirical KTG93 MLR
\citep{KTG93} provide the best-matching MLRs. Note that the KTG93 MLR
was derived from mass--luminosity data prior to 1980, but by using the
shape of the peak in $\Psi_{\rm phot}(M_V)$ as an additional
constraint the constructed MLR became robust.  The lower panel
demonstrates that the deviations of observational data from the model
MLRs are typically much smaller than 30~per cent, excluding the
putatively metal-rich stars (1--4).  }}
\label{fig:dev}
\end{center}
\end{figure}

Eq.~\ref{eq:mf_lf} shows that any non-linear structure in the MLR is
mapped into observable structure in the LF, provided the MF does not
have compensating structure. Such a conspiracy is implausible because
the MF is defined through the star-formation process, but the MLR is a
result of the internal constitution of stars.  The MLR, its derivative
and deviations of models from observational data are shown in
Figs.~\ref{fig:mlr} and~\ref{fig:dev}, respectively.  It is apparent
that the slope is very small at faint luminosities leading to large
uncertainties in the MF near the hydrogen burning mass limit.

The physics underlying the non-linearities of the MLR are due to an
interplay of changing opacities, the internal stellar structure and
the equation of state of the matter deep inside the stars.  Starting
at high masses ($m\simgreat\,{\rm few}\,M_\odot$), as the mass of a
star is reduced H$^-$ opacity becomes increasingly important through
the short-lived capture of electrons by H-atoms resulting in reduced
stellar luminosities for intermediate and low-mass stars. The $m(M_V)$
relation becomes less steep in the broad interval $3<M_V<8$ leading to
the Wielen dip (Fig.~\ref{fig:MWlf}).  The $m(M_V)$ relation steepens
near $M_V=10$ because the formation of H$_2$ in the very outermost
region of main-sequence stars causes the onset of convection up to and
above the photo-sphere leading to a flattening of the temperature
gradient and therefore to a larger effective temperature as opposed to
to an artificial case without convection but the same central
temperature.  This leads to brighter luminosities and full convection
for $m\le0.35\,M_\odot$.  The modern Delfosse data beautifully confirm
the steepening in the interval $10<M_V<13$ predicted in 1990, and the
dotted MLR demonstrates the effects of suppressing the formation of
the H$_2$ molecule by lowering it's dissociation energy from 4.48~eV
to 1~eV \citep{KTG90}.  The $m(M_V)$ relation flattens again for
$M_V>14$, $m<0.2\,M_\odot$ as degeneracy in the stellar core becomes
increasingly important for smaller masses limiting further contraction
\citep{HN63,ChB97}.  Therefore, owing to the changing conditions
within the stars with changing mass, a pronounced local maximum in
$-dm/dM_V$ results at $M_V\approx11.5$, postulated in 1990 to be the
origin of the maximum in $\Psi_{\rm phot}$ near $M_V=12$
\citep{KTG90}.  

The implication that the LFs of all stellar populations should show
such a feature, although realistic metallicity-dependent stellar
models were not available yet, was noted \citep{KTG93}.  The subsequent
finding that all known stellar populations have such a maximum in the
LF (Figs.~\ref{fig:MWlf} and~\ref{fig:cllf}) constitutes one of the
{\it most impressive achievements of stellar-structure theory}.
Different theoretical $m(M_V)$ relations have the maximum in
$-dm/dM_V$ at different $M_V$, suggesting the possibility of testing
stellar structure theory near the critical mass
$m\approx0.35\,M_\odot$, where stars become fully convective
\citep{KT97,BCC98}. But since the IMF also defines the LF the shape and
location cannot be unambiguously used for this purpose unless it is
postulated that the IMF is invariant. 

\begin{figure}
\begin{center}
\rotatebox{0}{\resizebox{0.8 \textwidth}{!}{\includegraphics{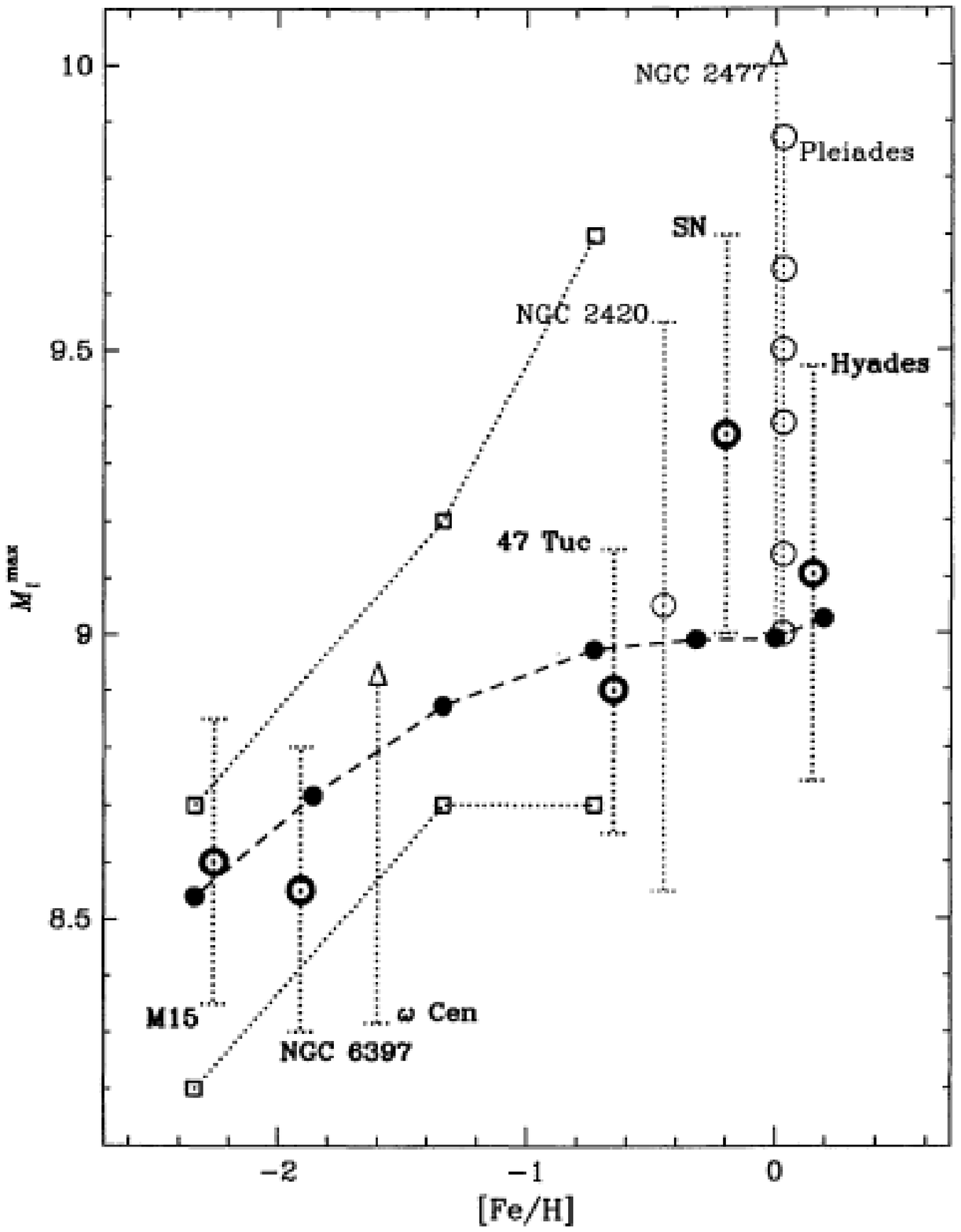}}}
\vskip 0mm
\caption {\small{ The position of the maximum in $-dm/dM_I$ as a
function of metallicity of the theoretical mass--luminosity data of
\citet{KT97} is shown as solid dots. The open squares represent bounds
by the stellar-structure models of \citet{dAM96}, and the open circles
are observational constraints for different populations (e.g. SN for
the composite solar-neighbourhood population, Pleiades for the simple
population of an intermediate-age cluster). Thick circles are more
certain than the thin circles, and for the Pleiades a sequence of
positions of the LF-maximum is given, from top to bottom, with the
following combinations of (distance modulus, age): (5.5, 70~Myr),
(5.5, 120~Myr), (5.5, main sequence), (6, 70~Myr), (6, 120~Myr), (6,
main sequence). For more details see \citet{KT97}.  }}
\label{fig:LFpeak}
\end{center}
\end{figure}

\begin{figure}
\begin{center}
\rotatebox{0}{\resizebox{0.7 \textwidth}{!}{\includegraphics{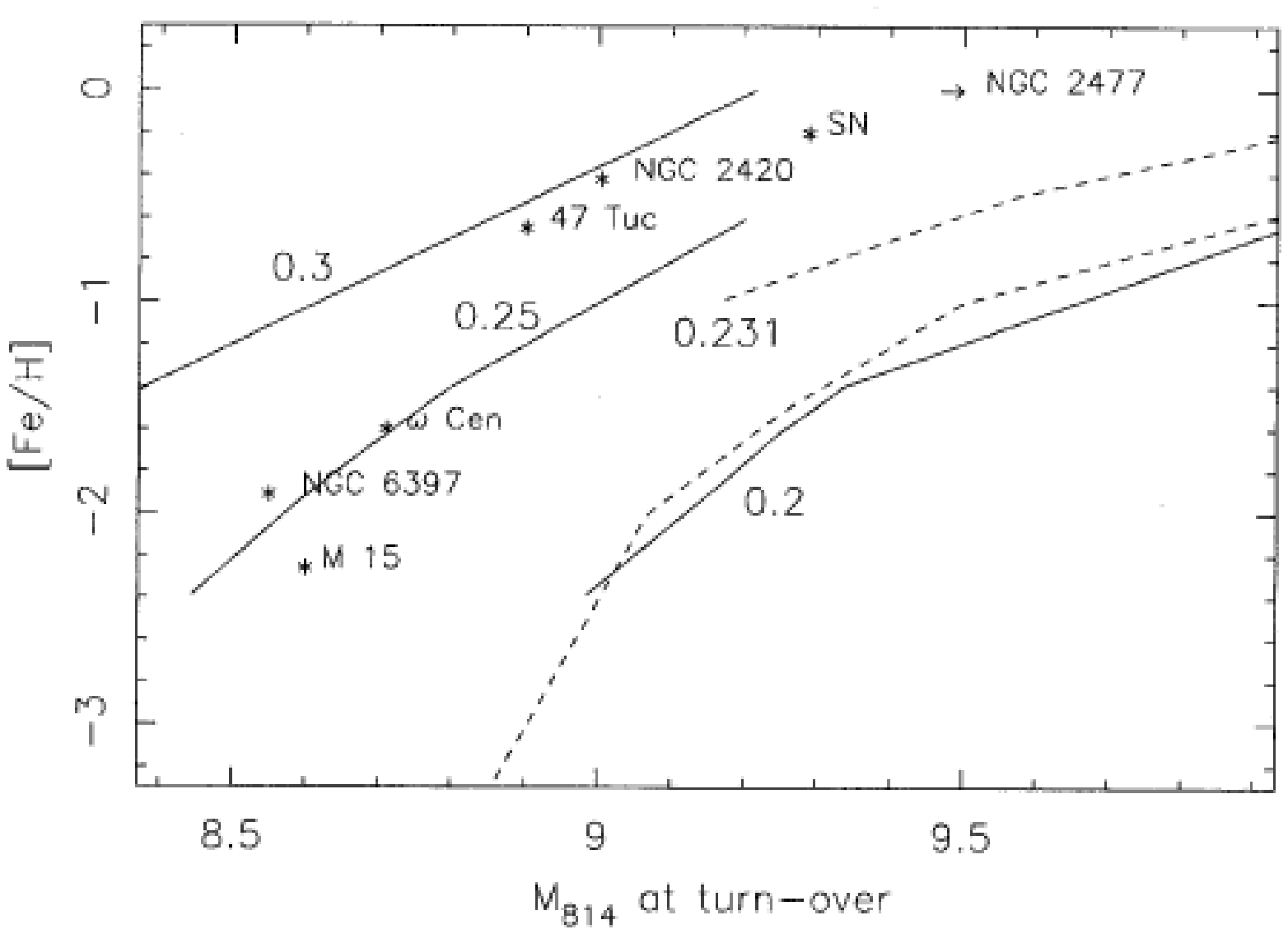}}}
\vskip 0mm
\caption {\small{Similar to Fig.~\ref{fig:LFpeak} but from
\citet{vonHippelGil96}, their fig.5: The absolute $I$-band-equivalent
magnitude of the maximum in the LF as a function of metallicity for
different populations.  The solid and dashed lines are loci of
constant mass ($0.2, 0.231, 0.3\,M_\odot$) according to theoretical
stellar structure calculations.  See \citet{vonHippelGil96} for more
details.  }}
\label{fig:LFpeak2}
\end{center}
\end{figure}

A study of the position of the maximum in the $I$-band LF has been
undertaken by \citet{vonHippelGil96} and \citet{KT97} finding that the
observed position of the maximum shifts to brighter magnitude with
decreasing metallicity, as expected from theory
(Figs.~\ref{fig:LFpeak} and~\ref{fig:LFpeak2}).


In addition to the non-linearities in the $m(M_P)$ relation,
unresolved multiple systems affect the MF derived from the photometric
LF, in particular since no stellar population is known to exist that
has a binary proportion smaller than 50~per cent, apart possibly from
dynamically highly evolved globular clusters.  Suppose an observer
sees 100~systems. Of these~40, 15~and 5~are binary, triple and
quadruple, respectively, these being realistic proportions. There are
thus 85~companion stars which the observer is not aware of if none of
the multiple systems are resolved. Since the distribution of secondary
masses for a given primary mass is not uniform but typically increases
with decreasing mass \citep{MZ01}, the bias is such that low-mass
stars are significantly underrepresented in any survey that does not
detect companions \citep{KTG91,Hetal98,L98,MZ01}.  Also, if the
companion(s) are bright enough to affect the system luminosity
noticeably then the estimated photometric distance will be too small,
artificially enhancing inferred space densities which are, however,
mostly compensated for by the larger distances sampled by binary
systems in a flux-limited survey, together with an exponential density
fall-off perpendicular to the Galactic disk \citep{K01b}.  A faint
companion will also be missed if the system is formally resolved but
the companion lies below the flux limit of the survey.

\begin{figure}
\begin{center}
\rotatebox{0}{\resizebox{0.75
\textwidth}{!}{\includegraphics{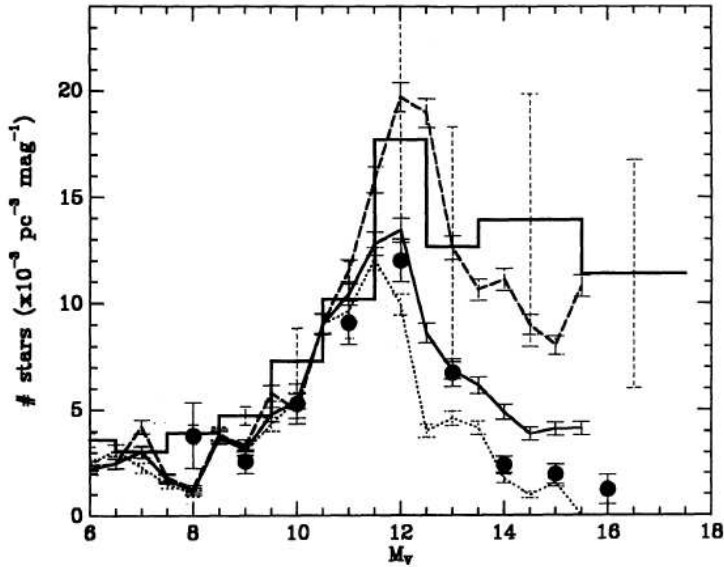}}}
\vskip 0cm
\caption
{\small{ Comparison of the model field luminosity function (curves) of
a single-metallicity and single-age population that is without
measurement errors, with observations in the photometric V-band (a
comparison of the corresponding LFs in bolometric magnitudes can be
found in \citealt{K95e}). {\it The model} assumes the standard stellar
IMF, eq.~\ref{eq:imf} below.  The model single star luminosity
function is normalised to the nearby luminosity function at $M_{\rm
V}\approx10, M_{\rm bol}\approx9$, giving the normalisation constant
in the MF $k$, and the plot shows $k\,\Psi_{\rm mod,sing}$ (long
dashed curve), $k\,\Psi_{\rm mod,sys}(t=0)$ (dotted curve, without
pre-main sequence brightening) and $k\,\Psi_{\rm mod,sys}(t=1\,{\rm
Gyr})$ (solid curve).  Note that the solid curve is the luminosity
function for a realistic model of the Galactic field population of
systems consisting of 48~per~cent binaries which have a period
distribution consistent with the empirical G-, K-, and M-dwarf period
distribution, the mass ratio distributions for G-dwarf systems as
observed \citep{DM91}, and the overall mass-ratio distribution given
by fig.~2 in \cite{Kr_etal03}, where a concise description of the
``standard star-formation model'' can be found.  {\it The observed}
nearby stellar luminosity function, $\Psi_{\rm near}$, which is not
corrected for Malmquist-type bias (tables~2 and~8 in \cite{K95a} and
which is smoothed by using larger bin widths at the faint end, as
detailed in section~4 of that paper, is plotted as the solid-line
histogram.  The filled circles represent the best-estimate Malmquist
corrected photometric luminosity function, ${\overline\Psi}_{\rm
phot}$, Fig.~\ref{fig:MWlf}.  By correcting for Malmquist bias \citep{SIP} 
the LF becomes that of a single-age, single-metallicity population.
Taken from \citet{K95e}.  }}
\label{fig:lfmods}
\end{center}
\end{figure}

Comprehensive star-count analysis of the solar neighbourhood that
incorporate unresolved binary systems, metallicity and age spreads and
the density fall-off perpendicular to the Galactic disk show that the
IMF can be approximated by a two-part power-law with
$\alpha_1=1.3\pm0.7, 0.08 <m/M_\odot \le 0.5$, $\alpha_2=2.2,
0.5<m/M_\odot \le 1$, a result obtained for two different MLRs
\citep{K01b}. The index $\alpha_2$ is constrained tightly owing to the
well-constrained $\Psi_{\rm near}$, the well-constrained empirical MLR
in this mass range and because unresolved binary systems do not
significantly affect the solar-neighbourhood LF in this mass range
because primaries with $m\simgreat1\,M_\odot$ are rare and are not
sampled.  

Fig.~\ref{fig:lfmods} demonstrates models of the single-star and
system LFs for the KTG93 MLR shown in Fig.~\ref{fig:mlr}.  The
significant difference between the single-star and system LFs is
evident, being most of the explanation of the disputed\footnote{The
discrepancy evident in Fig.~\ref{fig:lfmods} between the nearby LF,
$\Psi_{\rm near}$, and the photometric LF, $\Psi_{\rm phot}$, invoked
a significant dispute as to the nature of this discrepancy. On the one
hand \citep{K95a} the difference is thought to be due to unseen
companions in the deep but low-resolution surveys used to construct
$\Psi_{\rm phot}$, with the possibility that photometric calibration
for VLMSs may remain problematical so that the exact shape of
$\Psi_{\rm phot}$ for $M_V\simgreat 14$ is probably uncertain.  On the
other hand \citep{RG97} the difference is thought to come from
non-linearities in the $V-I, M_V$ colour--magnitude relation used for
photometric parallax. Taking into account such structure it can be
shown that the photometric surveys underestimate stellar space
densities so that $\Psi_{\rm phot}$ moves closer to the extended
estimate of $\Psi_{\rm near}$ using a sample of stars within 8~pc or
further.  While this is an important point, the extended $\Psi_{\rm
near}$ is incomplete (see footnote~\ref{fn:compl}) and theoretical
colour-magnitude relations do not have the required degree of
non-linearity. The observational colour--magnitude data also do not
conclusively suggest a feature with the required strength
\citep{BCAH98}. Furthermore, $\Psi_{\rm phot}$ agrees almost perfectly
with the LFs measured for star clusters of solar and population~I
metallicity (Fig.~\ref{fig:cllf}) so that it appears unlikely that
non-linearities in the colour--magnitude relation are the dominant
source of the discrepancy.}  discrepancy between the observed
$\Psi_{\rm near}$ and $\Psi_{\rm phot}$. Note though that the observed
photometric LF contains triple and quadruple systems that are not
accounted for by the model. Note also that the photometric LF has been
corrected for Malmquist bias and so constitutes the system LF in which
the broadening due to a metallicity and age spread and photometric
errors has been largely removed. It is therefore directly comparable
to the model system LF, and both indeed show a very similar KTG peak.
The observed nearby LF, on the other hand, has not been corrected for
the metallicity and age spread nor for trigonometric distance errors,
and so it appears broadened.  The model single-star LF, in contrast,
does not, by construction, incorporate these, and thus appears with a
more pronounced maximum.  Such observational effects can be
incorporated rather easily into full-scale star-count modelling
\citep{KTG93}. The deviation of the model system LF from the observed
photometric LF for $M_V\simgreat 14$ may indicate a change of the
pairing properties of the VLMS or BD population (\S~\ref{sec:bds}).

Since the nearby LF is badly defined statistically for $M_V\simgreat
13$, the resulting model, such as shown in Fig.~\ref{fig:lfmods}, is a
{\it prediction} of the true, single-star LF that should become
apparent once the immediate solar-neighbourhood sample has been
enlarged significantly through the planned space-based astrometric
survey {\sc Gaia} \citep{Gil98}, followed by an intensive follow-up
imaging and radial-velocity observing programme scrutinising every
nearby candidate for unseen companions \citep{K01b}.  Despite such a
monumental effort, the structure in $\Psi_{\rm near}^{\rm GAIA}$ will
be smeared out due to the metallicity and age spread of the local
stellar sample, a factor to to be considered seriously.

\subsection{Star clusters} 
\label{sec:stcl}

Star clusters offer populations that are co-eval, equi-distant and
that have the same chemical composition, but as a compensation of
these advantages the extraction of faint cluster members is very
arduous because of contamination by the background Galactic-field
population. The first step is to obtain photometry of everything
stellar in the vicinity of a cluster and to select only those stars
that lie near one or a range of isochrones, taking into account that
unresolved binaries are brighter than single stars. The next step is
to measure proper motions and radial velocities of all candidates to
select only those high-probability members that have coinciding space
motion with a dispersion consistent with the a priori unknown but
estimated internal kinematics of the cluster. Since nearby clusters
for which proper-motion measurements are possible appear large on the
sky, the observational effort is horrendous. For clusters such as
globulars that are isolated the second step can be omitted, but in
dense clusters stars missed due to crowding need to be corrected for.
The stellar LFs in clusters turn out to have the same general shape as
$\Psi_{\rm phot}$ (Fig.~\ref{fig:cllf}), with the maximum being
slightly off-set depending on the metallicity of the population
(Figs.~\ref{fig:LFpeak} and~\ref{fig:LFpeak2}).  A 100~Myr isochrone
(the age of the Pleiades) is also plotted in Fig.~\ref{fig:mlr} to
emphasise that for young clusters additional structure (in this case
another maximum near $M_V=8$ in the LF is expected via
eq.~\ref{eq:mf_lf} . This is verified for the Pleiades cluster
\citep{Belikov98}, and is due to stars with $m<0.6\,M_\odot$ not
having reached the main-sequence yet \citep{ChB00}.

LFs for star clusters are, like $\Psi_{\rm phot}$, system LFs because
binary systems are not resolved in the typical star-count survey. The
binary-star population evolves due to encounters, and after a few
initial crossing times only those binary systems survive that have a
binding energy larger than the typical kinetic energy of stars in the
cluster.  A further complication with cluster LFs is that star
clusters preferentially loose single low-mass stars across the tidal
boundary as a result of ever-continuing re-distribution of energy
during encounters while the retained population has an increasing
binary proportion and increasing average stellar mass. The global PDMF
thus flattens with time with a rate proportional to the fraction of
the cluster lifetime and, for highly evolved initially rich open
clusters, it evolves towards a delta function near the turnoff mass,
the mass-loss rate being a function of Galactocentric distance.  This
is a major problem for aged open clusters (initially $N<10^4$~stars)
with life-times of only a few~100~Myr \citep{K95b}.

\begin{figure}
\begin{center}
\rotatebox{0}{\resizebox{0.75 \textwidth}{!}{\includegraphics{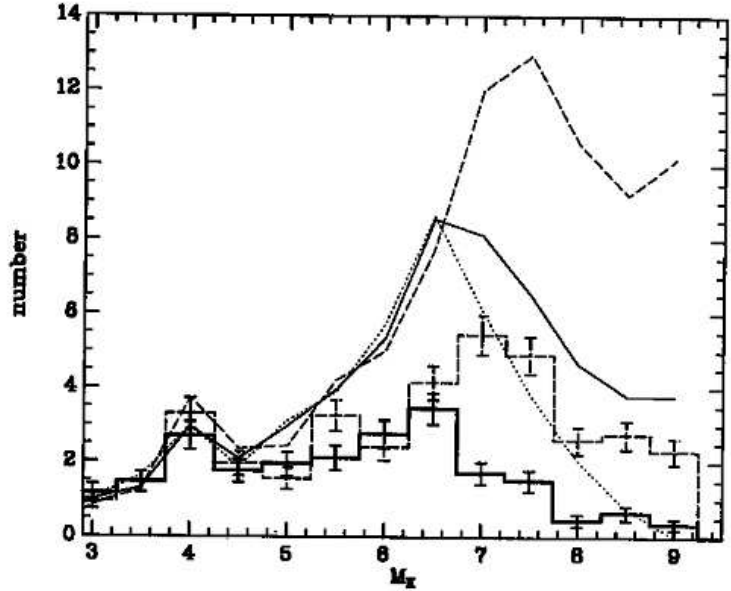}}}
\vskip -4mm
\caption{\small{ Models of the $K$-band single-star and system LFs in
an ensemble of 20~dynamically highly evolved clusters (thin and thick
histograms, respectively). Each cluster model consists initially of
200~binaries with a half-mass radius of 0.85~pc, and the LFs are shown
at an age of 480~Myr (44 initial crossing times) and count all stars
and systems within the central sphere with a radius of 2~pc. The
clusters are random renditions from the same parent distributions
(binary-star orbital parameters, IMF, stellar positions and
velocities) and are initially in dynamical equilibrium. The upper
dashed curve is the initial single-star LF (KTG93 MLR and standard
IMF, eq.~\ref{eq:imf} below) and the solid curve is the model
Galactic-field LF of systems. This is an accurate representation of
the Galactic-field population in terms of the IMF and mixture of
single and binary stars, and is derived by stars forming in clusters
such as shown here that dissolve with time.  Both of these LFs are
identical to the ones shown in Fig.~\ref{fig:lfmods}. The dotted curve
is the initial system LF (100~\% binaries).  From \citet{K95b}.  }}
\label{fig:LFclmod}
\end{center}
\end{figure}

These processes are now well quantified \citep{K01c, Port_etal01}, and
Fig.~\ref{fig:LFclmod} shows that a dynamically very evolved cluster
such as the Hyades has been depleted significantly in low-mass
stars. Even so, the binary-star correction that needs to be applied to
the LF in order to arrive at the single-star present-day LF is
significant\footnote{Note that a {\it system} can be a binary star or a
single star, and that the {\it single-star} LF is the LF obtained by
counting all stars independently of whether they are in systems or
single.}.

\begin{figure}
\begin{center}
\rotatebox{0}{\resizebox{0.65
\textwidth}{!}{\includegraphics{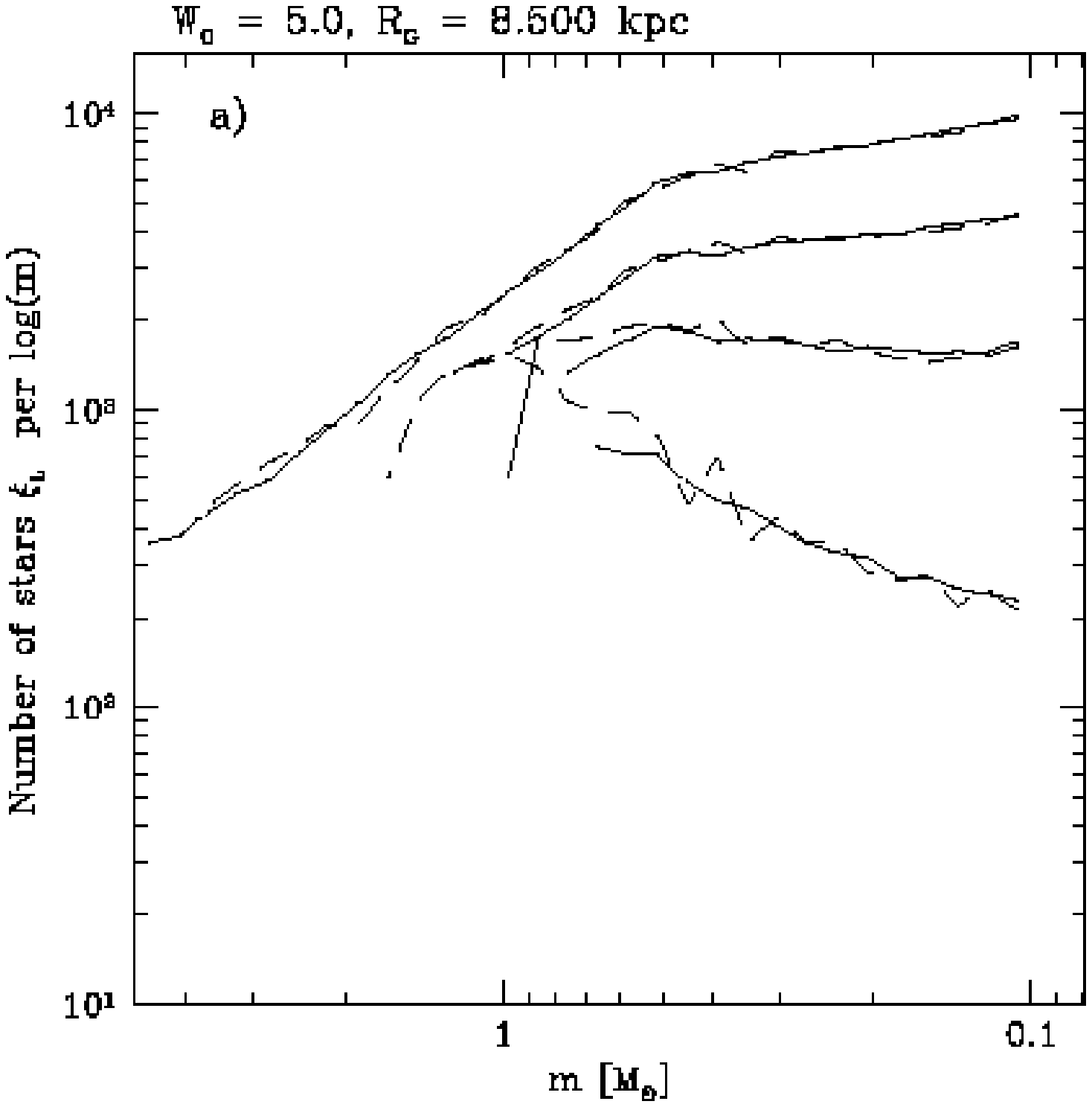}}}
\vskip -4mm
\caption{\small{Present-day mass functions in a King-model cluster
with concentration $W_0=5$ on a circular orbit about the MW centre
with radius 8.5~kpc. Shown are the MFs of all bound stars at ages
corresponding to $\tau_f=\,$0~\%, 30~\%, 60~\% and 90~\% of the
cluster life-time (from top to bottom). For each age the solid line
represents one computation with $1.28\times 10^5$~stars, the dashed
lines show the sum of four clusters each with 8000~stars (scaled to
the same number of stars as the massive computation). Results for
other circular and eccentric orbits and cluster concentrations are
virtually indistinguishable. From \citet{BaumMakin03}.  }}
\label{fig:MFBaumg}
\end{center}
\end{figure}

A computationally challenging investigation of the systematic changes
of the MF in evolving clusters of different masses has been published
by \citet{BaumMakin03}. Baumgardt \& Makino quantify the depletion of
the clusters of low-mass stars through stellar-dynamical processes
with a hitherto not available quality and conclusively show that
highly evolved clusters have a very substantially skewed present-day
MF (Fig.~\ref{fig:MFBaumg}). If the cluster ages are expressed in
fractions, $\tau_f$, of the overall cluster lifetime, which depends on
the initial cluster mass, its concentration and orbit, then different
clusters on different orbits lead to virtually the same present-day
MFs at the same $\tau_f$.  Their results were obtained for clusters
that are initially in dynamical equilibrium and that do not contain
binary stars (these are computationally highly demanding), so that
future analysis, including initially non-virialised clusters and a
high primordial binary fraction \citep{GK05}, will be required to
further refine these excellent results.

\begin{figure}
\begin{center}
\rotatebox{0}{\resizebox{0.99 \textwidth}{!}{\includegraphics{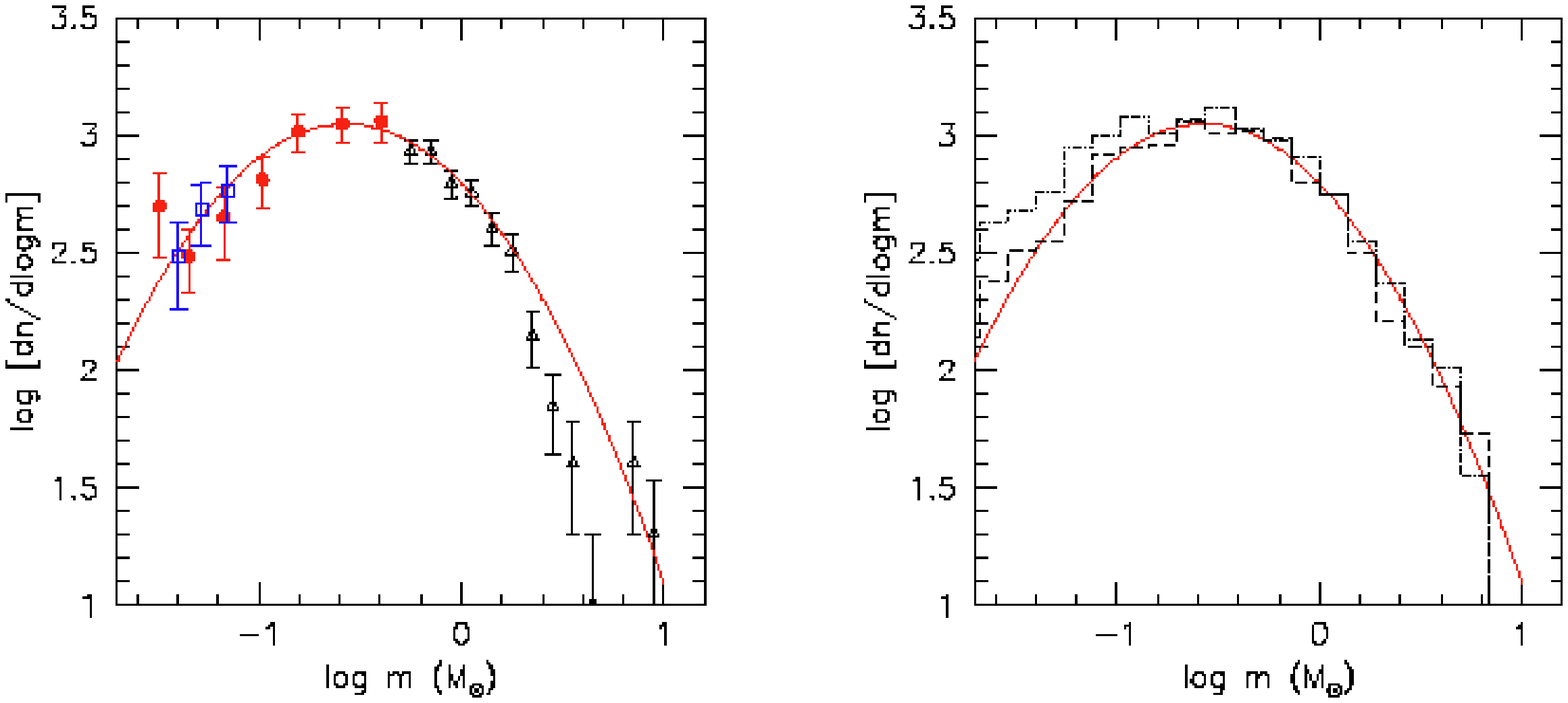}}}
\vskip -4mm
\caption{\small{ The observed MF in the Pleiades cluster. {\it Left
panel}: The symbols are observational data (for details see
\citealt{MKB04}) and the curve is a log-normal fit. {\it Right panel}:
The solid curve is the same log-normal fit.  Theoretical system MFs
for two initial models of the Pleiades cluster according to
\citet{KAH} are plotted at an age of 100~Myr.  These models assume the
young cluster to be deeply embedded in gas, with a star-formation
efficiency of 33~\%, a gas-expulsion time-scale shorter than the
crossing time and to contain $10^4$ stars and BDs. Model~A (dashed
histogram) has an initial central number density $\rho_{\rm
C}=10^{4.8}$~stars/pc$^3$, while model~B (dotted histogram) has
$\rho_{\rm C}=10^{5.8}$~stars/pc$^3$.  The embedded phase lasts
0.6~Myr, and during this time mass segregation develops in the
initially denser cluster model~B. Note that these models are not a fit
but a prediction of the Pleiades MF, assuming it had a standard IMF
(eq.~\ref{eq:imf}).  From \citet{MKB04}.  }}
\label{fig:MorauxPl}
\end{center}
\end{figure}

The first realistic calculations of the formation of an open star
cluster such as the Pleiades demonstrate that the binary properties of
stars remaining in the cluster are comparable to those observed even
if all stars initially form in binary systems with Taurus-Auriga-like
properties \citep{KAH, Kr_etal03}.  They also demonstrate the complex
and counter-intuitive interplay between the initial concentration,
mass segregation at the time of residual gas expulsion, and the final
ratio of the number of BDs to stars (Fig.~\ref{fig:MorauxPl}). Thus,
the modelling by \citet{KAH} shows that an initially denser cluster
evolves to significant mass segregation when the gas explosively
leaves the system. Contrary to naive expectation, according to which a
mass-segregated cluster should loose more of its least massive members
during expansion after gas expulsion, the ensuing violent relaxation
of the cluster retains more free-floating BDs than the less-dense
model. This comes about because BDs are split from the stellar
binaries more efficiently in the denser cluster.
These issues remain an active area of research, because at least two
changes need to be made to the modelling: on the one hand, BDs need to
be treated as an extra population \citep{Kr_etal03, Kr_Bouv03b} so
that the above discovered free-floating BDs that result from the
disruption of star--BD binaries will not be available, and on the
other hand some observations suggest that star clusters may form
highly mass-segregated (but see Fig.~\ref{fig:mfn2}).  The
mass-dependent loss of stars thus definitely remains an issue to be
studied.

From the above work it is now established that even clusters as young
as the Pleiades are significantly evolved because clusters of all
masses form from highly concentrated embedded morphologies
\citep{Kr_paris05}.  Also, the low-mass stars in clusters as young as
the Pleiades or M35 (Fig.~\ref{fig:mfn1} below) have not yet reached
the main sequence, so that pre-main sequence stellar-evolution
calculations have to be resorted to when transforming measured
luminosities to stellar masses via the MLR.

For ages younger than a few~Myr this becomes a serious problem:
Classical pre-main sequence theory, which assumes hydrostatic
contraction of spherical non-, sometimes slowly-rotating stars from
idealised initial states breaks down because of the overlap with the
star formation processes that defies detailed treatment. Stars this
young remember their accretion history, invalidating the application
of classical pre-main sequence stellar evolution tracks, a point made
explicitly clear by the excellent work of Wuchterl and collaborators
\citep{WK01, WT03}, and are in any case rotating rapidly and
non-spherical. Such realistic pre-main sequence tracks are not
available yet. 

Nevertheless, owing to the lack of an alternative, research of the IMF
in very young clusters has to resort to spectroscopic classification
of individual stars to place them on a theoretical isochrone of
existing classical theory to estimate masses (e.g. \citealt{Meyer00,
Luhman04, Barrado_etal04, Slesnick_etal04}). In such cases the age
spread becomes comparable to the age of the cluster
(eq.~\ref{eq:imf_pdmf}). Binary systems are mostly not resolved.

A few results are shown in Figs.~\ref{fig:mfn1}
and~\ref{fig:mfn2}. While the usual argument is for an invariant IMF
\citep{K01c,K02,Chrev03}, as is apparent for most population~II stars
(e.g. fig.5 in \citealt{Chrev03}), Fig.~\ref{fig:mfn1} shows that some
appreciable differences in measured MFs are evident. The M35 MF
appears to be highly deficient in low-mass stars. This clearly needs
further study because M35 and the Pleiades appear to be otherwise
fairly similar in terms of age, metallicity (M35 is somewhat less
metal-rich than the Pleiades) and the size of the survey volume.

Taking the ONC as the best-studied example of a very young and nearby
rich cluster (age$\;\approx 1$~Myr, distance$\;\approx 450$~pc;
$N\approx 5000-10000$~stars and BDs; \citealt{HC00,L00,MLL00, K00,
Slesnick_etal04}), Fig.~\ref{fig:mfn1} shows how the shape of the
deduced IMF varies with improving (but still classical) pre-main
sequence contraction tracks. This demonstrates that any sub-structure
cannot, at present, be relied upon to reflect possible underlying
physical mechanisms of star formation.

For the much more massive and long-lived globular clusters
($N\simgreat 10^5$~stars) theoretical stellar-dynamical work shows
that the MF measured for stars near the cluster's half-mass radius is
approximately similar to the global PDMF, while inwards and outwards
of this radius the MF is flatter (smaller $\alpha_1$) and steeper
(larger $\alpha_1$), respectively, owing to dynamical mass segregation
\citep{VH97}. However, mass loss from the cluster flattens the global
PDMF such that it no longer resembles the IMF anywhere
(Fig.~\ref{fig:MFBaumg}), for which evidence has been found in some
cases \citep{PZ99}. The MFs measured for globular clusters must
therefore generally be flatter than the IMF, which is indeed born-out
by observations (Fig.~\ref{fig:apl} below). However, again the story
is by no means straightforward, because globular clusters have
significantly smaller binary fractions than population~II clusters
\citep{Ivanova_etal05}, and so the binary-star correction is smaller
for globular cluster MFs. Therefore, and as already pointed out by
\citet{K01a}, it appears quiet realistically possible that
population~II IMFs were in fact flatter (smaller $\alpha_1$) than
population~I IMFs, as would be qualitatively expected from simple
fragmentation theory (\S~\ref{sec:var_th} below). Clearly, this issue
needs detailed investigation which, however, is computationally
extremely highly demanding, requiring the use of state-of-the art
$N$-body codes and special-purpose hardware.

\begin{figure}
\begin{center}
\rotatebox{0}{\resizebox{0.99 \textwidth}{!}{\includegraphics{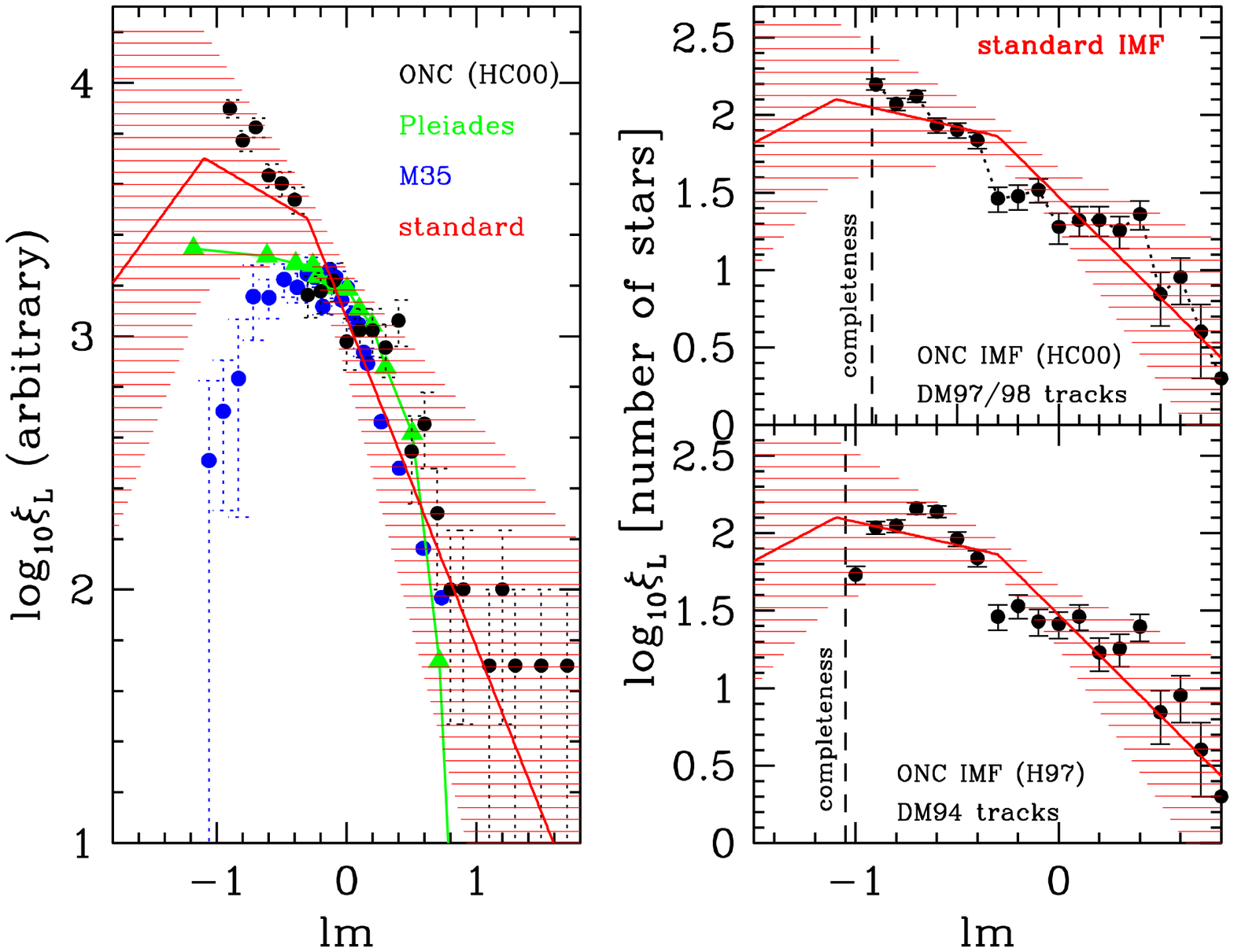}}}
\vskip -35mm
\caption
{\small{\it Left panel:} The measured system mass functions ($lm\equiv
{\rm log}_{10}(m/M_\odot)$) in the ONC \citep{HC00}: optical data,
$r\le2.5$~pc, $\tau_{\rm cl}<2$~Myr, [Fe/H]$=-0.02$
\citep{Estetal98} in the Pleiades \citep{Hambetal99}: $r\le6.7$~pc,
$\tau_{\rm cl}\approx100$~Myr, [Fe/H]$=+0.01$, and in M35
\citep{Netal01}: $r\le4.1$~pc, $\tau_{\rm cl}\approx160$~Myr,
[Fe/H]$=-0.21$, where $r$ is the approximate radius of the survey
around the cluster centre and $\tau_{\rm cl}$ the nuclear age. The
strong decrease of the M35 MF below $m\approx0.5\,M_\odot$ remains
present despite using different MLRs (e.g. DM97, as in the right
panel).  None of these MFs are corrected for unresolved binary
systems.  The standard single-star IMF (eq.~\ref{eq:imf}) is plotted
as the three straight lines. {\it Right panel:} The shape of the ONC
IMF differs significantly for $m<0.22\,M_\odot$ if different pre-main
sequence evolution tracks, and thus essentially different theoretical
MLRs, are employed (DM stands for tracks calculated by D'Antona \&
Mazzitelli, see \citealt{HC00} for details.)  }
\label{fig:mfn1}
\end{center}
\end{figure}

\begin{figure}
\begin{center}
\rotatebox{0}{\resizebox{0.99 \textwidth}{!}{\includegraphics{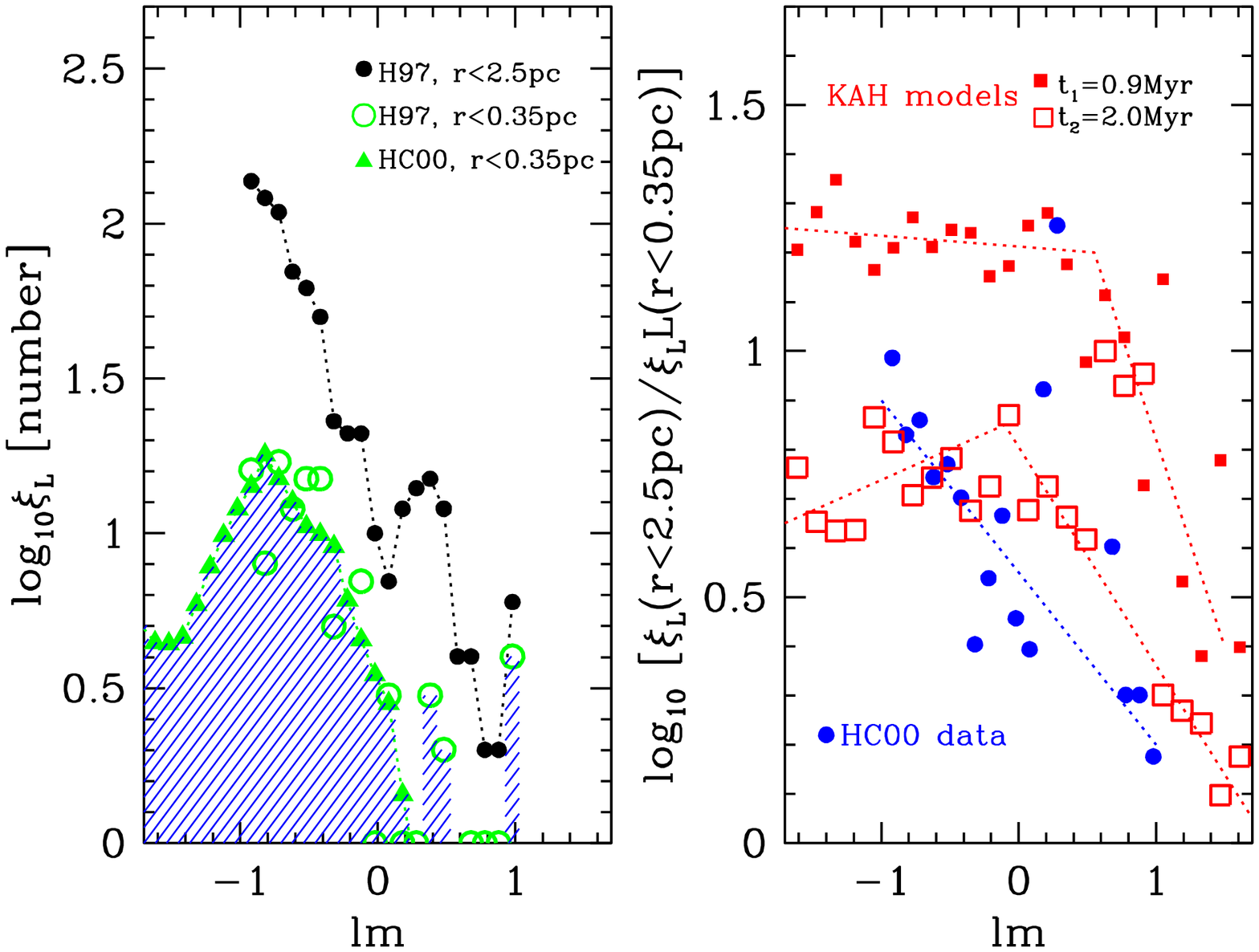}}}
\vskip -35mm
\caption
{\small{{\it Left panel:} Mass segregation is very pronounced in the
ONC, as is evident by comparing the MF for all stars within
$r=2.5$~pc with the MF for all stars with $r<0.35$~pc
(\citealt{HC00}, HC00) ($lm\equiv {\rm log}_{10}(m/M_\odot)$). For
both samples the reddening $A_V<2.5$~mag (\citealt{Hill97}, H97, is
for an optical and spectroscopic survey, whereas HC00 is a
near-infrared survey). {\it Right panel:} The ratio, $\xi_{\rm
L}(r<2.5\,{\rm pc})/\xi_{\rm L}(r<0.35\,{\rm pc})$, of the MFs shown
in the left panel increases significantly with decreasing mass,
demonstrating the significant depletion of low-mass stars in the
central region of the ONC. Stellar-dynamical models of the ONC
\citep{KAH} approximately reproduce this trend at an age of 2~Myr for
the standard IMF (eq.~\ref{eq:imf}, whereby the system masses of
surviving binary systems are counted instead of the individual stars
many of which are in unresolved binaries) even if no initial mass
segregation is assumed (at $t=0$, $\xi_{\rm L}(r<2.5\,{\rm
pc})/\xi_{\rm L}(r<0.35\,{\rm pc}) =\,$constant).  The model snapshots
shown are from model~B in \cite{KAH} under the assumption that prior
to gas-expulsion, 2--1.4~Myr ago, the central stellar density was
$\rho_{\rm C}=10^{5.8}$~stars/pc$^3$. The dotted lines are eye-ball
fits to the plotted data.  }}
\label{fig:mfn2}
\end{center}
\end{figure}

The IMFs obtained for nearby and young star clusters as well as for
globular clusters are summarised in Table~\ref{tab:apl}.

\section{Very low-mass stars and brown dwarfs}
\label{sec:bds}

These are stars near to the hydrogen-burning mass limit (VLMS) or
below (BD), in which case they are not massive enough to achieve
sufficiently high central pressures and temperatures to stabilise
against continued contraction by burning H and thus indefinitely cool
to unobservable luminosities and temperatures. Observationally it is
very difficult to distinguish between VLMSs and BDs, because a
sufficiently young BD may have colours and spectral features
corresponding to a VLMS.  BDs were studied as theoretical objects
in~1963 by \citet{HN63}, who performed the first truly self-consistent
estimate of the minimum hydrogen burning mass limit by computing the
luminosity at the surface and the energy release rate by nuclear
burning. Modern theory of the evolution and internal constitution of
BDs has advanced considerably owing to the inclusion of an improved
equation of state and realistic model-atmospheres that take into
account absorption by many molecular species as well as dust allowing
the identification of characteristic photometric signatures
\citep{ChB00}.  The first BDs were detected in~1995, and since then
they have been found in the solar neighbourhood and in young star
clusters \citep{Basri00} allowing increasingly sophisticated estimates
of their mass distribution.

For the solar neighbourhood, near-infrared large-scale surveys have
now identified many dozens of BDs probably closer than 25~pc
\citep{Allen_etal04}. Since these objects do not have reliable
distance measurements an ambiguity exists between their ages and
distances, and only statistical analysis that relies on an assumed
star-formation history for the solar neighbourhood can presently
constrain the IMF \citep{Ch02}, finding a 60~\% confidence interval
$\alpha_0 = 0.3\pm0.6$ for $0.04-0.08\,M_\odot$ approximately for the
Galactic-field BD IMF \citep{Allen_etal04}.

Surveys of young star clusters have also discovered BDs by finding
objects that extend the colour--magnitude relation towards the faint
locus while being kinematical members. Given the great difficulty of
this endeavour, only a few clusters now possess constraints on the
MF. The Pleiades star cluster has proven especially useful, given its
proximity ($d\approx127$~pc) and young age ($\tau_{\rm
cl}\approx100$~Myr). Results indicate $\alpha_0\approx0.5-0.6$
(Table~\ref{tab:apl}). Estimates for other clusters (ONC,
$\sigma$~Ori, IC~348; Table~\ref{tab:apl}) also indicate
$\alpha_0\simless0.8$. In their table~1, \citet{Allen_etal04}
summarise the available measurements for 11~populations finding that
$\alpha_0\approx 0-1$.

\subsection{BD binaries}

The above estimates of the IMF suffer under the same bias affecting
stars, namely unseen companions. BD--BD binary systems are known to
exist \citep{Basri00}, in the field \citep{Bouy_etal03,Close_etal03}
and in clusters \citep{Martin_etal03}. Their frequency is not yet very
well constrained since detailed scrutiny of individual objects is
time-intensive on large telescopes. But the results show conclusively
that the semi-major axis distribution of VLMSs and BDs is much more
compact than that of M~dwarfs, K~dwarfs and
G~dwarfs. \citet{Bouy_etal03, Close_etal03, Martin_etal03,
PhanBao_etal05} all find that BD binaries with semi-major axis
$a\simgreat 15$~AU are very rare. Using Monte-Carlo experiments on
published multiple-epoch radial-velocity data of VLMSs and BDs,
\citet{MF05} deduce an overall binary fraction between~32 and~45~\%
with a semi-major-axis distribution that peaks near~4~AU and is
truncated at about 20~AU.  In the Pleiades cluster where their offset
in the colour--magnitude diagram from the single-BD locus makes them
conspicuous, \citet{Pinfield_etal03} find the BD binary fraction may
be as high as 60~\%.

The truncated semi-major-axis distribution of BDs may be a result of
binary-disruption in dense clusters of an initial stellar-like
distribution. \citet{Kr_etal03} test this notion by setting-up the
hypothesis that BDs form according to the same pairing rules as stars,
the {\it star-like hypothesis}. This ought to be true since objects
with masses $0.04-0.072\,M_\odot$ should not have very different
pairing rules than stars that span a much larger range of masses
($0.1-1\,M_\odot$) but show virtually the same period-distribution
function (the M-, K- and G-dwarf samples,
\citealt{FiMa92,Mayor_etal92,DM91}, respectively). Thus, the
hypothesis is motivated by observed orbital distribution functions of
stellar binaries not being sensitive to the primary mass, which must
come about if the overall physics of the formation problem is similar.
Further arguments for a star-like origin of BDs comes from the
detection of accretion onto and disks about very young BDs, and that
the BDs and stars in Taurus-Auriga have indistinguishable spatial and
velocity distributions \citep{WhiteBasri03}.

\begin{figure}
\begin{center}
\rotatebox{0}{\resizebox{0.6 \textwidth}{!}{\includegraphics{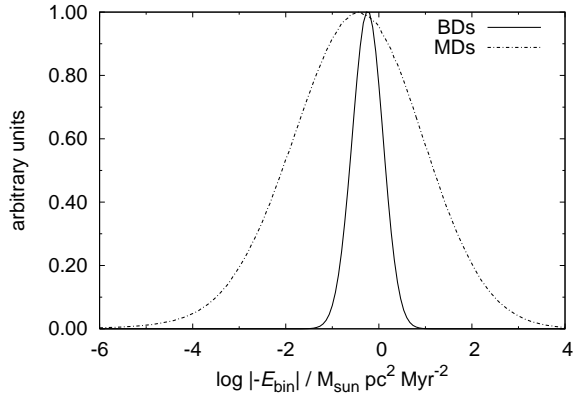}}}
\vskip -5mm
\caption
{\small{ The distribution of binding energies, $E_{\rm bin} =
    -G\,m_1\,m_2/(2\,a)$, of BDs (solid line) compared to those of
    M~dwarfs (dot-dashed line). The BD distribution is computed by
    adopting the semi-major-axis, $a$, distribution from
    \citet{Close_etal03} and choosing $10^7$ BD masses,
    $m_i\in(0.04-0.1\,M_\odot)$, from a power-law MF with
    $\alpha_0=0.3$. The M~dwarf energy distribution is computed by
    assuming the $a$-distribution from \citet{FiMa92} (which is
    practically identical to that of G~dwarfs) and choosing $10^7$
    masses, $m_i\in(0.1-0.5\,M_\odot)$, from a power-law MF with
    $\alpha_1=1.3$. The BD distribution is incomplete to the right of
    the solid line because tight BD pairs could not be discovered by
    the available surveys \citep{Bouy_etal03, Close_etal03, MF05}.  }}
\label{fig:BDbindEn}
\end{center}
\end{figure}

\citet{KAH, Kr_etal03} perform $N$-body calculations of ONC- and
Taurus-Auriga-like stellar aggregates to predict the semi-major-axis
distribution functions of BD--BD, star--BD and star--star binaries.
These calculations demonstrate that the binary proportion among BDs is
smaller than among low-mass stars after a few crossing times, owing to
their weaker binding energies.  The distribution of separations,
however, extends to similar distances as for stellar systems
($a\approx 10^3$~AU), disagreeing completely with the observed
distribution.  The star-like hypothesis thus predicts far too many
wide BD--BD binaries.  This can also be seen from the distribution of
binding energies of real BD binaries.  It is very different to that of
stars by having a low-energy cutoff, $^{\rm BD}E_{\rm bin, cut}\approx
-10^{-0.9}\,M_\odot\,$(pc/Myr)$^2$, that is much higher than that of
the M~dwarfs, $^{\rm M}E_{\rm bin, cut}\approx
-10^{-3}\,M_\odot\,$(pc/Myr)$^2$ (Fig.~\ref{fig:BDbindEn}).  This is a
very strong indicator for some fundamental difference in the dynamical
history of BDs which sets a different energy scale than for stars.

Furthermore, the theoretical distributions contain a substantial
number of star--BD pairs, which also disagrees with the observational
constrains that have found very few BD companions to nearby stars
\citep{Basri00, PhanBao_etal05}.  Basically, if BDs form exactly like
stars, then the number of star--BD binaries would be significantly
larger than is observed, since for example G-dwarfs prefer to pair
with M-dwarfs.  The observed general absence of BD companions is
referred to as the {\it BD desert} \citep{ZuMaz01}, since stellar
companions and planets are found at such separations
\citep{Halbw00,Vogt01}.  A few very wide systems with BD companions
can form during the final stages of dissolution of a small cluster
\citep{Fuente98}, and three such common proper-motion pairs have
perhaps been found \citep{Gizis01}.

Finally, the star-like hypothesis predicts far too few star--star
binaries in Taurus-Auriga, where binary disruption has not been
active.  \citet{Kr_etal03} thus conclude that the observed BD
population is incompatible with their hypothesis, such that BDs need
to be treated as a separate, or extra, population.

\subsection{The number of BDs per star}
\label{sec:BDnr}

\citet{Briceno_etal02} report that Taurus-Auriga appears to form
significantly fewer BDs per star than the ONC. Both systems are very
different physically but have similar ages of about 1~Myr.  This
finding was interpreted to be the first possible direct evidence of a
variable IMF, being nicely consistent qualitatively with the Jean-mass,
\begin{equation}
M_{\rm J}\propto \rho^{-1/2}T^{3/2},
\label{eq:MJeans}
\end{equation}
being larger in Taurus-Auriga than in the ONC because its gas density,
$\rho$, is smaller by one--two orders of magnitude, while the
temperatures, $T$, are similar to within a factor of few.

Given this exciting finding, \citet{Kr_etal03} computed $N$-body
models of the stellar aggregates in Taurus-Auriga in order to
investigate the hypothesis that BDs form star-like. They find that the
same initial number of BDs per star in Taurus-Auriga and in the ONC
actually leads to different observed ratios because BD--BD and star--BD
binaries are disrupted more efficiently in the ONC; the observer thus
sees many more BDs there than in the comparatively dynamically
unevolved Taurus-Auriga groups. But, as already noted above, this
approach fails because the \citet{Kr_etal03} model has too many wide
BD--BD binaries, and also it predicts too many star--BD
binaries. Given this problem, \citet{Kr_Bouv03b} study the production
rate of BDs per star assuming BDs are a separate population, such as
ejected embryos \citep{RC01}.  Again they find that both, the
physically very different environments of Taurus-Auriga and the ONC,
can have produced the same ratios (about one BD per 4--5 stars) if BDs
are ejected embryos with a dispersion of ejection velocities of about
2~km/s.

Based on some additional observations, \citet{Luhman04} revised the
\citet{Briceno_etal02} results by finding that the number of BDs per
star had been underestimated in Taurus-Auriga. Since the new
spectroscopic study of \citet{Slesnick_etal04} also revised the number
of BDs per star in the ONC downwards, \citet{Luhman04} retracts the
significance of the claimed difference of the ratio in Taurus-Auriga
and the ONC. Is a universal, invariant, BD production scenario still
consistent with the updated numbers?

Let the true ratio of the number of BDs per late-type star be
\begin{equation}
R \equiv { N(0.02-0.08\,M_\odot) \over N(0.15-1.0\,M_\odot)} 
  \equiv { N_{\rm BD,tot} \over N_{\rm st,tot}}.
\end{equation}
Note that here stars more massive than $10\,M_\odot$ are not counted
because this would not make much sense given that Taurus-Auriga is
mostly producing late-type stars given the limited gas mass available.
But the observed ratio is
\begin{equation}
R = { N_{\rm BD,obs} \over N_{\rm st,obs}} = N_{\rm BD,tot}({\cal B}+{\cal U}) 
    {(1+f) \over N_{\rm st,tot}},
\end{equation}
since the observed number of BDs, $N_{\rm BD,obs}$ is the total number
produced multiplied by the fraction of BDs that are gravitationally
bound to the population (${\cal B}$) plus the unbound fraction, ${\cal
U}$, which did not yet have enough time to leave the survey
area. These fractions can be computed for dynamical models of the
Taurus-Auriga and ONC and depend on their mass and the dispersion of
velocities of the BDs.  This velocity dispersion can either be the
same as that of the stars if BDs form like stars, or larger if they
are ejected embryos \citep{RC01}.  The observed number of ``stars'' is
actually the number of systems such that the total number of
individual stars is $N_{\rm st,tot} = (1+f)\,N_{\rm st,obs}$, where $f
\equiv N_{\rm st, bin} / N_{\rm st,sys}$ is the binary fraction and
$N_{\rm st, bin}$ is the number of binary systems in a sample of
$N_{\rm st, sys}$ stellar systems. Note that here no distinction is
made between single or binary BDs.  For Taurus-Auriga,
\citet{Luhman04} observes $R_{\rm TA, obs} = 0.25$ from which follows
\begin{equation}
R_{\rm TA} = 0.18 \;\; {\rm since} \;\;  f_{\rm TA}=1, {\cal B} + {\cal U} = 0.35+0.35
\end{equation}
\citep{Kr_etal03}.  According to \citet{Slesnick_etal04},
the revised ratio for the ONC is $R_{\rm ONC,obs} = 0.28$ so that
\begin{equation}
R_{\rm ONC}=0.19 \;\; {\rm because} \;\; f_{\rm ONC} = 0.5, {\cal B} + {\cal U} = 1+0
\end{equation}
\citep{Kr_etal03}. Note that the regions around the stellar
groupings in Taurus-Auriga not yet surveyed should contain about 30~\%
of all BDs, while all BDs are retained in the ONC.

Therefore, the updated numbers imply that about {\it one BD is
produced per five late-type stars}, and that the dispersion of
ejection velocities is $\sigma_{\rm ej} \approx 1.3$~km/s. These
numbers are an update of those given in \citet{Kr_etal03}, but the
results have not changed much. Note that a BD with a mass of
$0.06\,M_\odot$ and a velocity of 1.3~km/s has a kinetic energy of
$10^{-1.29}\,M_\odot\,$(pc/Myr)$^2$ which is rather comparable to the
cut-off in BD--BD binding energies (Fig.~\ref{fig:BDbindEn}). {\it
This supports the notion that most BDs may be mildly ejected embryos}.

BDs can come in different flavours \citep{Kr_Bouv03b}: star-like BDs,
ejected embryos, collisional BDs and photo-evaporated BDs. As seen
above, star-like BDs appear to be very rare, because BDs do not mix
with stars in terms of pairing properties; some mechanism ensures the
two types of object are kept separated. Ejected embryos appear to be a
promising production channel, as noted above. \citet{Kr_Bouv03b}
discount the collisional removal of accretion envelopes for the
production of unfinished stars because this process is far too slow,
or rare. The removal of accretion envelopes through photo-evaporation
can occur, but only within the immediate vicinity of an O~star and
never in Taurus-Auriga \citep{WhitZinn04}. However, \citet{Kr_Bouv03b}
show that the radius within which photo-evaporation may be able to
remove substantial fractions of an accretion envelope within $10^5$~yr
is comparable to the cluster size in star-burst clusters that contain
thousands of O~stars. In such clusters photo-evaporated BDs may be
very common. Globular clusters may then be full of BDs.

Returning to the local star-forming regions, it thus now appears that
the very different physical environments evident in Taurus-Auriga and
the ONC produce the same number of BDs per late-type star, so that
there is no evidence for differences in the IMF across the hydrogen
burning mass limit. There is also no substantial evidence for a
difference in the stellar IMF in these two star-forming regions,
contrary to the assertion by \citet{Luhman04}. Fig.~\ref{fig:TAmfn}
shows the IMF in Taurus-Auriga: it is similar to the standard IMF
(eq.~\ref{eq:imf} below), provided corrections for the high
multiplicity are made. And in turn, the standard IMF is comparable to
that seen in the ONC (Fig.~\ref{fig:mfn1}). The caveat that the
classical pre-main sequence evolution tracks, upon which the
observational mass determinations rely, are not really applicable for
such young ages (\S~\ref{sec:stcl}) needs to be kept in mind though.
\begin{figure}
\begin{center}
\rotatebox{0}{\resizebox{0.75 \textwidth}{!}{\includegraphics{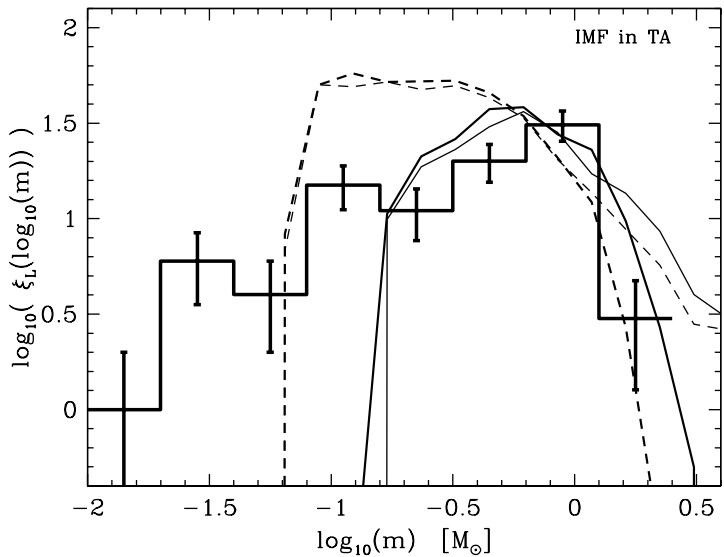}}}
\vskip -20mm
\caption
{\small{ The thick histogram shows the Taurus-Auriga MF observed by
\citet{Luhman_etal03}, and the dashed curves are the standard IMF
(eq.~\ref{eq:imf}). The solid curves are the model from
\citet{Kr_etal03} assuming all stars are in unresolved binary systems.
The model shown by thick curves assumes stars only form in the mass
interval $0.07-1.5\,M_\odot$, while the thin curves are for stars in
the mass range $0.07-50\,M_\odot$. BDs would be part of the model as
an extra population matched by the thick histogramme but are not
explicitly shown in the model. The binary-star model (solid lines) is
in good agreement with the observed MF in the stellar regime.  For
details see \citet{Kr_etal03}. Note the discontinuity in the IMF
evident between the dashed curve and the solid histogramme at
$10^{-1.2}\,M_\odot$.  }}
\label{fig:TAmfn}
\end{center}
\end{figure}

\section{The Shape of the IMF}
From the above discourse it thus becomes apparent that the
distribution of stars by mass is a power-law with exponent or index
$\alpha_2\approx 2.35$ for stellar masses $m\simgreat
0.5\,M_\odot$. There exists a physical stellar mass limit, $m_{\rm max
*}\approx 150\,M_\odot$ such that $m\le m_{\rm max *}$
(\S~\ref{sec:maxlim}).  The distribution of stars below the K/M~dwarf
transition mass, $0.5\,M_\odot$, can also be described by a power law,
but with exponent $\alpha_1\approx 1.3$ (\S~\ref{sec:lmst}).  Given
the latest results described in \S~\ref{sec:bds}, the
mass-distribution below the mass $m_1 \in (0.072-0.14)\,M_\odot$ is
uncertain, but measurements indicate a power-law with exponent $0 <
\alpha_0 < 1$. Because the binary properties of VLMSs and BDs differ
substantially from those in the low-mass star regime, it can be argued
that VLMSs and BDs need to be considered as a separate population that
is linked-to, but different from stars. This would then open the
likely-hood that the IMF is discontinuous near $m_1$, as may already
be evident in Fig.~\ref{fig:TAmfn}.  And so fitting a functional
description of the mass distribution with the continuity constraint
would be wrong. For this reason it would appear to be dangerous, and
in the worst case unphysical, to use one single function such as the
log-normal form to attempt to describe the mass distribution from the
fragmentation limit\footnote{When a cloud collapses its density
increases but its temperature remains constant as long as the opacity
remains low enough to enable the contraction work to be radiated
away. The Jeans mass (eq.~\ref{eq:MJeans}) consequently decreases and
further fragments with smaller masses form. When, however, the density
increases to a level such that the cloud core becomes optically thick,
then the temperature increases, and the Jeans mass follows suit. Thus
an opacity-limited minimum fragmentation mass is arrived at
\citep{LowLyndenBell76,Boss86, Kumar03}.}, $m_0\approx 0.01\,M_\odot$,
through to the physical stability limit $m_{\rm max*}$. Indeed,
\citet{Chrev03} already noted~(i) that a single log-normal function
over the entire stellar mass range is not compatible with the data, a
power-law form for $m>1\,M_\odot$ being required, and~(ii) that the
continuous IMF predicts too many VLMSs and BDs by a factor of about
three.

With these recent insights (power-law IMF over two orders of magnitude
in mass and probable discontinuity near the sub-stellar mass limit),
little of the argument for the advantages of a log-normal or any other
mathematical form (Table~\ref{tab:imfs} below) remains. Indeed, any
such other mathematical form has the little nastiness that the tails
of the distribution react to changes in the parametrisation in a way
perhaps not wanted when testing models. To give an example: a single
log-normal form would change the slope of the IMF at large masses even
if only the LF for late-type stars is to be varied. The standard
(eq.~\ref{eq:imf}) two-part power-law stellar IMF, on the other hand,
would allow changes to the index at low masses without affecting the
high-mass end, and the addition of further power-law segments is
mathematically convenient. The standard two-part power-law stellar IMF
also captures the essence of the physics of star formation, namely a
featureless power-law form for the largest range of stellar masses and
a turnover near some fraction of a solar mass which probably is a
result of increasing inefficiency of star formation
(\S~\ref{sec:var_th}).  The overall shape of the IMF then sets the
characteristic or average stellar mass.

\subsection{The standard or average IMF}
The various constraints arrived at above are summarised by a
multi-part power-law,
\begin{equation}
\xi (m) = k\left\{
          \begin{array}{l@{\quad\quad,\quad}l@{\quad,\quad}l}
   \left({m\over m_1}\right)^{-\alpha_0}  &m_0 < m \le m_1 &n=0,\\
   \left({m\over m_1}\right)^{-\alpha_1}  &m_1 < m \le m_2 &n=1,\\
   \left[
       \prod\limits^{n\ge2}_{i=2}\left({m_i\over
          m_{i-1}}\right)^{-\alpha_{i-1}}
       \right] 
        \left({m\over m_n}\right)^{-\alpha_n} 
        &m_n < m \le m_{n+1}  &n\ge2,\\
          \end{array}\right.
\label{eq:imf_mult}
\end{equation}
where $k$ contains the desired scaling and the mass-ratios ensure
continuity, with
\begin{equation}
          \begin{array}{l@{\quad\quad,\quad}ll@{\quad,\quad}l}
\alpha_0 = +0.3\pm0.7   &0.01 &\le m/M_\odot < 0.08 &n=0, \\
\alpha_1 = +1.3\pm0.5   &0.08 &\le m/M_\odot < 0.50 &n=1, \\
\alpha_2 = +2.3\pm0.7   &0.5  &\le m/M_\odot \simless 150  &n=2.
          \end{array}
\label{eq:imf}
\end{equation}
The uncertainties are discussed in \S~\ref{sec:apl}

This two-part power-law stellar IMF is referred to as the {\it
standard} or {\it canonical stellar IMF}.  It is also the {\it average
IMF} since it rests upon a derivation from local star-counts for
$m<1\,M_\odot$ that sample star-formation over the age of the MW disk
(about 10~Gyr), and upon an averaging of the results obtained by the
surveys of many OB associations and massive clusters for
$>\,$few$\,M_\odot$ (Fig.~\ref{fig:kroupa_figmassey}).
Eq.~\ref{eq:imf} is therefore a good bench-mark against which
individually-derived IMFs can be compared to study possible systematic
changes of the IMF with changing physical conditions. 

A larger value for $\alpha_3\approx 2.7$ in the {\it stellar IMF} may
be closer to the truth than the above canonical Salpeter/Massey value
($\alpha_3=2.3$) when binary-companions are corrected for
(\S~\ref{sec:massst}, \citealt{SR91}), in this case arriving at the
{\it KTG93 IMF} \citep{KTG93},
\begin{equation}
          \begin{array}{l@{\quad\quad,\quad}ll@{\quad,\quad}l}
\alpha_0 = +0.3\pm0.7   &0.01 &\le m/M_\odot < 0.08 &n=0, \\
\alpha_1 = +1.3\pm0.5   &0.08 &\le m/M_\odot < 0.50 &n=1, \\
\alpha_2 = +2.3\pm0.3   &0.5  &\le m/M_\odot  <1    &n=2, \\
\alpha_3 = +2.7\pm0.7   &1    &\le m/M_\odot        &n=3.
          \end{array}
\label{eq:KTGimf}
\end{equation}
Note that in this case the observational data would not constrain the
physical upper stellar mass limit \citep{WK04}. A detailed study of
systematic biases affecting power-law index measurements based on
binned data is under way and will allow deeper insights into these
issues \citep{Jesus05}.

Note that $\alpha_3 = 2.7$ (the Scalo index) was also derived by
\citet{Sc86} for the MW disk population of massive stars such that the
three-part-power law KTG93 IMF needs to be identified with the {\it
composite IMF} introduced in \S~\ref{sec:comppop}. Whether the stellar
IMF needs to be changed away from the Salpeter/Massey value
($\alpha_3=2.3$) because of corrections for unresolved multiples, and
by how much the composite IMF then differs from the stellar IMF above
$1\,M_\odot$, thus needs to be studied in more detail.

\subsection{The alpha plot}
\label{sec:apl}
\begin{figure}
\begin{center}
\rotatebox{0}{\resizebox{0.99 \textwidth}{!}{\includegraphics{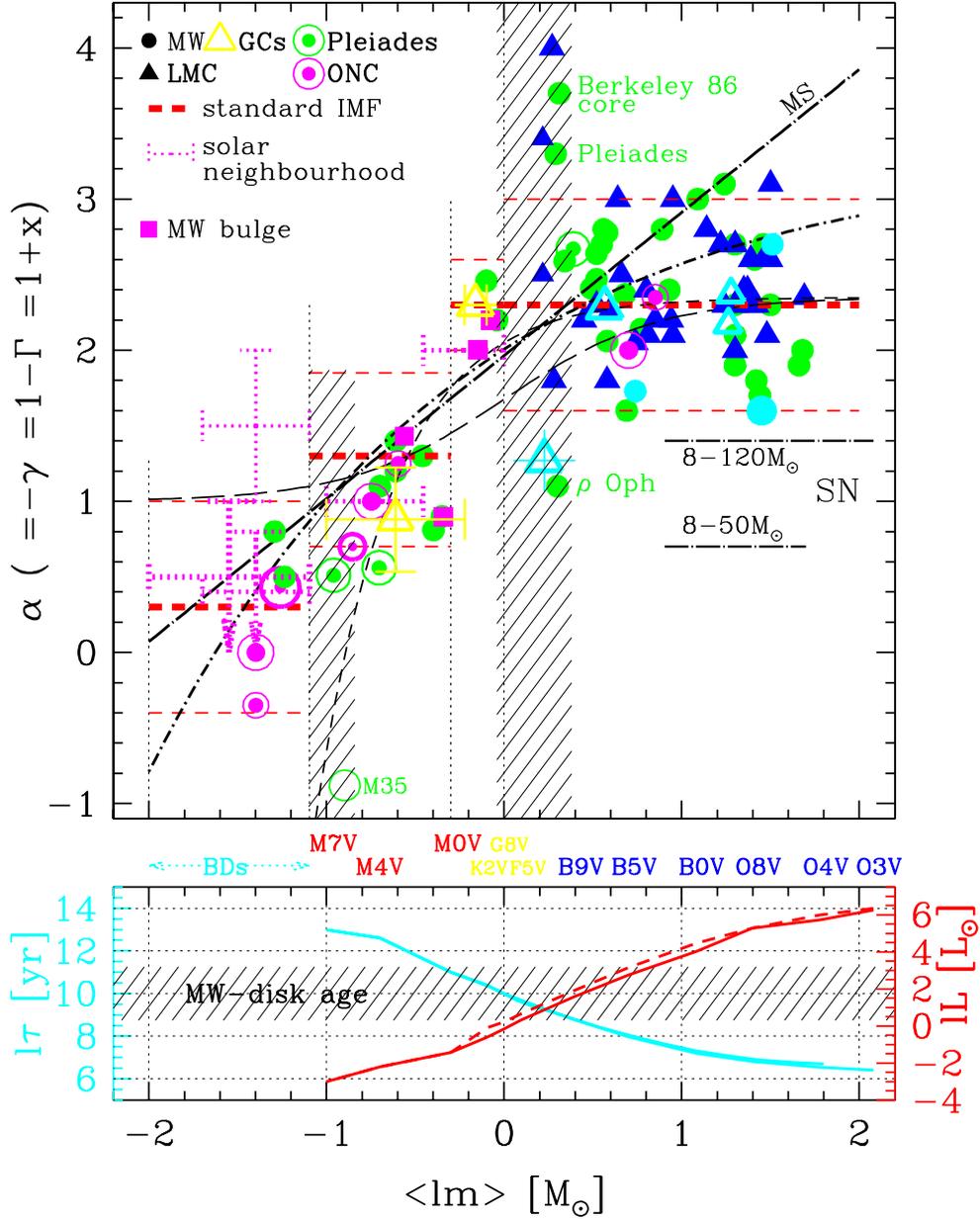}}}
\vskip -1mm
\caption[]{\small{ The alpha plot (upper panel), and bolometric MLR,
$lL(lm)$, stellar main-sequence life-time, $l\tau$, and possible range
of Milky-Way (MW) disk ages shown as the shaded region (lower
panel). Notation: $lm\equiv {\rm log}_{10}(m/M_\odot)$, $l\tau\equiv
{\rm log}_{10}(\tau/{\rm yr})$, $lL\equiv {\rm log}_{10}(L/L_\odot)$.
For more details see text. }}
\label{fig:apl}
\end{center}
\end{figure}

A convenient way for summarising various studies of the IMF is by
plotting independently-derived power-law indices in dependence of the
stellar mass over which they are fitted
\citep{Sc98,K01a,K02,Hillrev04}.  Fig.~\ref{fig:apl} plots such data:
The shape of the IMF is mapped by plotting measurements of $\alpha$ at
$<\!lm\!>=(lm_2-lm_1)/2$ obtained by fitting power-laws,
$\xi(m)\propto m^{-\alpha}$, to logarithmic mass ranges $lm_1$ to
$lm_2$ (not indicated here for clarity).  Circles and triangles are
data compiled by \citet{Sc98,K01a} for MW and Large-Magellanic-Cloud
(LMC) clusters and OB associations, as well as newer data, some of
which are emphasised using different symbols (and colours)
(Table~\ref{tab:apl}).  Unresolved multiple systems are not corrected
for in all these data including the MW-bulge data.  The standard
stellar IMF (eq.~\ref{eq:imf}, corrected for unseen binary-star
companions for $m<1\,M_\odot$) is the two-part power-law (thick
short-dashed lines).  

Note that the M~dwarf ($0.1-0.5 \,M_\odot$) MFs for the various
clusters are systematically flatter (smaller $\alpha_1$) than the
standard IMF, which is mostly due to unresolved multiple systems in
the observed values.  Some of the data do coincide with the standard
IMF though, and \citet{K01a} argues that on correcting these for
unresolved binaries the underlying true single-star IMF ought to have
$\alpha_1\approx 1.8$. This may indicate a systematic variation of
$\alpha_1$ with metallicity because the data are young clusters that
are typically more metal-rich than the average Galactic field
population for which $\alpha_1=1.3$ (cf. eq.~\ref{eq:systemvar} below)

A power-law extension into the BD regime with a smaller index
($\alpha_0=+0.3$) is shown as a third thick short-dashed segment, but
this part of the mass distribution may not be a continuous extension
of the stellar distribution, as noted above.  The upper and lower thin
short-dashed lines are the estimated 99~\% confidence range on
$\alpha_i$ (eq.~\ref{eq:imf}).  Other binary-star-corrected
solar-neighbourhood-IMF measurements are indicated as (magenta) dotted
error-bars (Table~\ref{tab:apl}).  Note that for $m>1\,M_\odot$
correction for unseen companions may steepen the standard IMF to
$\alpha\approx2.7$.

The long-dash-dotted horizontal lines in Fig.~\ref{fig:apl} labelled
``SN'' are those IMFs with $\alpha_3=0.70 (1.4)$ but $\alpha_0,
\alpha_1, \alpha_2$ as in eq.~\ref{eq:imf}, for which 50~\% of the
stellar (including BD) mass is in stars with $8 - 50 (8 -
120)\,M_\odot$, respectively. It is noteworthy that none of the
available clusters, not even including the star-burst clusters
(Table~\ref{tab:apl}), has such a top-heavy IMF. Any hypothetical
cluster so dominated by massive stars would disperse due to the mass
loss from the supernova explosions.

The vertical dotted lines in Fig.~\ref{fig:apl} delineate the four
mass ranges (eq.~\ref{eq:imf}), and the shaded areas highlight those
stellar mass regions where the derivation of the IMF is additionally
complicated especially for Galactic field stars: for $0.08<m/M_\odot<
0.15$ long pre-main sequence contraction times \citep{ChB00} make the
conversion from an empirical LF to an IMF (eq.~\ref{eq:mf_lf})
dependent on the precise knowledge of stellar ages and the SFH, and
for $0.8< m/M_\odot<2.5$ uncertain main-sequence evolution,
Galactic-disk age and the SFH of the MW disk do not allow accurate IMF
determinations \citep{Binney00}.  

\subsection{The distribution of data points in the alpha-plot}
\label{sec:ahist}
The first thing to note about the data distribution in the alpha-plot
is that there is (unfortunately) no clear systematic difference in IMF
determinations with metallicity nor density of the population
(cf. Fig.~\ref{fig:kroupa_figmassey}).

\begin{figure}
\begin{center}
\rotatebox{0}{\resizebox{0.8\textwidth}{!}{\includegraphics{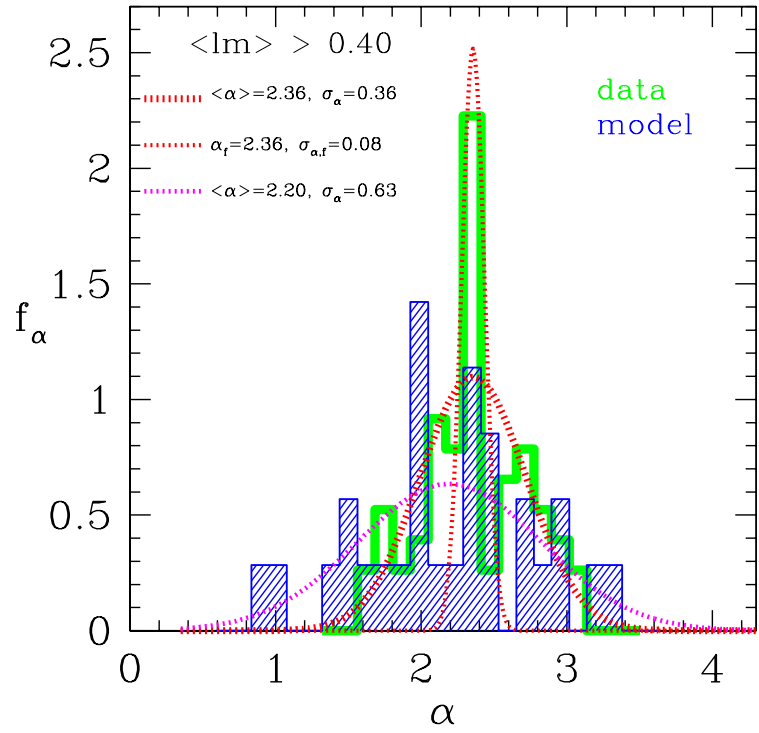}}}
\vskip -15mm
\caption
{\small{Distribution of $\alpha$ values for massive stars. See text
for details.}}
\label{fig:ahist}
\end{center}
\end{figure}

In order to understand the origin and nature of the dispersion of
power-law indices evident in the alpha plot, \citet{K01a}
investigates the dispersion of $\alpha$ values for a given mass
interval theoretically. The result is that the dispersion can be
understood in terms of statistical sampling from a universal IMF (as
also found by \citealt{Elm97,Elm99}) together with stellar-dynamical
biases. Given such a theoretical investigation, it is now possible to
compare the theoretical distribution of $\alpha$ values for an
ensemble of star clusters with the observational data. This is done
for stars with $m>2.5\,M_\odot$ in Fig.~\ref{fig:ahist} where the open
(green) histogram shows the distribution of observational data from
Fig.~\ref{fig:apl}. The (blue) shaded histogram is the theoretical
ensemble of 12 star clusters containing initially 800 to $10^4$~stars
that are ``observed'' at 3 and 70~Myr: stellar companions are merged
to give the system MFs, which are used to measure $\alpha$, but the
input single-star IMF is in all cases the standard form
(eq.~\ref{eq:imf}).  The dotted curves are Gaussians with the same
average $\alpha$ and standard deviation in $\alpha$ obtained from the
histograms. Fixing $\alpha_{\rm f}=<\!\!\alpha\!\!>$ and using only
$\!\mid \alpha \!\mid \le 2\,\sigma_\alpha$ for the observational data
gives the narrow thin (red) dotted Gaussian distribution which
describes the Salpeter peak extremely well (not by construction).

The interesting finding is thus that the observational data have a
very pronounced Salpeter/Massey peak, with broad near-symmetric
wings. This indicates that there are no serious biases that should
skew the distribution. For example, if the observational sample
contained clusters or associations within which the OB stars have a
low binary fraction compared to others that have a very high
multiplicity fraction, we would expect the binary-deficient cases to
deviate towards low $\alpha$ values. Likewise, energetic dynamical
ejections from cluster cores would skew the IMF towards more massive
stars (reduced $\alpha$) while mass segregation has a similar effect.

In contrast, the theoretical data show (i) a distribution with a mean
shifted to smaller $\alpha\approx 2.2$ that has (ii) a larger width
than the observational one. The input standard Salpeter/Massey index
is not really evident in the theoretical data, and if these were the
observational data then it is likely that the astronomical community
would strongly argue for the case that the IMF shows appreciable
variations. It is peculiar that the empirical data are so much better
behaved, since all the additional complications (theoretical stellar
models, rotating/non-rotating stars) are not modelled in the
theoretical data.  The elucidation of the reason for the difference
between the much more ``well-behaved'' observational data and the
theoretical data will need further theoretical work which will have to
attempt to re-produce the observational procedure as exactly as is
possible.

\subsection{Mass density and some other numbers}
The nearby Hipparcos LF, $\Psi_{\rm near}({\rm Hipp})$, has
$\rho=(5.9\pm0.3)\times10^{-3}$~stars/pc$^3$ in the interval
$M_V=5.5-7.5$ corresponding to the mass interval $m_2=0.891 -
0.687\,M_\odot$ \citep{K01b} using the KTG93 MLR
(Fig.~\ref{fig:mlr}). $\int_{m_1}^{m_2}\xi(m)\,dm=\rho$ yields
$k=0.877\pm0.045$~stars/(pc$^3\,M_\odot$).  The number fractions, mass
fractions and Galactic-field mass densities contributed by stars in
different mass-ranges are summarised in Table~\ref{tab:frac}. 

The local mass density made up of interstellar matter, $\rho^{\rm
gas}\approx0.04\pm0.02\,M_\odot$/pc$^3$, and stellar remnants,
$\rho^{\rm rem}\approx0.003\,M_\odot$/pc$^3$ \citep{Weide}, or
$\rho^{\rm rem}\approx0.005\,M_\odot$/pc$^3$ (\citealt{Chrev03} and
references therein).  Giant stars contribute about $0.6\times
10^{-3}\,M_\odot$pc$^{-3}$ \citep{Haywood_etal97}, so that
main-sequence stars make up about half of the baryonic matter density
in the local Galactic disk (Table~\ref{tab:frac}).  BDs, which for
some time were regarded as candidates for contributing to the
dark-matter problem, do not constitute a dynamically important mass
component of the Galaxy, even when eq.~\ref{eq:imf} is extrapolated to
$0.0\,M_\odot$ giving $\rho^{\rm BD} = 3.3\times 10^{-3}\,
M_\odot$/pc$^3$. This is corroborated by dynamical analysis of local
stellar space motions that imply there is no dark matter in the
Milky-Way disk \citep{FF94}, and the revision of the thick-disk mass
density to larger values \citep{Fuhrmann04,Soubiran_etal03} further
reduces the need for dark matter within the solar circle ({cf
\citealt{Chrev03}).

Table~\ref{tab:frac} also shows that a star cluster looses about
13~per cent of its mass through stellar evolution within 10~Myr if
$\alpha_3=2.3$ (turnoff-mass $m_{\rm to}\approx20\,M_\odot$), or
within 300~Myr if $\alpha_3=2.7$ (turnoff-mass $m_{\rm
to}\approx3\,M_\odot$). After 5~Gyr the mass loss through stellar
evolution alone amounts to about 45~per cent if $\alpha_3=2.3$ or
29~per cent if $\alpha_3=2.7$.  Mass loss through stellar evolution
therefore poses no risk for the survival of star clusters for the IMFs
discussed here, since the mass-loss rate is small enough for the
cluster to adiabatically re-adjust.  A star-cluster would be
threatened through mass loss from supernova explosions if
$\alpha\approx1.4$ for $8<m/\,M_\odot\le 120$ which would mean a
mass-loss of 50~per cent within about 40~Myr when the last supernova
explodes.  It is remarkable that none of the measurements has found
such a low $\alpha$ for massive stars (Fig.~\ref{fig:apl}).

\subsection{Other IMF forms}
As already discussed at the beginning of this section, the standard or
canonical power-law IMF (eq.~\ref{eq:imf}) provides a good description
of the data combined with mathematical ease and physical meaning. A
strong advantage of this parametrisation is that each section can be
changed without affecting another part of the IMF, as stated above. As
an explicit example, should the BD MF be revised, $\alpha_0$ can be
adopted accordingly without affecting the rest of the mass
distribution in the well-tested stellar regime.

But additional forms are in use and are preferred for some
investigations: In Fig.~\ref{fig:apl} the quasi-diagonal (black) lines
are alternative analytical forms summarised in Table~\ref{tab:imfs}.
Of these the generalised Rosin-Rammler function (eq.~{\it Ch} in
Table~\ref{tab:imfs}, thick short-dash-dotted curve) best represents
the data, apart from a deviation for $m\simgreat 10\,M_\odot$, which
\citet{Chrev03} fixes by adopting a Salpeter/\-Mas\-sey power-law
extension for $m>1\,M_\odot$.  Interpretation of $m_o$ in terms of a
characteristic mass poses difficulties.  The popular Miller-Scalo
log-normal IMF (eq.~{\it MS} in Table~\ref{tab:imfs}) deviates even
more strongly from the empirical data at high masses. Larson's
eq.~{\it Lb} in Table~\ref{tab:imfs} fits rather well, except that it
may predict too many BDs.  Finally, the {\it effective initial mass
function for galactic disks} proposed by \citet{Hollenbach_etal05}
(eq.~{\it Holl}) reproduces the data in the alpha-plot quite well
(fig.~1 in \citealt{Hollenbach_etal05}), and is not incorporated into
Fig.~\ref{fig:apl} here. Note however that a {\it composite IMF} ought
to be steeper (have a larger $\alpha$) at high masses
(\S~\ref{sec:comppop}).

The closed functional IMF formulations (eqs.~{\it MS, La, Lb, Ch,
Holl}) have the advantage that possible variations of the IMF with
physical conditions can be studied more naturally than with a
multi-power-law form, because they typically have a characteristic
mass that can be varied directly.

\section{Variation of the IMF and theoretical aspects}
\label{sec:var_th}

Is the scatter of data points in the alpha-plot (Fig.~\ref{fig:apl}) a
result of IMF variations? For this to be conclusively convincing would
require a measurement to lie further from the standard IMF than the
conservative uncertainty range adopted in eq.~\ref{eq:imf} and shown
in the figure.  However, the adopted uncertainties on $\alpha_i$ in
eq.~\ref{eq:imf} stem from the scatter in the alpha plot, so that this
argument is circular.

An independent indication of the uncertainties inherent to IMF
determinations can be obtained by comparing IMF estimates of the same
cluster by different authors. This is demonstrated for the
well-studied Pleiades, ONC and for 30~Dor (Table~\ref{tab:apl},
Fig.~\ref{fig:apl}). Overall, the uncertainties in $\alpha$ are about
$\pm0.5$ which is also about the scatter evident in all the data, so
that there is no indication of significant outliers (except in the
shaded regions, see below).  Substantial differences for VLMSs and BDs
are evident for the extremely young ONC allowing an estimate of likely
non-physical variations in the alpha plot. Data reduction at these low
masses is hampered by variability, variable reddening, spurious
detections \citep{Slesnick_etal04}.  It is clear that because the
procedure of measuring $\alpha(lm)$ is not standardised and because
the IMF is not a single power-law, author--author variations occur
simply due to the use of different mass ranges when fitting
power-laws. Here the work by \citet{Jesus05}, who is addressing such
issues, is important.

Significant departures from the standard IMF only occur in the shaded
areas of the alpha plot. These are, however, not reliable. The upper
mass range in the shaded area near $1\,M_\odot$ poses the problem that
the star-clusters investigated have evolved such that the turn-off
mass is near to this range.  Some clusters such as $\rho$~Oph are so
sparse that more massive stars did not form. In these cases the shaded
range is close to the upper mass limit leading to possible stochastic
stellar-dynamical biases since the most massive stars meet near the
core of a cluster due to mass segregation, but three-body or
higher-order encounters there can cause expulsions from the
cluster. Furthermore, $\rho$~Oph is still forming, leading to unknown
effects that are likely to enhance variations in the first derivative
of the IMF (i.e. in $\alpha$ values).  Dynamical ejections are
probably the cause for the ``OB field-star'' MF which has
$\alpha\approx4.5$ and is interpreted to be the result of isolated
high-mass star-formation in small clouds \citep{M98}. Precise
proper-motion measurements have, however, shown that even the best
candidates for this isolated population have very high space motions
\citep{Ram01} which are the result of energetic stellar-dynamical
ejections when massive binary systems interact in the cores of
star-clusters in normal but intense star-forming regions, thus posing
important constraints on the properties of OB binary systems
\citep{ClarkePringle92,K01c,P-AK06a}.

The shaded area near $0.1\,M_\odot$ poses the problem that the VLMSs
are not on the main sequence for most of the clusters studied, and are
again prone to bias through mass-segregation by being underrepresented
within the central cluster area that is easiest to study
observationally. Especially the latter is probably biasing the M35
datum, but some effect with metallicity may be operating, especially
so since M35 appears to have a smaller $\alpha$ near the H-burning
mass limit than the Pleiades cluster which has a similar age but has a
larger abundance of metals (Fig.~\ref{fig:mfn1}). The M35 cluster
ought to be looked at again with telescopes.

To address stellar-dynamical biases, an extensive theoretical library
of binary-rich star clusters has been assembled \citep{K01a} covering
150~Myr of stellar-dynamical evolution taking into account stellar
evolution and assuming the standard IMF (eq.~\ref{eq:imf}) in all
cases. This extends the notion raised by \citet{Elm99} that the
scatter seen in the empirical alpha-plot may be due to finite$-N$
sampling from a universal IMF that can be interpreted as a probability
density distribution.  Evaluating the MF within and outside of the
clusters, at different times and for clusters containing initially
$800-10^4$ stars, leads to a theoretical alpha-plot which reproduces
the spread in $\alpha(lm)$ values evident in the empirical alpha-plot
quite well. This verifies the conservative uncertainties adopted in
the standard IMF (eq.~\ref{eq:imf}) but also implies that the scatter
in the empirical alpha-plot around the standard IMF cannot be
interpreted as true variations. This work also addresses the bias due
to unresolved binary systems, with the result that typical open
clusters ought to be underrepresented in VLMSs. This bias is
incorporated in the definition of the standard IMF, as is evident in
the empirical alpha-plot.

One peculiar feature of the empirical alpha-plot that merits further
study is the distribution of $\alpha$ data for $m>2.5\,M_\odot$
(\S~\ref{sec:ahist}). A histogram of the empirical data
(Fig.~\ref{fig:ahist}) shows a narrow peak positioned at the Salpeter
value, with symmetric broad wings.  The empirical data are not
distributed normally.  The aforementioned theoretical alpha-plot shows
a very different distribution, the width being {\it larger} than for
the empirical data.  Interestingly, the spread,
$\sigma_{\alpha,f}=0.08$, of the narrow peak in the empirical data is
very similar to the uncertainties quoted by Massey in an extensive and
consistent observational determination of the IMF for massive stars,
$\alpha=2.2\pm0.1$ (Fig.~\ref{fig:kroupa_figmassey}). If $\alpha=
2.3\pm0.1$ is adopted for massive stars, then the measurement
$\alpha=1.6\pm0.1$ for the massive but difficult-to-observe Arches
cluster near the Galactic centre (Table~\ref{tab:apl}) would
definitely mean an IMF that is top-heavy for this extreme population.
But as discussed in \S~\ref{sec:intro}, the Arches is suffering
heavily from the tidal forces that strip its less massive members so
that the PDMF is likely skewed towards massive stars. \cite{Kimetal06}
indeed verify the near-Salpeter/Massey IMF in the Arches cluster.
{\it The question thus remains if the non-Gaussian distribution of
empirical $\alpha$ values contains information on a possible physical
variation of the IMF. }

Two other well-studied massive star-burst clusters have $\alpha
\approx\,${\it Salpeter/Massey} (30~Dor and NGC~3603,
Table~\ref{tab:apl}) implying no clear evidence for a bias that
resolved star-burst clusters prefer smaller $\alpha$ and thus more
massive stars relatively to the number of low-mass stars. Low-mass
stars are known to form in 30~Dor \citep{Sirianni_etal00}, although
their MF has not been measured yet due to the large distance of about
55~kpc. From the ONC we know that the entire mass spectrum
$0.05\simless m/M_\odot\simless 60$ is represented roughly following
the standard IMF (Fig.~\ref{fig:mfn1}, \citealt{P-AK06a}). The
Pleiades appears to have had an IMF very similar to the standard one
(Fig.~\ref{fig:MorauxPl}), although for massive stars a steeper IMF
with $\alpha_3\approx 2.7$ appears to be suggested by the theoretical
work \citep{MKB04}.

The available evidence is thus that low-mass stars and massive stars
form together even in extreme environments without, as yet, convincing
demonstration of a variation of the number ratio.  This is also
supported by an impressive observational study \citep{L00,Luhman04} of
many close-by star-forming regions using one consistent methodology to
avoid author--author variations.  The result is that the IMF does not
show measurable differences from low-density star-forming regions in
small molecular clouds ($n= 0.2-1$~stars/pc$^3$ in $\rho$~Oph) to
high-density cases in giant molecular clouds ($n= (1-5)\times
10^4$~stars/pc$^3$ in the ONC). This result extends to the populations
in the truly exotic ancient and metal-poor dwarf-spheroidal satellite
galaxies which are speculated to be severely dominated by dark matter
but definitely constitute star-forming conditions very different from
present-day events.  Two such close companions to the Milky Way have
been observed \citep{Grill98,Felt99} finding the same MF as in
globular clusters for $0.5\simless m/M_\odot \simless 0.9$ and thus no
very evident differences to the standard IMF. 

However, there are peculiar indications of top-heavy IMFs such as in
some massive star-burst clusters in the M82 galaxy. Spectroscopy of
the unresolved M82-F cluster derives, via the inferred velocity
dispersion, a mass and from the luminosity a mass-to-light ratio that
is significantly smaller than the ratio expected from the standard IMF
for such a young (about 60~Myr) population. The implication is that
the M82-F population may be significantly depleted in low-mass stars,
provided the velocity dispersion is representative of the entire
cluster. A possibility that will have to be addressed using
stellar-dynamical modelling of forming star clusters is that M82-F may
have lost low-mass stars due to tidal shocking \citep{SmGa01}.  Highly
pronounced mass segregation which leads to a dynamically decoupled
central core of OB stars is an important mechanism for reducing the
measured mass-to-light ratio \citep{Boily_etal05}, while rapid
expulsion of residual gas from forming clusters enhances the measured
mass-to-light ratios \citep{GB06}.

The proclamation by \citet{Briceno_etal02} that Taurus-Auriga is
producing significantly fewer BDs per star than the ONC, and thus for
a possible first robust indication of a variable IMF, was found to not
be the correct interpretation by \citet{Kr_etal03,Kr_Bouv03b} and was
retracted by \citet{Luhman04} based on newer data. A new analysis of
these actually shows that Taurus-Auriga and the ONC are producing the
same number of BDs per star, about~0.2
(\S~\ref{sec:bds}). Nevertheless, the discussion in \S~\ref{sec:bds}
shows that BDs probably need to be treated as a separate population
such that the IMF may be discontinuous near the stellar/sub-stellar
mass limit. Since BDs may come in four flavours, \citet{Kr_Bouv03b}
suggest that the photo-evaporated sort could far out-weigh the others
in star-burst clusters, so that globular clusters could contain a
larger fraction of BDs per star than the current modest Galactic
clusters. 

Differences in the metallicity, $Z$, of the population also do not
lead to observable variations of the IMF for massive stars
(\citealt{Massey03}, Fig.~\ref{fig:kroupa_figmassey}).  The mass of the
most massive star, $m_{\rm max*} \approx 150\,M_\odot$, is independent
of $Z$. Thus the distribution of masses for massive stars does not
appear to be affected by the metallicity of the star-forming gas and
therefore radiation pressure on dust grains during star-assembly
cannot be a physical mechanism establishing $m_{\rm max*}$
(\S~\ref{sec:massst}).

However, there may be an effect for VLMSs.  Present-day star-forming
clouds typically have somewhat higher metal-abundances
([Fe/H]$\approx+0.2$) compared to 5~Gyr ago ([Fe/H]$\approx-0.3$)
\citep{BM98} which is the mean age of the population defining the
standard IMF. The data in the empirical alpha-plot indicate that some
of the younger clusters may have a single-star MF that is somewhat
steeper (larger $\alpha_1$) than the standard IMF if unresolved
binary-stars are corrected for (Fig.~\ref{fig:apl}). This may mean
that clouds with a larger [Fe/H] produce relatively more low-mass
stars \citep{K01a} which is tentatively supported by the typically but
not significantly flatter MFs in globular clusters \citep{PZ99} that
have [Fe/H]$\,\approx -1.5$, and the recent finding that the old and
metal-poor ([Fe/H]$=-0.6$) thick-disk population also has a flatter MF
below $0.3\,M_\odot$ with $\alpha_1\approx0.5$ \citep{RR01}.  If such
a systematic effect is present, then for $m\simless 0.7\,M_\odot$ and
to first order,
\begin{equation}
\alpha \approx 1.3 + \Delta\alpha\, {\rm[Fe/H]},
\label{eq:systemvar}
\end{equation}
with $\Delta\alpha \approx0.5$, being similar to the adopted
uncertainty on $\alpha$.  Theoretical considerations do suggest that
for sufficiently small metallicity a gas cloud cannot cool efficiently
causing the Jeans mass required for gravitational collapse to be
larger. In particular, the first stars ought to have large masses
owing to this effect and the generally higher ambient temperatures at
early cosmological epochs \citep{Larson98,Bromm}. Finding the remnants
of these poses a major challenge.

An easier target is measuring the IMF for low-mass and VLMSs in
metal-poor environments, such as young star-clusters in the Small
Magellanic Cloud. That metallicity does play a role is becoming
increasingly evident in the planetary-mass regime in that the detected
exo-planets occur mostly around stars that are more metal-rich than
the Sun \citep{Vogt01,ZuMaz01}.

While the Jeans-mass argument (eq.~\ref{eq:MJeans}) should be useful as
a general indication of the rough mass scale where fragmentation of a
contracting gas cloud occurs, the concept clearly breaks down when
considering the stellar masses that form in star clusters. The central
regions of these are denser, formally leading to smaller Jeans masses
which is the opposite of the observed trend, where even in very young
clusters massive stars tend to be located in the inner regions. More
complex physics is clearly involved. Stars may, to a certain extend,
regulate their own mass by outflows \citep{AL96}, and the coagulation
of proto-stars probably plays a significant role in the densest
regions where the cloud-core collapse time, $\tau_{\rm coll}$, is
longer than the fragment collision time-scale, $\tau_{\rm cr}$. The
collapse of a fragment to a proto-star takes no longer than $\tau_{\rm
coll}\approx 0.1$~Myr \citep{WK01}, so that 
\begin{equation}
t_{\rm cr}/{\rm Myr} \approx 42\,\left( (R/{\rm pc})^3 \over
(M_{\rm ecl}/M_\odot) \right)^{1\over2} < 0.1\;{\rm Myr},
\end{equation}
where $R$ is the half-mass radius of a Plummer-sphere model, implies
$M/R^3 > 10^5\, M_\odot$/pc$^{-3}$. Such densities, where
proto-stellar interactions are expected to affect the emerging stellar
mass-spectrum, are found in the centres of very dense and rich
embedded star clusters such as the ONC before they expand as a result
of gas expulsion \citep{KAH,VB03}. Thus, only for massive stars is the form
of the IMF probably affected by coagulation, which may explain why
massive stars are usually centrally concentrated in very young
clusters \citep{Bonn98,Kl01}. However, the observed mass segregation
in very young clusters cannot as yet be taken as conclusive evidence
for primordial mass segregation and coagulation, unless precise
$N$-body computations of the embedded cluster are performed for each
case in question. For example, models of the ONC show that the degree
of observed mass segregation can be established dynamically within
about 2~Myr (Fig.~\ref{fig:mfn2}) despite the embedded and much denser
configuration having no initial mass segregation. The notion behind
such an assumption is that star clusters fragment heavily
sub-clustered \citep{Meg,Bont01,Kl01}, and each sub-cluster may form a
few OB stars with a few hundred associated lower-mass stars
(Table~\ref{tab:frac}), so that the overall morphology may resemble a
system without significant initial mass segregation. The theoretical
time-scale, $t_2-t_1$ in Fig.~\ref{fig:mfn2}, for mass segregation to
be established can be shortened by decreasing the relaxation
time. This can be achieved by reducing the number of stars in the
model, for example. But it may prove impossible to find agreement at
the same time with the density profile and kinematics because the ONC
is probably expanding rapidly now. Clearly this issue requires more
study.

The origin of most stellar masses is indicated by a remarkable
discovery for the low-mass $\rho$~Oph cluster in which star-formation
is on-going. Here the pre-stellar and proto-star MF are
indistinguishable and both are startlingly similar to the standard
IMF, even in showing the same flattening of the power-law at
$0.55\,M_\odot$ \citep{Motte,Bont01}. The pre-stellar cores have sizes
and densities that are in agreement with the Jeans-instability
argument for the conditions in the $\rho$~Oph cloud, so that
cloud-fragmentation due to the collapse of density fluctuations in a
dissipating turbulent interstellar medium
\citep{NordlundPadoan02,PadoanNordlund02,
MacLowKlessen04,TilleyPudritz04,PadoanNordlund04} appears to be the
most-important mechanism shaping the stellar IMF for masses
$0.5\simless m/M_\odot\simless\,$a~few $\,M_\odot$, the shape of the
IMF being determined by the spectrum of density fluctuations in the
molecular cloud. Similar results have been obtained for the Serpens
clouds and for the clouds in Taurus--Auriga
(\citealt{TestiSargent98,Onishi_etal02}, respectively). The majority
of stellar masses making up the standard IMF thus do not appear to
suffer significant subsequent modifications such as competitive
accretion \citep{Bonn01b}, proto-stellar mergers or even
self-limitation through feedback processes. The work of
\citet{PadoanNordlund02} demonstrates that, under certain reasonable
assumptions, the mass function of gravitationally unstable cloud cores
deriving from the power-spectrum of a super-sonic turbulent medium
leads to the observed standard IMF above $1\,M_\odot$. The flattening
at lower masses is a result of a reduction of the star-formation
efficiency because at small masses only the densest cores can survive
sufficiently long to collapse.

The intriguing result from $\rho$~Oph, Serpens and Taurus--Auriga, in
which the stellar, proto-stellar and pre-stellar clump mass spectra
are similar to the stellar IMF (eq.~\ref{eq:imf}), is consistent with
the independent finding that the properties of binary systems in the
Galactic field can be understood if most stars formed in modest,
$\rho$~Oph-type clusters with primordial binary properties as observed
in Taurus-Auriga \citep{K95d}, and with the independent result derived
from an analysis of the distribution of local star clusters that most
stars appear to stem from such modest clusters \citep{AdamsMyers01}.
However, the standard IMF is also similar to the IMF in the ONC
(Fig.~\ref{fig:mfn1}) implying that fragmentation of the pre-cluster
cloud proceeded similarly.

The impressive computations by \citet{BonnellBate02} and collaborators
of dense clusters indeed not only predict the observed $m_{\rm
max}(M_{\rm ecl})$ relation (Fig.~\ref{fig:pk_mmaxf}), but they also
show that a Salpeter/Massey power-law IMF is obtained as a result of
competitive accretion and the merging of stars near the cluster core
driven by accretion onto it. The reason as to why the IMF is so
invariant above a few $M_\odot$ may thus be that the various physical
processes all conspire to give the same overall scale-free result.

This still leaves the origin and nature of VLMSs and BDs unclear. As
the discussion in \S~\ref{sec:bds} suggests, VLMSs and BDs probably
need to be considered a different or extra population which does not
mix well, in terms of pairing, with stars. \citet{MF05} demonstrate
that virtually {\it all} currently available theoretical work on the
formation of BDs is excluded by the data. Together with the
realisation that existing star-formation theory fails to reproduce the
binary properties of young stars \citep{Goodwin_etal04,GK05}, this
suggests that our theories lack major ingredients that probably are
related to stellar feedback processes.  That VLMS and BD binaries have
a significantly higher energy scale than stellar binaries does,
however, suggest that their formation may be linked to a dense
environment which they probably leave abruptly.

\section{Composite stellar population}
\label{sec:comppop}

We have thus seen that while a conclusive theoretical formulation of
the IMF is still wanting, we do have some good ideas about its origin
and a good impression of its shape, and empirically there is not a
strong case for systematic variation of the IMF with physical
conditions. It is thus reasonable to assume that the stellar IMF is
invariant.

The natural assumption has often been made that independent of the
star-formation mode, the stellar distribution is sampled randomly from
the invariant IMF (e.g. \citealt{El04}).  Thus, for example, $10^{5}$
clusters, each with $20$~stars, would have the same composite
(i.e. combined) IMF as one cluster with $2\times 10^{6}$~stars.

However, the existence of the $m_{\rm max}(M_{\rm ecl})$ relation
(\S~\ref{sec:stmass_clmass}) has profound consequences for {\it
composite populations}.  It immediately implies, for example, that
$10^{5}$ clusters, each with $20$~stars, {\it cannot} have the same
composite (i.e. combined) IMF as one cluster with $2\times
10^{6}$~stars, because the small clusters can never make stars more
massive than about $1.5\,M_\odot$ (Fig.~\ref{fig:pk_mmaxf}).  Thus,
galaxies, that are composite stellar populations consisting of many
star clusters, most of which may be dissolved, would have steeper
composite, or integrated galactic IMFs (IGIMFs), than the stellar IMF
in each individual cluster \citep{Vanbev82, KW03}.

The IGIMF is an integral over all star-formation events in a given
star-formation ``epoch'' $t, t+\delta t$,
\begin{equation} 
\xi_{\rm IGIMF}(m;t) = \int_{M_{\rm ecl,min}}^{M_{\rm
ecl,max}(SFR(t))} \xi\left(m\le m_{\rm max}\left(M_{\rm
ecl}\right)\right)~\xi_{\rm ecl}(M_{\rm ecl})~dM_{\rm ecl}.
\label{eq:igimf_t}
\end{equation}
Here $\xi(m\le m_{\rm max})~\xi_{\rm ecl}(M_{\rm ecl})~dM_{\rm ecl}$
is the stellar IMF contributed by $\xi_{\rm ecl}~dM_{\rm ecl}\propto
M_{\rm ecl}^{-\beta}\,dM_{\rm ecl}$ clusters with mass near $M_{\rm
ecl}$.  $M_{\rm ecl,max}$ follows from the maximum star-cluster-mass
{\it vs} global-star-formation-rate-of-the-galaxy relation,
\begin{equation}
M_{\rm ecl,max}={\rm fn}(SFR)
\label{eq:maxcl_sfr}
\end{equation}
(eq.~1 in \citealt{WK05b}, as derived by \citealt{WKL04}).  
$M_{\rm ecl,min}\,=\,5\,M_{\odot}$ is adopted in the standard modelling and
corresponds to the smallest star-cluster units observed.
At time $t$ the SFR is
\begin{equation}
SFR(t) = {M_{\rm tot} \over \delta t},
\end{equation}
where 
\begin{equation}
M_{\rm tot} = \int_{M_{\rm ecl,min}}^{M_{\rm ecl,max}}\,M_{\rm
ecl}\,\xi_{\rm ecl}(M_{\rm ecl})\, dM_{\rm ecl}
\end{equation}
is the total stellar mass assembled in time $\delta t$ which
\citet{WKL04} define to be a ``star-formation epoch'', within which the
ECMF is sampled to completion. This formulation leads naturally to the
observed $M_{\rm ecl,max}(SFR)$ correlation if the ECMF is invariant,
$\beta\approx2.35$ and if the ``epoch'' lasts about $\delta t=10$~Myr.
Thus, the embedded cluster mass function is fully sampled in 10~Myr
intervals, independent of the SFR. This time-scale compares very well
indeed to the star-formation time-scale in normal galactic disks
measured by \citet{Egusa_etal04} using an entirely independent method,
namely from the offset of HII regions from the molecular clouds in
spiral-wave patterns.  The time-integrated IGIMF then follows from
\begin{equation} 
\label{eq:igimf}
\xi_{\rm IGIMF}(m) = \int_0^{\tau_{\rm G}} \xi_{\rm IGIMF}(m;t)\,dt,
\end{equation}
where $\tau_{\rm G}$ is the age of the galaxy under scrutiny.

Note that $\xi_{\rm IGIMF}(m)$ is the mass function of all stars ever
to have formed in a galaxy, and can be used to estimate the total
number of supernovae ever to have occurred, for example. $\xi_{\rm
IGIMF}(m;t)$, on the other hand, includes the time-dependence through
a dependency on $SFR(t)$ of a galaxy and allows one to compute the
time-dependent evolution of a stellar population over the life-time of
a galaxy.

Because more-massive stellar clusters are observed to form for higher
SFRs (eq.~\ref{eq:maxcl_sfr}), the ECMF is sampled to larger masses in
galaxies that are experiencing high SFRs, leading to IGIMFs that are
flatter than for low-mass galaxies that have had only a low-level of
star-formation activity.  \citet{WK05b} show that the sensitivity of
the IGIMF power-law index for $m\simgreat 1\,M_\odot$ towards $SFR$
variations increases with decreasing $SFR$.

Thus, galaxies with a small mass in stars can either form with a very
low continuous SFR (appearing today as low-surface-brightness but
gas-rich galaxies) or with a brief initial SF burst (dE or dSph
galaxies). The IGIMF ought to vary significantly among such galaxies
(Fig.~\ref{fig:alphaIGIMF}).
\begin{figure}
\begin{center}
\rotatebox{0}{\resizebox{0.6 \textwidth}{!}{\includegraphics{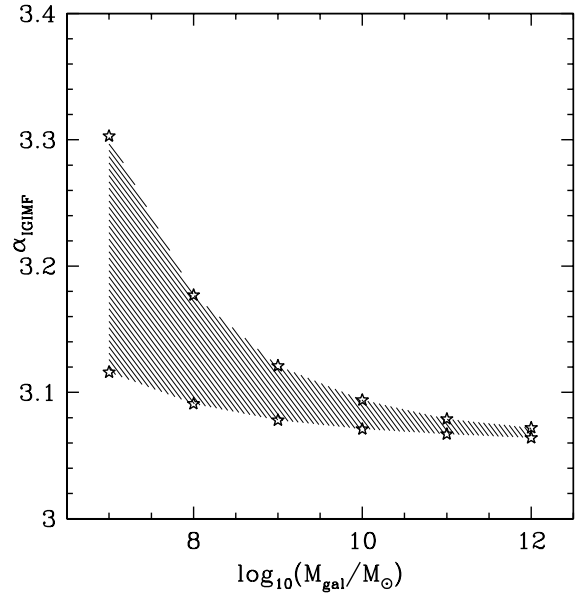}}}
\vskip -15mm
\caption{\small{The power-law index $\alpha_{\rm IGIMF}$ of the IGIMF
above $1.3\,M_\odot$ as a function of the stellar mass of a galaxy,
which determines the average SFR over a Hubble time.  The lower bound
is for an initial SF burst that forms the entire stellar galaxy, while
the upper bound is for a constant SFR over a Hubble time. In all cases
the input stellar IMF is the standard form (eq.~\ref{eq:imf}).  Note
that the detailed results depend on the adopted form of the ECMF
(cf. Fig.~\ref{fig:SN}).  For further details see \citet{WK05b}.  }}
\label{fig:alphaIGIMF}
\end{center}
\end{figure}
In all cases, however, the IGIMFs are invariant for $m\simless
1.3\,M_\odot$ which is the maximal stellar mass in $5\,M_\odot$
``clusters'' (Fig.~\ref{fig:pk_mmaxf}).  Low-surface-brightness
galaxies would therefore appear chemically young, while the dispersion
in chemical properties ought to be larger for dwarf galaxies than for
larger galaxies \citep{GP05,WK05b,KWK07}.  Another interesting
implication is that the number of supernovae per star would be
significantly smaller over cosmological times than predicted by an
invariant Salpeter IMF (\citealt{GP05},Fig.~\ref{fig:SN}).
\begin{figure}
\begin{center}
\rotatebox{0}{\resizebox{0.6 \textwidth}{!}{\includegraphics{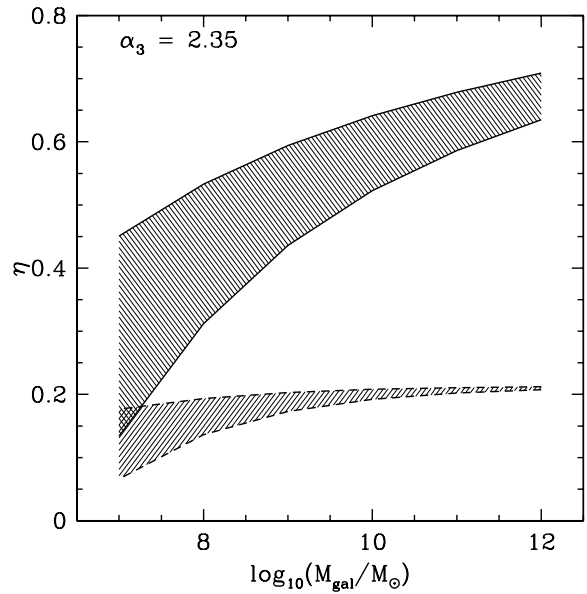}}}
\vskip -15mm
\caption{\small{ The number of supernovae per star in the IGIMF
divided by the number of supernovae per star in the standard IMF,
$\eta$, as a function of the stellar galaxy mass. The upper shaded
area is for an ECMF with $\beta=2$, while the lower shaded area
assumes $\beta=2.35$. The upper bound for each shaded region is for an
initial SF burst model, while the lower bounds are for a constant SFR
over a Hubble time.  For further details see \citet{WK05b}.  }}
\label{fig:SN}
\end{center}
\end{figure}

These new insights should lead to a revision of theoretical work on
galaxy formation that typically until now relied on an invariant IMF.
Empirical evidence in favour of or against the notion of a
galaxy-variable IGIMF is being studied and will ultimately lead to a
refinement of the ideas.

\section{Concluding comments}

Spectacular advances have been achieved over the past decade in the
field of IMF research and this affects a vast area of astrophysics.

The stellar IMF appears to be extra-ordinarily universal. It is
described by eq.~\ref{eq:imf}.  The evidence for top-heavy IMFs comes
from either unresolved clusters or clusters that are very difficult to
observe. The most significant uncertainty that remains at high masses
despite the presence of exquisite data is the exact shape of the IMF
for massive stars, because the biases due to unresolved multiple
systems and due to stellar rotation have not been studied in
sufficient detail. The true IMF may be closer to the Scalo value
$\alpha_3\approx2.7$ rather than being Salpeter/Massey. For
sub-stellar masses more data are necessary to improve constraints on
$\alpha_0$ and to better quantify any claims for a variation of the
IMF.  Further problems are inadequate stellar models for stars younger
than a~few~Myr, and the under-representation of stellar-dynamical
modelling which, however, is an absolute necessity in the search for
IMF variations among clusters.

Among intriguing recent results are that the BDs appear to be a
distinct population from that of low-mass stars; their pairing
properties have a different energy scale. This sets the stage for a
probable discontinuity in the IMF near the stellar/sub-stellar
boundary which has already probably been detected. This then questions
the validity of fitting continuous log-normal functions across the
VLMS/BD barrier. Furthermore, the IMF does appear to have a physical
maximum stellar mass that has now been found empirically. Stars with
$m\simgreat150\,M_\odot$ do not appear to exist, unless they implode
invisibly shortly after being formed. In addition, by realising that
star clusters are the true fundamental building blocks of a galaxy, we
merely need to add up all clusters and their (invariant) IMFs to
arrive at the integrated galactic initial mass function. The IGIMF
varies according to the star-formation history of the galaxy. This
formulation now allows us to compute the IGIMF, as a function of time,
for galaxies that are converting their gas supply only slowly to
stars, or for galaxies that like to eat it up all at once. The
implications are that equally-old galaxies can have very different
chemical compositions ranging from young to evolved, and that the
cosmological supernova~II rate may be significantly smaller and dependent
on galaxy type than if an invariant Salpeter IMF is
assumed. Galaxy-formation and evolution computations with this latter
assumption are not likely to be correct.

Thus, the IMF research as presented here has advanced onto
cosmological issues, whereby many details still need to be worked
out. But the most intimately connected research field, namely star
formation, is being tested rather dramatically.  The brilliant
computational results that have been becoming available have lead to a
beautiful agreement with the observed maximum-stellar-mass {\it vs}
cluster-mass correlation and reproduce the general shape of the
IMF. However, different ideas (competitive accretion, coagulation and
simply the distribution of gravitationally unstable regions in
turbulent clouds) all lead to virtually the same type of theoretical
IMF. Detailed computations of the formation of stars actually get it
all wrong - the primordial-binary properties cannot yet be described,
and the failure of modern theory is especially evident in the BD
regime, where virtually all current ideas are being excluded by the
data. The failures of modern theory are not surprising though, because
the primary process driving star-formation is injection of feedback
energy, the consistent inclusion of which goes beyond currently
available computational resources.  On the bright side, the
convergence of different mechanisms to the same Salpeter/Massey IMF
may be interpreted to simply mean that the shape of the IMF for
$m\simgreat 0.5\,M_\odot$ is indeed invariant to the physics, as is
observed.

\acknowledgements I would like to thank the organisers for this
splendid meeting, Carsten Weidner, Christopher Tout and Gerry Gilmore
for very stimulating and important contributions, and Ingo Thies and
J\"org Dabringhausen for help with some of the material. I am
especially indebted to Sverre Aarseth whose immense tutoring allowed
me to do the numerical dynamics work.  I would also like to thank
M.R.S.~Hawkins who had introduced me to this field in about~1987
whilst I visited the Siding-Spring Observatory as a summer vacation
scholar at the ANU. Mike gave me a delightful lecture on the low-mass
LF one night when I visited his observing run to learn about modern,
state-of-the-art Schmidt-telescope surveying {\it before} I embarked
on post-graduate work.  This research was much later supported through
DFG grants KR1635/2, KR1635/3 and a Heisenberg fellowship, KR1635/4,
and currently KR1635/13.


\begin{landscape}
\begin{table}
{\small
\begin{tabular}{lccc}

\hline

&$\alpha$   &$\alpha$   &$\alpha$\\
&mass range [$M_\odot$] &mass range [$M_\odot$] &mass range [$M_\odot$]\\

\hline\hline

{\bf Orion Nebula cluster, ONC}\\ 

\cite{MLL00}
&$-0.35$      &$+1.25$      &$+2.35$\\
{\it magenta small open circles with central dot} 
&$0.02-0.08$  &$0.08-0.80$  &$0.80-63.1$\\
{\it magenta large open circles with central dot} 

&$+0.00$      &$+1.00$      &$+2.00$\\
&$0.02-0.08$  &$ 0.08-0.40$ &$0.4-63.10$\\ 

\cite{HC00} (HC00)
&$+0.43$\\
{\it magenta large thick open circle}
&$0.02-0.15$\\
\hspace{5mm}{\it  with central dot}\\

\cite{L00}
&$+0.70$\\
{\it magenta small thick open circle}
&$0.035-0.56$\\
\hspace{5mm}{\it  with central dot}\\

\hline

{\bf Pleiades}\\
\cite{MBS01} &$+0.51\pm0.15$\\
{\it green circles with central dot}
&$0.04-0.30$\\

\cite{Hambetal99}, from \cite{Netal01}
&$+0.56$             &$+2.67$\\
{\it green circles with central dot}
&$0.065-0.60$        &$0.6-10.0$ \\

\hline

{\bf $\mathbf{\sigma}$ Ori}\\
\cite{Betal01}
&$0.8\pm0.4$\\
{\it green solid circle}
&$0.013-0.20$\\

\hline

{\bf M35}\\
\cite{Netal01}
&$-0.88\pm0.12$     &$0.81\pm0.02$   &$2.59\pm0.04$\\
{\it green solid circle}$^1$
&$0.08-0.2$         &$0.2-0.8$       &$0.8-6.0$\\

\hline

{\bf IC 348}\\
\cite{Netal00} for MLR from \cite{BCAH98}
&$+0.5$\\
{\it green solid circle}
&$0.015-0.22$\\

\hline

{\bf NGC~2264}\\
\cite{Petal00}
&&&$+2.7$\\
{\it green solid circle}
&&&$2.0-6.3$\\

\hline

{\bf 5~LMC regions}\\
\cite{Parker_etal01}
&&&$+2.3\pm0.2$\\
{\it blue solid triangle}
&&&$5-60$\\

\hline

{\bf NGC~1818} in LMC\\
\cite{Santetal01}, outer region
&&$+2.5$\\
{\it blue solid triangle}
&&$0.9-3$\\

{\bf NGC~1805} in LMC\\
\cite{Santetal01}, outer region
&&$+3.4$\\
{\it blue solid triangle}
&&$0.9-3$\\

\hline

\end{tabular}
}
\caption{\small{\it continued}}
\end{table}
\end{landscape}

\addtocounter{table}{-1}

\begin{landscape}
\begin{table}
{\small
\begin{tabular}{lccc}

\hline

&$\alpha$   &$\alpha$   &$\alpha$\\
&mass range [$M_\odot$] &mass range [$M_\odot$] &mass range [$M_\odot$]\\

\hline\hline

{\bf 30~Dor}$^\star$ in LMC\\
\cite{Setal99}, $r>3.6$~pc
&&&$+2.37\pm0.08$\\
{\it cyan small open triangle}
&&&$3-120$\\

\cite{Setal99}, $1.1<r/{\rm pc}<4.5$
&&&$+2.17\pm0.05$\\
{\it cyan small open triangle}
&&&$2.8-120$\\

\cite{Sirianni_etal00}
&&$+1.27\pm0.08$   &$+2.28\pm0.05$\\
{\it cyan large open triangle}$^2$
&&$1.35-2.1$       &$2.1-6.5$\\

\hline

{\bf Arches cluster}$^\star$\\
\cite{Fetal99}, all radii
&&&$+1.6\pm0.1$\\
{\it cyan large solid circle} 
&&&$6.3-125$\\

\hline

{\bf NGC~3603}$^\star$\\
\cite{Eetal98} 
&&$+1.73$    &$+2.7$\\
{\it cyan small solid circle}
&&$1-30$     &$15-70$\\

\hline

{\bf Globular clusters}\\
\cite{PZ99}
&$+0.88\pm0.35$   &$+2.3$\\
{\it yellow open triangles} &$0.1-0.6$        &$0.6-0.8$\\

\hline

{\bf Galactic bulge}\\
\cite{Hetal98}
&$+0.9$         &$+2.2$\\
{\it magenta filled square}
&$0.3-0.7$      &$0.7-1.0$\\

\cite{Zetal00}
&$+1.43\pm0.13$ &$+2.0\pm0.23$\\
{\it magenta filled square}
&$0.15-0.5$     &$0.5-1.0$\\

\hline

\multicolumn{4}{l}{
{\bf Solar Neighbourhood} ({\it magenta dotted lines})} \\
\cite{Retal99}
&$+1.5\pm0.5$\\
&$0.02-0.08$\\

\cite{Hetal99}
&$\le +0.8$\\
&$0.02-0.08$\\

\cite{Ch01,Ch02}
&$\le+1$      &$+1$        \quad / \quad $+2$\\
&$0.01-0.08$  &$0.10-0.35$ \quad / \quad $0.35-1.0$\\

\hline

\end{tabular}
}

\caption{\small{$\alpha(<\!lm\!>)$ data obtained since 1998 and until
    2002. The data are shown in Fig.~\ref{fig:apl} in addition to the
    previously available data set compiled by \cite{Sc98}. Each
    $\alpha$ value is obtained at $<\!lm\!>=(lm_2-lm_1)/2$, $lm\equiv
    {\rm log}_{10}m$, by the respective authors by fitting a power-law
    MF over the logarithmic mass range given by $m_1$ and $m_2$ listed
    above.  Some authors do not quote uncertainties on their $\alpha$
    values.  Notes: $^\star$ are star-burst clusters; $^1$ thin green
    open circle emphasises the low-mass M35 datum which may be
    incomplete; $^2$ the mass range $1.35<m/M_\odot<2.1$ may be
    incomplete and is emphasised by the cross through the cyan large
    open triangle.  }}
\label{tab:apl}
\end{table}
\end{landscape}


\begin{table}
{\small
\begin{tabular}{l|ccc|ccc|c|c}

\hline\hline

mass range     
&\multicolumn{3}{c|}{$\eta_N$}
&\multicolumn{3}{c|}{$\eta_M$}    
&$\rho^{\rm st}$
&$\Sigma^{\rm st}$  \\

[$M_\odot$]    
&\multicolumn{3}{c|}{[per cent]}
&\multicolumn{3}{c|}{[per cent]}
&[$M_\odot/{\rm pc}^3$]  
&[$M_\odot/{\rm pc}^2$] \\

&\multicolumn{3}{c|}{$\alpha_3$}
&\multicolumn{3}{c|}{$\alpha_3$}    
&$\alpha_3$
&$\alpha_3$  \\

&2.3 &2.7 &4.5 &2.3 &2.7 &4.5 &4.5 &4.5\\

\hline

0.01--0.08 
&37.15
&37.69 
&38.63

&4.08
&5.39
&7.39

&$3.21\times10^{-3}$
&1.60

\\ 

0.08--0.5
&47.81
&48.50
&49.71

&26.61
&35.16
&48.21

&$2.09\times10^{-2}$
&10.45

\\ 

0.5--1
&8.94
&9.07
&9.30

&16.13
&21.31
&29.22

&$1.27\times10^{-2}$
&6.35

\\ 

1 -- 8
&5.70
&4.60
&2.36

&32.38
&30.30
&15.09

&$6.54\times10^{-3}$
&1.18

\\ 

8 -- 120 
&0.40
&0.14
&0.00

&20.80
&7.83
&0.08

&$3.63\times10^{-5}$
&$6.53\times10^{-3}$

\\ 

\hline

$\overline{m}/M_\odot=$
&$0.380$
&$0.292$
&$0.218$

&
&
&

&$\rho_{\rm tot}^{\rm st}=0.043$
&$\Sigma_{\rm tot}^{\rm st}=19.6$

\\

\hline \hline

&\multicolumn{1}{r|}{}
&\multicolumn{2}{c|}{$\alpha_3=2.3$} 
&\multicolumn{2}{c||}{$\alpha_3=2.7$}
&
&\multicolumn{2}{c}{$\Delta M_{\rm cl}/M_{\rm cl}$}
\\ 

&\multicolumn{1}{r|}{$m_{\rm max}$} 
&$N_{\rm cl}$  &$M_{\rm cl}$  
&$N_{\rm cl}$  &\multicolumn{1}{r||}{$M_{\rm cl}$}
&$m_{\rm to}$  &\multicolumn{2}{c}{[per cent]}
\\

&\multicolumn{1}{r|}{[$M_\odot$]} & &[$M_\odot$] & 
&\multicolumn{1}{r||}{[$M_\odot$]} &[$M_\odot$] 
&$\alpha_3=2.3$ &$\alpha_3=2.7$\\

\cline{2-9}

&\multicolumn{1}{r|}{1}      &16    &2.9     &21                 
&\multicolumn{1}{r||}{3.8}
&80 &3.2 &0.7
\\
								     
&\multicolumn{1}{r|}{8}      &245   &74     &725                
&\multicolumn{1}{r||}{195}
&60 &4.9 &1.1
\\
								      
&\multicolumn{1}{r|}{20}     &806   &269     &3442               
&\multicolumn{1}{r||}{967}
&40 &7.5 &1.8
\\
								      
&\multicolumn{1}{r|}{40}     &1984  &703     &$1.1\times10^4$    
&\multicolumn{1}{r||}{2302}
&20 &13 &4.7
\\

&\multicolumn{1}{r|}{60}     &3361 &1225     &$2.2\times10^4$    
&\multicolumn{1}{r||}{6428}
&8 &22 &8.0
\\

&\multicolumn{1}{r|}{80}     &4885 &1812     &$3.6\times10^4$    
&\multicolumn{1}{r||}{$1.1\times10^4$}
&3 &32 &15
\\
  
&\multicolumn{1}{r|}{100}    &6528 &2451     &$5.3\times10^4$    
&\multicolumn{1}{r||}{$1.5\times10^4$}
&1 &44 &29
\\

&\multicolumn{1}{r|}{120}    &8274 &3136     &$7.2\times10^4$    
&\multicolumn{1}{r||}{$2.1\times10^4$}
&0.7 &47 &33
\\

\hline\hline

\end{tabular}
}
\caption{\small{The number fraction $\eta_N=100\,\int_{m_1}^{m_2}
\xi(m)\,dm/ \int_{m_l}^{m_u}\xi(m)\,dm$, and the mass fraction
$\eta_M=100\,\int_{m_1}^{m_2} m\,\xi(m)\,dm/ M_{\rm cl}$, $M_{\rm cl}=
\int_{m_l}^{m_u} m\,\xi(m)\,dm$, in per cent of BDs or main-sequence
stars in the mass interval $m_1$ to $m_2$, and the stellar
contribution, $\rho^{\rm st}$, to the Oort limit and to the
Galactic-disk surface mass-density, $\Sigma^{\rm st}=2\,h\rho^{\rm
st}$, near to the Sun, taking $m_l=0.01\,M_\odot$, $m_u=120\,M_\odot$
and the Galactic-disk scale-height $h=250$~pc ($m<1\,M_\odot$
\citealt{KTG93}) and $h=90$~pc ($m>1\,M_\odot$,
\citealt{Sc86}). Results are shown for the standard IMF
(eq.~\ref{eq:imf}), for the high-mass-star IMF approximately corrected
for unresolved companions ($\alpha_3=2.7, m>1\,M_\odot$), and for the
PDMF ($\alpha_3=4.5$, \citealt{Sc86,KTG93}) which describes the
distribution of stellar masses now populating the Galactic disk. For
gas $\Sigma^{\rm gas}=13\pm3\,M_\odot$/pc$^2$ and remnants
$\Sigma^{\rm rem}\approx3\,M_\odot$/pc$^2$ \citep{Weide}.  The average
stellar mass is $\overline{m}= \int_{m_l}^{m_u} m\,\xi(m)\,dm/
\int_{m_l}^{m_u}\xi(m)\,dm$.  $N_{\rm cl}$ is the number of stars that
have to form in a star cluster such that the most massive star in the
population has the mass $m_{\rm max}$. The mass of this population is
$M_{\rm cl}$, and the condition is $\int_{m_{\rm
max}}^{\infty}\xi(m)\,dm=1$ with $\int_{0.01}^{m_{\rm max}} \xi(m)\,dm
= N_{\rm cl}-1$.  $\Delta M_{\rm cl}/M_{\rm cl}$ is the fraction of
mass lost from the cluster due to stellar evolution, assuming that for
$m\ge8\,M_\odot$ all neutron stars and black holes are kicked out due
to an asymmetrical supernova explosion, but that white dwarfs are
retained \citep{Wetal92} and have masses $m_{\rm WD} = 0.084\,m_{\rm
ini}+0.444\,[M_\odot]$. This is a linear fit to the data in
\cite[][their table~3]{W2000} for progenitor masses $1\le m/M_\odot
\le 7$ and $m_{\rm WD}=0.5\,M_\odot$ for $0.7\le m/M_\odot <1$. The
evolution time for a star of mass $m_{\rm to}$ to reach the turn-off
age is available in Fig.~\ref{fig:apl}.  }}
\label{tab:frac}
\end{table}

\begin{landscape}
\begin{table}
{\small
\begin{tabular}{l|l|l|l}

\hline\hline
general
&\multicolumn{2}{l}{$dN = \xi(m)\,dm = \xi_{\rm L}(m)dlm$}\\
&\multicolumn{2}{l}{$\xi_{\rm L}(m) = (m\,{\rm ln}10)\,\xi(m)$}
&{\it gen}\\

Scalo's IMF index \citep{Sc86} 
&\multicolumn{2}{l}{$\Gamma(m)\equiv {d\over dlm}
\left({\rm log}_{10}\xi_{\rm L}(lm)\right)$}
&{\it Gam}\\

&\multicolumn{2}{l}{$\Gamma = -x = 1+\gamma = 1-\alpha$}
&{\it ind}\\
&e.g.  for power-law form:
&$\xi_{\rm L} = A\,m^\Gamma = A\,m^{-x}$\\
&&$\xi = A'\,m^\alpha = A'\,m^{-\gamma}$\\
&&$A' = A/{\rm ln}10$\\

\hline\hline

\citet{S55} 
&$\xi_{\rm L}(lm) = A\,m^\Gamma$
&$\Gamma=-1.35 \, (\alpha=2.35)$
&{\it S}\\

&\multicolumn{2}{l}{$A = 0.03\,{\rm pc}^{-3}\,{\rm
log}_{10}^{-1}M_\odot; \quad 0.4\le m/M_\odot \le 10$}\\

\hline

\citet{MS79} 
&$\xi_{\rm L}(lm) = A\,{\rm exp}\left[-{\left(lm-lm_o\right)^2 \over 
2\,\sigma_{lm}^2 } \right]$
&$\Gamma(lm) = -{\left(lm-lm_o\right) \over \sigma_{lm}^2}\,{\rm log}_{10}e$
&{\it MS}\\

{\it thick long-dash-dotted line}
&\multicolumn{2}{l}{$A = 
106\,{\rm pc}^{-2}\,{\rm log}_{10}^{-1}M_\odot;
\quad lm_o=-1.02; \quad \sigma_{lm}=0.68$}\\

\hline

\citet{Larson98}
&$\xi_{\rm L}(lm) = A\, m^{-1.35} {\rm exp}\left[-{m_o\over
                    m}\right]$
&$\Gamma(lm) = -1.35  + {m_o\over m}$
&{\it La}\\
{\it thin short-dashed line}
&\multicolumn{2}{l}{$A=-\,; \quad\quad m_o=0.3\,M_\odot$}\\

\hline

\citet{Larson98}
&$\xi_{\rm L}(lm) = A\,\left[1 + {m\over m_o}\right]^{-1.35}$ 
&$\Gamma(lm) = -1.35\left(1 + {m_o\over m}\right)^{-1}$
&{\it Lb}\\
{\it thin long-dashed line}
&\multicolumn{2}{l}{$A=-\,; \quad\quad m_o=1\,M_\odot$}\\

\hline

\citet{Ch01,Ch02}
&$\xi(m) = A\,m^{-\delta}\,{\rm exp}
         \left[-\left({m_o\over m}\right)^\beta \right]$
&$\Gamma(lm) = 1 - \delta + \beta\left(m_o\over m\right)^\beta$
&{\it Ch}\\

{\it thick short-dash-dotted line}
&\multicolumn{2}{l}{$A=3.0\,{\rm pc}^{-3}\,M_\odot^{-1};
\quad m_o=716.4\,M_\odot; \quad \delta=3.3;  \quad
\beta=0.25 $}\\

\hline

\citet{Hollenbach_etal05}
&$\xi_{\rm L}(m) = k\,m^{-\Gamma} ( 1-{\rm exp} [-(m/m_{\rm ch})^{\gamma+\Gamma}] )$
&
&{\it Holl}\\

{\it not plotted in Fig.~\ref{fig:apl}}

&\multicolumn{2}{l}{$\gamma = 0.4, \Gamma=1.35, m_{\rm ch}=0.18\,M_\odot$}\\

\hline\hline

\end{tabular}
}
\caption{\small{Summary of different proposed analytical IMF forms
(the modern power-law form, the standard IMF, is presented in
eq.~\ref{eq:imf}).  Notation: $lm\equiv{\rm log}_{10}(m/M_\odot)={\rm
ln}(m/M_\odot)/{\rm ln}10$; $dN$ is the number of single stars in the
mass interval $m$ to $m+dm$ and in the logarithmic-mass interval $lm$
to $lm+dlm$.  The mass-dependent IMF indices, $\Gamma(m)$ (eq.~{\it
Gam}), are plotted in Fig.~\ref{fig:apl} using the line-types defined
here.  Eq.~{\it MS} was derived by Miller\&Scalo assuming a constant
star-formation rate and a Galactic disk age of 12~Gyr (the uncertainty
of which is indicated in the lower panel of
Fig.~\ref{fig:apl}a). \citet{Larson98} does not fit his forms
(eqs.~{\it La} and~{\it Lb}) to solar-neighbourhood star-count data
but rather uses these to discuss general aspects of likely systematic
IMF evolution; the $m_o$ in eq.~{\it La} and~{\it Lb} given here are
approximate eye-ball fits to the standard IMF.  }}
\label{tab:imfs}
\end{table}
\end{landscape}


%
%
%
\bibliographystyle{/Users/pavel/BIBTEX_LIB/mn2e}
\bibliography{/Users/pavel/BIBTEX_LIB/quotes_IMF}

\begin{thebibliography}{}

\bibitem[\protect\citeauthoryear{{Aarseth}}{{Aarseth}}{1999}]{Aarseth99}
{Aarseth} S.~J.,  1999, \pasp, 111, 1333

\bibitem[\protect\citeauthoryear{{Adams} \& {Fatuzzo}}{{Adams} \&
  {Fatuzzo}}{1996}]{AF96}
{Adams} F.~C.,  {Fatuzzo} M.,  1996, \apj, 464, 256

\bibitem[\protect\citeauthoryear{{Adams} \& {Laughlin}}{{Adams} \&
  {Laughlin}}{1996}]{AL96}
{Adams} F.~C.,  {Laughlin} G.,  1996, \apj, 468, 586

\bibitem[\protect\citeauthoryear{{Adams} \& {Myers}}{{Adams} \&
  {Myers}}{2001}]{AdamsMyers01}
{Adams} F.~C.,  {Myers} P.~C.,  2001, \apj, 553, 744

\bibitem[\protect\citeauthoryear{{Allen}, {Koerner}, {Reid} \&
  {Trilling}}{{Allen} et~al.}{2005}]{Allen_etal04}
{Allen} P.~R.,  {Koerner} D.~W.,  {Reid} I.~N.,    {Trilling} D.~E.,  2005,
  \apj, 625, 385

\bibitem[\protect\citeauthoryear{{Andersen}}{{Andersen}}{1991}]{A91}
{Andersen} J.,  1991, \aapr, 3, 91

\bibitem[\protect\citeauthoryear{{B{\' e}jar}, {Mart{\'{\i}}n}, {Zapatero
  Osorio}, {Rebolo}, {Barrado y Navascu{\' e}s}, {Bailer-Jones}, {Mundt},
  {Baraffe}, {Chabrier} \& {Allard}}{{B{\' e}jar} et~al.}{2001}]{Betal01}
{B{\' e}jar} V.~J.~S.,  {Mart{\'{\i}}n} E.~L.,  {Zapatero Osorio} M.~R.,
  {Rebolo} R.,  {Barrado y Navascu{\' e}s} D.,  {Bailer-Jones} C.~A.~L.,
  {Mundt} R.,  {Baraffe} I.,  {Chabrier} C.,    {Allard} F.,  2001, \apj, 556,
  830

\bibitem[\protect\citeauthoryear{{Baes}, {Stamatellos}, {Davies}, {Whitworth},
  {Sabatini}, {Roberts}, {Linder} \& {Evans}}{{Baes}
  et~al.}{2005}]{Baes_etal05}
{Baes} M.,  {Stamatellos} D.,  {Davies} J.~I.,  {Whitworth} A.~P.,  {Sabatini}
  S.,  {Roberts} S.,  {Linder} S.~M.,    {Evans} R.,  2005, New Astronomy, 10,
  523

\bibitem[\protect\citeauthoryear{{Bahcall}}{{Bahcall}}{1984}]{B84}
{Bahcall} J.~N.,  1984, \apj, 287, 926

\bibitem[\protect\citeauthoryear{{Baraffe}, {Chabrier}, {Allard} \&
  {Hauschildt}}{{Baraffe} et~al.}{1998}]{BCAH98}
{Baraffe} I.,  {Chabrier} G.,  {Allard} F.,    {Hauschildt} P.~H.,  1998, \aap,
  337, 403

\bibitem[\protect\citeauthoryear{{Barrado y Navascu{\' e}s}, {Stauffer},
  {Bouvier} \& {Mart{\'{\i}}n}}{{Barrado y Navascu{\' e}s}
  et~al.}{2001}]{Netal01}
{Barrado y Navascu{\' e}s} D.,  {Stauffer} J.~R.,  {Bouvier} J.,
  {Mart{\'{\i}}n} E.~L.,  2001, \apj, 546, 1006

\bibitem[\protect\citeauthoryear{{Barrado y Navascu{\' e}s}, {Stauffer} \&
  {Jayawardhana}}{{Barrado y Navascu{\' e}s} et~al.}{2004}]{Barrado_etal04}
{Barrado y Navascu{\' e}s} D.,  {Stauffer} J.~R.,    {Jayawardhana} R.,  2004,
  \apj, 614, 386

\bibitem[\protect\citeauthoryear{{Basri}}{{Basri}}{2000}]{Basri00}
{Basri} G.,  2000, \araa, 38, 485

\bibitem[\protect\citeauthoryear{{Baumgardt} \& {Makino}}{{Baumgardt} \&
  {Makino}}{2003}]{BaumMakin03}
{Baumgardt} H.,  {Makino} J.,  2003, \mnras, 340, 227

\bibitem[\protect\citeauthoryear{{Beech} \& {Mitalas}}{{Beech} \&
  {Mitalas}}{1994}]{BM94}
{Beech} M.,  {Mitalas} R.,  1994, \apjs, 95, 517

\bibitem[\protect\citeauthoryear{{Belikov}, {Hirte}, {Meusinger}, {Piskunov} \&
  {Schilbach}}{{Belikov} et~al.}{1998}]{Belikov98}
{Belikov} A.~N.,  {Hirte} S.,  {Meusinger} H.,  {Piskunov} A.~E.,
  {Schilbach} E.,  1998, \aap, 332, 575

\bibitem[\protect\citeauthoryear{{Beuzit}, {S{\' e}gransan}, {Forveille},
  {Udry}, {Delfosse}, {Mayor}, {Perrier}, {Hainaut}, {Roddier}, {Roddier} \&
  {Mart{\'{\i}}n}}{{Beuzit} et~al.}{2004}]{Beuzit01}
{Beuzit} J.-L.,  {S{\' e}gransan} D.,  {Forveille} T.,  {Udry} S.,  {Delfosse}
  X.,  {Mayor} M.,  {Perrier} C.,  {Hainaut} M.-C.,  {Roddier} C.,  {Roddier}
  F.,    {Mart{\'{\i}}n} E.~L.,  2004, \aap, 425, 997

\bibitem[\protect\citeauthoryear{{Binney}, {Dehnen} \& {Bertelli}}{{Binney}
  et~al.}{2000}]{Binney00}
{Binney} J.,  {Dehnen} W.,    {Bertelli} G.,  2000, \mnras, 318, 658

\bibitem[\protect\citeauthoryear{{Binney} \& {Merrifield}}{{Binney} \&
  {Merrifield}}{1998}]{BM98}
{Binney} J.,  {Merrifield} M.,  1998, {Galactic astronomy}.
Galactic astronomy / James Binney and Michael Merrifield.~ Princeton, NJ :
  Princeton University Press, 1998.~ (Princeton series in astrophysics) QB857
  .B522 1998 (\$35.00)

\bibitem[\protect\citeauthoryear{{Boily}, {Lan{\c c}on}, {Deiters} \&
  {Heggie}}{{Boily} et~al.}{2005}]{Boily_etal05}
{Boily} C.~M.,  {Lan{\c c}on} A.,  {Deiters} S.,    {Heggie} D.~C.,  2005,
  \apjl, 620, L27

\bibitem[\protect\citeauthoryear{{Bonnell} \& {Bate}}{{Bonnell} \&
  {Bate}}{2002}]{BonnellBate02}
{Bonnell} I.~A.,  {Bate} M.~R.,  2002, \mnras, 336, 659

\bibitem[\protect\citeauthoryear{{Bonnell}, {Bate} \& {Vine}}{{Bonnell}
  et~al.}{2003}]{BBV03}
{Bonnell} I.~A.,  {Bate} M.~R.,    {Vine} S.~G.,  2003, \mnras, 343, 413

\bibitem[\protect\citeauthoryear{{Bonnell}, {Bate} \& {Zinnecker}}{{Bonnell}
  et~al.}{1998}]{Bonn98}
{Bonnell} I.~A.,  {Bate} M.~R.,    {Zinnecker} H.,  1998, \mnras, 298, 93

\bibitem[\protect\citeauthoryear{{Bonnell}, {Clarke}, {Bate} \&
  {Pringle}}{{Bonnell} et~al.}{2001}]{Bonn01b}
{Bonnell} I.~A.,  {Clarke} C.~J.,  {Bate} M.~R.,    {Pringle} J.~E.,  2001,
  \mnras, 324, 573

\bibitem[\protect\citeauthoryear{{Bonnell}, {Vine} \& {Bate}}{{Bonnell}
  et~al.}{2004}]{BVB04}
{Bonnell} I.~A.,  {Vine} S.~G.,    {Bate} M.~R.,  2004, \mnras, 349, 735

\bibitem[\protect\citeauthoryear{{Bontemps}, {Andr{\' e}} \& {Kaas}
  A.~A.}{{Bontemps} et~al.}{2001}]{Bont01}
{Bontemps} S.,  {Andr{\' e}} P.,    {Kaas} A.~A. e.~a.,  2001, \aap, 372, 173

\bibitem[\protect\citeauthoryear{{Bosch}, {Selman}, {Melnick} \&
  {Terlevich}}{{Bosch} et~al.}{2001}]{Bosch_etal01}
{Bosch} G.,  {Selman} F.,  {Melnick} J.,    {Terlevich} R.,  2001, \aap, 380,
  137

\bibitem[\protect\citeauthoryear{{Boss}}{{Boss}}{1986}]{Boss86}
{Boss} A.~R.,  1986, \apjs, 62, 519

\bibitem[\protect\citeauthoryear{{Bouy}, {Brandner}, {Mart{\'{\i}}n},
  {Delfosse}, {Allard} \& {Basri}}{{Bouy} et~al.}{2003}]{Bouy_etal03}
{Bouy} H.,  {Brandner} W.,  {Mart{\'{\i}}n} E.~L.,  {Delfosse} X.,  {Allard}
  F.,    {Basri} G.,  2003, \aj, 126, 1526

\bibitem[\protect\citeauthoryear{{Brice{\~ n}o}, {Luhman}, {Hartmann},
  {Stauffer} \& {Kirkpatrick}}{{Brice{\~ n}o} et~al.}{2002}]{Briceno_etal02}
{Brice{\~ n}o} C.,  {Luhman} K.~L.,  {Hartmann} L.,  {Stauffer} J.~R.,
  {Kirkpatrick} J.~D.,  2002, \apj, 580, 317

\bibitem[\protect\citeauthoryear{{Brocato}, {Cassisi} \&
  {Castellani}}{{Brocato} et~al.}{1998}]{BCC98}
{Brocato} E.,  {Cassisi} S.,    {Castellani} V.,  1998, \mnras, 295, 711

\bibitem[\protect\citeauthoryear{{Bromm}, {Ferrara}, {Coppi} \&
  {Larson}}{{Bromm} et~al.}{2001}]{Bromm}
{Bromm} V.,  {Ferrara} A.,  {Coppi} P.~S.,    {Larson} R.~B.,  2001, \mnras,
  328, 969

\bibitem[\protect\citeauthoryear{{Casassus}, {Bronfman}, {May} \&
  {Nyman}}{{Casassus} et~al.}{2000}]{Cass00}
{Casassus} S.,  {Bronfman} L.,  {May} J.,    {Nyman} L.-{\AA}.,  2000, \aap,
  358, 514

\bibitem[\protect\citeauthoryear{{Chabrier}}{{Chabrier}}{2001}]{Ch01}
{Chabrier} G.,  2001, \apj, 554, 1274

\bibitem[\protect\citeauthoryear{{Chabrier}}{{Chabrier}}{2002}]{Ch02}
{Chabrier} G.,  2002, \apj, 567, 304

\bibitem[\protect\citeauthoryear{{Chabrier}}{{Chabrier}}{2003a}]{Chrev03}
{Chabrier} G.,  2003a, \pasp, 115, 763

\bibitem[\protect\citeauthoryear{{Chabrier}}{{Chabrier}}{2003b}]{Ch03}
{Chabrier} G.,  2003b, \apjl, 586, L133

\bibitem[\protect\citeauthoryear{{Chabrier} \& {Baraffe}}{{Chabrier} \&
  {Baraffe}}{1997}]{ChB97}
{Chabrier} G.,  {Baraffe} I.,  1997, \aap, 327, 1039

\bibitem[\protect\citeauthoryear{{Chabrier} \& {Baraffe}}{{Chabrier} \&
  {Baraffe}}{2000}]{ChB00}
{Chabrier} G.,  {Baraffe} I.,  2000, \araa, 38, 337

\bibitem[\protect\citeauthoryear{{Chini}, {Hoffmeister}, {Kimeswenger},
  {Nielbock}, {N{\" u}rnberger}, {Schmidtobreick} \& {Sterzik}}{{Chini}
  et~al.}{2004}]{CHK04}
{Chini} R.,  {Hoffmeister} V.,  {Kimeswenger} S.,  {Nielbock} M.,  {N{\"
  u}rnberger} D.,  {Schmidtobreick} L.,    {Sterzik} M.,  2004, \nat, 429, 155

\bibitem[\protect\citeauthoryear{{Clarke} \& {Pringle}}{{Clarke} \&
  {Pringle}}{1992}]{ClarkePringle92}
{Clarke} C.~J.,  {Pringle} J.~E.,  1992, \mnras, 255, 423

\bibitem[\protect\citeauthoryear{{Close}, {Siegler}, {Freed} \&
  {Biller}}{{Close} et~al.}{2003}]{Close_etal03}
{Close} L.~M.,  {Siegler} N.,  {Freed} M.,    {Biller} B.,  2003, \apj, 587,
  407

\bibitem[\protect\citeauthoryear{{Corbelli}, {Palla} \& {Zinnecker}}{{Corbelli}
  et~al.}{2005}]{IMF50}
{Corbelli} E.,  {Palla} F.,    {Zinnecker} H.,  eds, 2005, {The Initial Mass
  Function 50 years later}

\bibitem[\protect\citeauthoryear{{D'Antona} \& {Mazzitelli}}{{D'Antona} \&
  {Mazzitelli}}{1996}]{dAM96}
{D'Antona} F.,  {Mazzitelli} I.,  1996, \apj, 456, 329

\bibitem[\protect\citeauthoryear{{de Boer}, {Fitzpatrick} \& {Savage}}{{de
  Boer} et~al.}{1985}]{deBoer_etal85}
{de Boer} K.~S.,  {Fitzpatrick} E.~L.,    {Savage} B.~D.,  1985, \mnras, 217,
  115

\bibitem[\protect\citeauthoryear{{de La Fuente Marcos}}{{de La Fuente
  Marcos}}{1997}]{Fuente97}
{de La Fuente Marcos} R.,  1997, \aap, 322, 764

\bibitem[\protect\citeauthoryear{{de La Fuente Marcos}}{{de La Fuente
  Marcos}}{1998}]{Fuente98}
{de La Fuente Marcos} R.,  1998, \aap, 333, L27

\bibitem[\protect\citeauthoryear{{de Marchi} \& {Paresce}}{{de Marchi} \&
  {Paresce}}{1995a}]{deMP95a}
{de Marchi} G.,  {Paresce} F.,  1995a, \aap, 304, 202

\bibitem[\protect\citeauthoryear{{de Marchi} \& {Paresce}}{{de Marchi} \&
  {Paresce}}{1995b}]{deMP95b}
{de Marchi} G.,  {Paresce} F.,  1995b, \aap, 304, 211

\bibitem[\protect\citeauthoryear{{Delfosse}, {Forveille}, {Beuzit}, {Udry},
  {Mayor} \& {Perrier}}{{Delfosse} et~al.}{1999}]{Detal99}
{Delfosse} X.,  {Forveille} T.,  {Beuzit} J.-L.,  {Udry} S.,  {Mayor} M.,
  {Perrier} C.,  1999, \aap, 344, 897

\bibitem[\protect\citeauthoryear{{Delfosse}, {Forveille}, {S{\' e}gransan},
  {Beuzit}, {Udry}, {Perrier} \& {Mayor}}{{Delfosse} et~al.}{2000}]{Detal00}
{Delfosse} X.,  {Forveille} T.,  {S{\' e}gransan} D.,  {Beuzit} J.-L.,  {Udry}
  S.,  {Perrier} C.,    {Mayor} M.,  2000, \aap, 364, 217

\bibitem[\protect\citeauthoryear{{Duch{\^ e}ne}, {Simon}, {Eisl{\" o}ffel} \&
  {Bouvier}}{{Duch{\^ e}ne} et~al.}{2001}]{Duchene01}
{Duch{\^ e}ne} G.,  {Simon} T.,  {Eisl{\" o}ffel} J.,    {Bouvier} J.,  2001,
  \aap, 379, 147

\bibitem[\protect\citeauthoryear{{Duquennoy} \& {Mayor}}{{Duquennoy} \&
  {Mayor}}{1991}]{DM91}
{Duquennoy} A.,  {Mayor} M.,  1991, \aap, 248, 485

\bibitem[\protect\citeauthoryear{{Eddington}}{{Eddington}}{1926}]{Edd26}
{Eddington} A.~S.,  1926, {The Internal Constitution of the Stars}.
The Internal Constitution of the Stars, Cambridge: Cambridge University Press,
  1926

\bibitem[\protect\citeauthoryear{{Egusa}, {Sofue} \& {Nakanishi}}{{Egusa}
  et~al.}{2004}]{Egusa_etal04}
{Egusa} F.,  {Sofue} Y.,    {Nakanishi} H.,  2004, \pasj, 56, L45

\bibitem[\protect\citeauthoryear{{Eisenhauer}}{{Eisenhauer}}{2001}]{Eisenh01}
{Eisenhauer} F.,  2001, in Starburst Galaxies: Near and Far {Evidence in Favour
  of IMF Variations}.
pp 24--+

\bibitem[\protect\citeauthoryear{{Eisenhauer}, {Quirrenbach}, {Zinnecker} \&
  {Genzel}}{{Eisenhauer} et~al.}{1998}]{Eetal98}
{Eisenhauer} F.,  {Quirrenbach} A.,  {Zinnecker} H.,    {Genzel} R.,  1998,
  \apj, 498, 278

\bibitem[\protect\citeauthoryear{{Elmegreen}}{{Elmegreen}}{1983}]{Elm83}
{Elmegreen} B.~G.,  1983, \mnras, 203, 1011

\bibitem[\protect\citeauthoryear{{Elmegreen}}{{Elmegreen}}{1997}]{Elm97}
{Elmegreen} B.~G.,  1997, \apj, 486, 944

\bibitem[\protect\citeauthoryear{{Elmegreen}}{{Elmegreen}}{1999}]{Elm99}
{Elmegreen} B.~G.,  1999, \apj, 515, 323

\bibitem[\protect\citeauthoryear{{Elmegreen}}{{Elmegreen}}{2000}]{Elm00}
{Elmegreen} B.~G.,  2000, \apj, 539, 342

\bibitem[\protect\citeauthoryear{{Elmegreen}}{{Elmegreen}}{2004}]{El04}
{Elmegreen} B.~G.,  2004, \mnras, 354, 367

\bibitem[\protect\citeauthoryear{{Esteban}, {Peimbert}, {Torres-Peimbert} \&
  {Escalante}}{{Esteban} et~al.}{1998}]{Estetal98}
{Esteban} C.,  {Peimbert} M.,  {Torres-Peimbert} S.,    {Escalante} V.,  1998,
  \mnras, 295, 401

\bibitem[\protect\citeauthoryear{{Fellhauer}, {Lin}, {Bolte}, {Aarseth} \&
  {Williams}}{{Fellhauer} et~al.}{2003}]{Fellhauer_etal03}
{Fellhauer} M.,  {Lin} D.~N.~C.,  {Bolte} M.,  {Aarseth} S.~J.,    {Williams}
  K.~A.,  2003, \apjl, 595, L53

\bibitem[\protect\citeauthoryear{{Feltzing}, {Gilmore} \& {Wyse}}{{Feltzing}
  et~al.}{1999}]{Felt99}
{Feltzing} S.,  {Gilmore} G.,    {Wyse} R.~F.~G.,  1999, \apjl, 516, L17

\bibitem[\protect\citeauthoryear{{Ferrini}, {Marchesoni} \&
  {Vulpiani}}{{Ferrini} et~al.}{1983}]{Ferrini_etal83}
{Ferrini} F.,  {Marchesoni} F.,    {Vulpiani} A.,  1983, \mnras, 202, 1071

\bibitem[\protect\citeauthoryear{{Figer}}{{Figer}}{2003}]{Figer02}
{Figer} D.~F.,  2003, in {van der Hucht} K.,  {Herrero} A.,   {Esteban} C.,
  eds, IAU Symposium {Massive stars and the creation of our Galactic Center}.
pp 487--+

\bibitem[\protect\citeauthoryear{{Figer}}{{Figer}}{2005}]{Fi05}
{Figer} D.~F.,  2005, \nat, 434, 192

\bibitem[\protect\citeauthoryear{{Figer}, {Kim}, {Morris}, {Serabyn}, {Rich} \&
  {McLean}}{{Figer} et~al.}{1999}]{Fetal99}
{Figer} D.~F.,  {Kim} S.~S.,  {Morris} M.,  {Serabyn} E.,  {Rich} R.~M.,
  {McLean} I.~S.,  1999, \apj, 525, 750

\bibitem[\protect\citeauthoryear{{Fischer} \& {Marcy}}{{Fischer} \&
  {Marcy}}{1992}]{FiMa92}
{Fischer} D.~A.,  {Marcy} G.~W.,  1992, \apj, 396, 178

\bibitem[\protect\citeauthoryear{{Fleck}}{{Fleck}}{1982}]{Fleck82}
{Fleck} R.~C.,  1982, \mnras, 201, 551

\bibitem[\protect\citeauthoryear{{Flynn} \& {Fuchs}}{{Flynn} \&
  {Fuchs}}{1994}]{FF94}
{Flynn} C.,  {Fuchs} B.,  1994, \mnras, 270, 471

\bibitem[\protect\citeauthoryear{{Fuhrmann}}{{Fuhrmann}}{2004}]{Fuhrmann04}
{Fuhrmann} K.,  2004, Astronomische Nachrichten, 325, 3

\bibitem[\protect\citeauthoryear{{Garc{\'{\i}}a} \&
  {Mermilliod}}{{Garc{\'{\i}}a} \& {Mermilliod}}{2001}]{GM01}
{Garc{\'{\i}}a} B.,  {Mermilliod} J.~C.,  2001, \aap, 368, 122

\bibitem[\protect\citeauthoryear{{Garcia-Segura}, {Langer} \& {Mac
  Low}}{{Garcia-Segura} et~al.}{1996}]{Garcia96a}
{Garcia-Segura} G.,  {Langer} N.,    {Mac Low} M.-M.,  1996, \aap, 316, 133

\bibitem[\protect\citeauthoryear{{Garcia-Segura}, {Mac Low} \&
  {Langer}}{{Garcia-Segura} et~al.}{1996}]{Garcia96b}
{Garcia-Segura} G.,  {Mac Low} M.-M.,    {Langer} N.,  1996, \aap, 305, 229

\bibitem[\protect\citeauthoryear{{Geyer} \& {Burkert}}{{Geyer} \&
  {Burkert}}{2001}]{GeyerBurkert01}
{Geyer} M.~P.,  {Burkert} A.,  2001, \mnras, 323, 988

\bibitem[\protect\citeauthoryear{{Gilmore} \& {Howell}}{{Gilmore} \&
  {Howell}}{1998}]{GilHow98}
{Gilmore} G.,  {Howell} D.,  eds, 1998, {The Stellar Initial Mass Function
  (38th Herstmonceux Conference)}

\bibitem[\protect\citeauthoryear{{Gilmore}, {Perryman}, {Lindegren}, {Favata},
  {Hoeg}, {Lattanzi}, {Luri}, {Mignard}, {Roeser} \& {de Zeeuw}}{{Gilmore}
  et~al.}{1998}]{Gil98}
{Gilmore} G.~F.,  {Perryman} M.~A.,  {Lindegren} L.,  {Favata} F.,  {Hoeg} E.,
  {Lattanzi} M.,  {Luri} X.,  {Mignard} F.,  {Roeser} S.,    {de Zeeuw} P.~T.,
  1998, in Proc. SPIE Vol. 3350, p. 541-550, Astronomical Interferometry,
  Robert D. Reasenberg; Ed. {GAIA: origin and evolution of the Milky Way}.
pp 541--550

\bibitem[\protect\citeauthoryear{{Gizis}, {Kirkpatrick}, {Burgasser}, {Reid},
  {Monet}, {Liebert} \& {Wilson}}{{Gizis} et~al.}{2001}]{Gizis01}
{Gizis} J.~E.,  {Kirkpatrick} J.~D.,  {Burgasser} A.,  {Reid} I.~N.,  {Monet}
  D.~G.,  {Liebert} J.,    {Wilson} J.~C.,  2001, \apjl, 551, L163

\bibitem[\protect\citeauthoryear{{Goodwin}}{{Goodwin}}{1997}]{Goodwin97}
{Goodwin} S.~P.,  1997, \mnras, 284, 785

\bibitem[\protect\citeauthoryear{{Goodwin} \& {Bastian}}{{Goodwin} \&
  {Bastian}}{2006}]{GB06}
{Goodwin} S.~P.,  {Bastian} N.,  2006, \mnras, 373, 752

\bibitem[\protect\citeauthoryear{{Goodwin} \& {Kroupa}}{{Goodwin} \&
  {Kroupa}}{2005}]{GK05}
{Goodwin} S.~P.,  {Kroupa} P.,  2005, ArXiv Astrophysics e-prints

\bibitem[\protect\citeauthoryear{{Goodwin} \& {Pagel}}{{Goodwin} \&
  {Pagel}}{2005}]{GP05}
{Goodwin} S.~P.,  {Pagel} B.~E.~J.,  2005, \mnras, 359, 707

\bibitem[\protect\citeauthoryear{{Goodwin}, {Whitworth} \&
  {Ward-Thompson}}{{Goodwin} et~al.}{2004}]{Goodwin_etal04}
{Goodwin} S.~P.,  {Whitworth} A.~P.,    {Ward-Thompson} D.,  2004, \aap, 414,
  633

\bibitem[\protect\citeauthoryear{{Gouliermis}, {Brandner} \&
  {Henning}}{{Gouliermis} et~al.}{2005}]{Goulier_etal05}
{Gouliermis} D.,  {Brandner} W.,    {Henning} T.,  2005, \apj, 623, 846

\bibitem[\protect\citeauthoryear{{Grillmair}, {Mould} \& {Holtzman}
  J.~A.}{{Grillmair} et~al.}{1998}]{Grill98}
{Grillmair} C.~J.,  {Mould} J.~R.,    {Holtzman} J.~A. e.~a.,  1998, \aj, 115,
  144

\bibitem[\protect\citeauthoryear{{Halbwachs}, {Arenou}, {Mayor}, {Udry} \&
  {Queloz}}{{Halbwachs} et~al.}{2000}]{Halbw00}
{Halbwachs} J.~L.,  {Arenou} F.,  {Mayor} M.,  {Udry} S.,    {Queloz} D.,
  2000, \aap, 355, 581

\bibitem[\protect\citeauthoryear{{Hambly}, {Hodgkin}, {Cossburn} \&
  {Jameson}}{{Hambly} et~al.}{1999}]{Hambetal99}
{Hambly} N.~C.,  {Hodgkin} S.~T.,  {Cossburn} M.~R.,    {Jameson} R.~F.,  1999,
  \mnras, 303, 835

\bibitem[\protect\citeauthoryear{{Hambly}, {Jameson} \& {Hawkins}}{{Hambly}
  et~al.}{1991}]{HJH91}
{Hambly} N.~C.,  {Jameson} R.~F.,    {Hawkins} M.~R.~S.,  1991, \mnras, 253, 1

\bibitem[\protect\citeauthoryear{{Hayashi} \& {Nakano}}{{Hayashi} \&
  {Nakano}}{1963}]{HN63}
{Hayashi} C.,  {Nakano} T.,  1963, Progress of Theor. Phys., 30, 460

\bibitem[\protect\citeauthoryear{{Haywood}, {Robin} \& {Creze}}{{Haywood}
  et~al.}{1997}]{Haywood_etal97}
{Haywood} M.,  {Robin} A.~C.,    {Creze} M.,  1997, \aap, 320, 440

\bibitem[\protect\citeauthoryear{{Henry}, {Ianna}, {Kirkpatrick} \&
  {Jahreiss}}{{Henry} et~al.}{1997}]{Henry97}
{Henry} T.~J.,  {Ianna} P.~A.,  {Kirkpatrick} J.~D.,    {Jahreiss} H.,  1997,
  \aj, 114, 388

\bibitem[\protect\citeauthoryear{{Herbst}, {Thompson}, {Fockenbrock}, {Rix} \&
  {Beckwith}}{{Herbst} et~al.}{1999}]{Hetal99}
{Herbst} T.~M.,  {Thompson} D.,  {Fockenbrock} R.,  {Rix} H.-W.,    {Beckwith}
  S.~V.~W.,  1999, \apjl, 526, L17

\bibitem[\protect\citeauthoryear{{Hillenbrand}}{{Hillenbrand}}{1997}]{Hill97}
{Hillenbrand} L.~A.,  1997, \aj, 113, 1733

\bibitem[\protect\citeauthoryear{{Hillenbrand}}{{Hillenbrand}}{2004}]{Hillrev0%
4}
{Hillenbrand} L.~A.,  2004, in The Dense Interstellar Medium in Galaxies {The
  Mass Function of Newly Formed Stars}.
pp 601--+

\bibitem[\protect\citeauthoryear{{Hillenbrand} \& {Carpenter}}{{Hillenbrand} \&
  {Carpenter}}{2000}]{HC00}
{Hillenbrand} L.~A.,  {Carpenter} J.~M.,  2000, \apj, 540, 236

\bibitem[\protect\citeauthoryear{{Hollenbach}, {Parravano} \&
  {McKee}}{{Hollenbach} et~al.}{2005}]{Hollenbach_etal05}
{Hollenbach} D.,  {Parravano} A.,    {McKee} C.~F.,  2005, in {Corbelli} E.,
  {Palla} F.,   {Zinnecker} H.,  eds, ASSL Vol. 327: The Initial Mass Function
  50 Years Later {An effective Initial Mass Function for galactic disks}.
pp 417--424

\bibitem[\protect\citeauthoryear{{Holtzman}, {Watson}, {Baum}, {Grillmair},
  {Groth}, {Light}, {Lynds} \& {O'Neil}}{{Holtzman} et~al.}{1998}]{Hetal98}
{Holtzman} J.~A.,  {Watson} A.~M.,  {Baum} W.~A.,  {Grillmair} C.~J.,  {Groth}
  E.~J.,  {Light} R.~M.,  {Lynds} R.,    {O'Neil} E.~J.,  1998, \aj, 115, 1946

\bibitem[\protect\citeauthoryear{{Hurley}, {Pols}, {Aarseth} \&
  {Tout}}{{Hurley} et~al.}{2005}]{Hurley_etal05}
{Hurley} J.~R.,  {Pols} O.~R.,  {Aarseth} S.~J.,    {Tout} C.~A.,  2005,
  \mnras, 363, 293

\bibitem[\protect\citeauthoryear{{Hurley}, {Pols} \& {Tout}}{{Hurley}
  et~al.}{2000}]{HPT00}
{Hurley} J.~R.,  {Pols} O.~R.,    {Tout} C.~A.,  2000, \mnras, 315, 543

\bibitem[\protect\citeauthoryear{{Ivanova}, {Belczynski}, {Fregeau} \&
  {Rasio}}{{Ivanova} et~al.}{2005}]{Ivanova_etal05}
{Ivanova} N.,  {Belczynski} K.,  {Fregeau} J.~M.,    {Rasio} F.~A.,  2005,
  \mnras, 358, 572

\bibitem[\protect\citeauthoryear{{Jahreiss}}{{Jahreiss}}{1994}]{J94}
{Jahreiss} H.,  1994, \apss, 217, 63

\bibitem[\protect\citeauthoryear{{Jahrei{\ss}} \& {Wielen}}{{Jahrei{\ss}} \&
  {Wielen}}{1997}]{JW97}
{Jahrei{\ss}} H.,  {Wielen} R.,  1997, in ESA SP-402: Hipparcos - Venice '97
  {The impact of HIPPARCOS on the Catalogue of Nearby Stars. The stellar
  luminosity function and local kinematics}.
pp 675--680

\bibitem[\protect\citeauthoryear{{Janka}}{{Janka}}{2001}]{Janka01}
{Janka} H.-T.,  2001, \aap, 368, 527

\bibitem[\protect\citeauthoryear{{Jao}, {Henry}, {Subasavage}, {Bean}, {Costa},
  {Ianna} \& {M{\' e}ndez}}{{Jao} et~al.}{2003}]{Hetal03}
{Jao} W.,  {Henry} T.~J.,  {Subasavage} J.~P.,  {Bean} J.~L.,  {Costa} E.,
  {Ianna} P.~A.,    {M{\' e}ndez} R.~A.,  2003, \aj, 125, 332

\bibitem[\protect\citeauthoryear{{Jijina} \& {Adams}}{{Jijina} \&
  {Adams}}{1996}]{JA96}
{Jijina} J.,  {Adams} F.~C.,  1996, \apj, 462, 874

\bibitem[\protect\citeauthoryear{{Kahn}}{{Kahn}}{1974}]{Kahn74}
{Kahn} F.~D.,  1974, \aap, 37, 149

\bibitem[\protect\citeauthoryear{{Kim}, {Figer}, {Kudritzki} \&
  {Najarro}}{{Kim} et~al.}{2006}]{Kimetal06}
{Kim} S.~S.,  {Figer} D.~F.,  {Kudritzki} R.~P.,    {Najarro} F.,  2006, \apjl,
  653, L113

\bibitem[\protect\citeauthoryear{{Kim}, {Figer}, {Lee} \& {Morris}}{{Kim}
  et~al.}{2000}]{Kimetal2000}
{Kim} S.~S.,  {Figer} D.~F.,  {Lee} H.~M.,    {Morris} M.,  2000, \apj, 545,
  301

\bibitem[\protect\citeauthoryear{{Klessen}}{{Klessen}}{2001}]{Kl01}
{Klessen} R.~S.,  2001, \apjl, 550, L77

\bibitem[\protect\citeauthoryear{{Koen}}{{Koen}}{2006}]{Koen06}
{Koen} C.,  2006, \mnras, 365, 590

\bibitem[\protect\citeauthoryear{{K{\"o}ppen}, {Weidner} \&
  {Kroupa}}{{K{\"o}ppen} et~al.}{2007}]{KWK07}
{K{\"o}ppen} J.,  {Weidner} C.,    {Kroupa} P.,  2007, \mnras, pp 1479--+

\bibitem[\protect\citeauthoryear{{Kroupa}}{{Kroupa}}{1995a}]{K95a}
{Kroupa} P.,  1995a, \apj, 453, 350

\bibitem[\protect\citeauthoryear{{Kroupa}}{{Kroupa}}{1995b}]{K95d}
{Kroupa} P.,  1995b, \mnras, 277, 1491

\bibitem[\protect\citeauthoryear{{Kroupa}}{{Kroupa}}{1995c}]{K95b}
{Kroupa} P.,  1995c, \mnras, 277, 1522

\bibitem[\protect\citeauthoryear{{Kroupa}}{{Kroupa}}{1995d}]{K95c}
{Kroupa} P.,  1995d, \mnras, 277, 1507

\bibitem[\protect\citeauthoryear{{Kroupa}}{{Kroupa}}{1995e}]{K95e}
{Kroupa} P.,  1995e, \apj, 453, 358

\bibitem[\protect\citeauthoryear{{Kroupa}}{{Kroupa}}{2000}]{K00}
{Kroupa} P.,  2000, New Astronomy, 4, 615

\bibitem[\protect\citeauthoryear{{Kroupa}}{{Kroupa}}{2001a}]{K01c}
{Kroupa} P.,  2001a, in IAU Symposium {Binary Stars in Young Clusters -- a
  Theoretical Perspective}.
pp 199--+

\bibitem[\protect\citeauthoryear{{Kroupa}}{{Kroupa}}{2001b}]{K01a}
{Kroupa} P.,  2001b, \mnras, 322, 231

\bibitem[\protect\citeauthoryear{{Kroupa}}{{Kroupa}}{2001c}]{K01b}
{Kroupa} P.,  2001c, in ASP Conf. Ser. 228: Dynamics of Star Clusters and the
  Milky Way {The Local Stellar Initial Mass Function}.
pp 187--+

\bibitem[\protect\citeauthoryear{{Kroupa}}{{Kroupa}}{2002}]{K02}
{Kroupa} P.,  2002, Science, 295, 82

\bibitem[\protect\citeauthoryear{{Kroupa}}{{Kroupa}}{2005}]{Kr_paris05}
{Kroupa} P.,  2005, in Proceedings of the Gaia Symposium "The Three-Dimensional
  Universe with Gaia" (ESA SP-576). Held at the Observatoire de Paris-Meudon,
  4-7 October 2004. Editors: C. Turon, K.S. O'Flaherty, M.A.C. Perryman {The
  Fundamental Building Blocks of Galaxies}.
pp 629--+

\bibitem[\protect\citeauthoryear{{Kroupa}, {Aarseth} \& {Hurley}}{{Kroupa}
  et~al.}{2001}]{KAH}
{Kroupa} P.,  {Aarseth} S.,    {Hurley} J.,  2001, \mnras, 321, 699

\bibitem[\protect\citeauthoryear{{Kroupa} \& {Bouvier}}{{Kroupa} \&
  {Bouvier}}{2003}]{Kr_Bouv03b}
{Kroupa} P.,  {Bouvier} J.,  2003, \mnras, 346, 369

\bibitem[\protect\citeauthoryear{{Kroupa}, {Bouvier}, {Duch{\^ e}ne} \&
  {Moraux}}{{Kroupa} et~al.}{2003}]{Kr_etal03}
{Kroupa} P.,  {Bouvier} J.,  {Duch{\^ e}ne} G.,    {Moraux} E.,  2003, \mnras,
  346, 354

\bibitem[\protect\citeauthoryear{{Kroupa}, {Gilmore} \& {Tout}}{{Kroupa}
  et~al.}{1991}]{KTG91}
{Kroupa} P.,  {Gilmore} G.,    {Tout} C.~A.,  1991, \mnras, 251, 293

\bibitem[\protect\citeauthoryear{{Kroupa} \& {Tout}}{{Kroupa} \&
  {Tout}}{1997}]{KT97}
{Kroupa} P.,  {Tout} C.~A.,  1997, \mnras, 287, 402

\bibitem[\protect\citeauthoryear{{Kroupa}, {Tout} \& {Gilmore}}{{Kroupa}
  et~al.}{1990}]{KTG90}
{Kroupa} P.,  {Tout} C.~A.,    {Gilmore} G.,  1990, \mnras, 244, 76

\bibitem[\protect\citeauthoryear{{Kroupa}, {Tout} \& {Gilmore}}{{Kroupa}
  et~al.}{1993}]{KTG93}
{Kroupa} P.,  {Tout} C.~A.,    {Gilmore} G.,  1993, \mnras, 262, 545

\bibitem[\protect\citeauthoryear{{Kroupa} \& {Weidner}}{{Kroupa} \&
  {Weidner}}{2003}]{KW03}
{Kroupa} P.,  {Weidner} C.,  2003, \apj, 598, 1076

\bibitem[\protect\citeauthoryear{{Kudritzki} \& {Puls}}{{Kudritzki} \&
  {Puls}}{2000}]{KP00}
{Kudritzki} R.,  {Puls} J.,  2000, \araa, 38, 613

\bibitem[\protect\citeauthoryear{{Kuijken}}{{Kuijken}}{1991}]{Kuijken}
{Kuijken} K.,  1991, \apj, 372, 125

\bibitem[\protect\citeauthoryear{{Kumar}}{{Kumar}}{2003}]{Kumar03}
{Kumar} S.~S.,  2003, in IAU Symposium {The Bottom of the Main Sequence and
  Beyond: Speculations, Calculations, Observations, and Discoveries
  (1958--2002)}.
pp~3--+

\bibitem[\protect\citeauthoryear{{Lada} \& {Lada}}{{Lada} \&
  {Lada}}{2003}]{Lada_Lada03}
{Lada} C.~J.,  {Lada} E.~A.,  2003, \araa, 41, 57

\bibitem[\protect\citeauthoryear{{Lada}, {Margulis} \& {Dearborn}}{{Lada}
  et~al.}{1984}]{Lada_etal84}
{Lada} C.~J.,  {Margulis} M.,    {Dearborn} D.,  1984, \apj, 285, 141

\bibitem[\protect\citeauthoryear{{Larson}}{{Larson}}{1982}]{Larson82}
{Larson} R.~B.,  1982, \mnras, 200, 159

\bibitem[\protect\citeauthoryear{{Larson}}{{Larson}}{1998}]{Larson98}
{Larson} R.~B.,  1998, \mnras, 301, 569

\bibitem[\protect\citeauthoryear{{Larson}}{{Larson}}{2003}]{Larson03}
{Larson} R.~B.,  2003, in ASP Conf. Ser. 287: Galactic Star Formation Across
  the Stellar Mass Spectrum {The Stellar Initial Mass Function and Beyond
  (Invited Review)}.
pp 65--80

\bibitem[\protect\citeauthoryear{{Low} \& {Lynden-Bell}}{{Low} \&
  {Lynden-Bell}}{1976}]{LowLyndenBell76}
{Low} C.,  {Lynden-Bell} D.,  1976, \mnras, 176, 367

\bibitem[\protect\citeauthoryear{{Luhman}}{{Luhman}}{2004}]{Luhman04}
{Luhman} K.~L.,  2004, \apj, 617, 1216

\bibitem[\protect\citeauthoryear{{Luhman}, {Brice{\~ n}o}, {Stauffer},
  {Hartmann}, {Barrado y Navascu{\' e}s} \& {Caldwell}}{{Luhman}
  et~al.}{2003}]{Luhman_etal03}
{Luhman} K.~L.,  {Brice{\~ n}o} C.,  {Stauffer} J.~R.,  {Hartmann} L.,
  {Barrado y Navascu{\' e}s} D.,    {Caldwell} N.,  2003, \apj, 590, 348

\bibitem[\protect\citeauthoryear{{Luhman}, {Rieke}, {Lada} \& {Lada}}{{Luhman}
  et~al.}{1998}]{L98}
{Luhman} K.~L.,  {Rieke} G.~H.,  {Lada} C.~J.,    {Lada} E.~A.,  1998, \apj,
  508, 347

\bibitem[\protect\citeauthoryear{{Luhman}, {Rieke}, {Young}, {Cotera}, {Chen},
  {Rieke}, {Schneider} \& {Thompson}}{{Luhman} et~al.}{2000}]{L00}
{Luhman} K.~L.,  {Rieke} G.~H.,  {Young} E.~T.,  {Cotera} A.~S.,  {Chen} H.,
  {Rieke} M.~J.,  {Schneider} G.,    {Thompson} R.~I.,  2000, \apj, 540, 1016

\bibitem[\protect\citeauthoryear{{Mac Low} \& {Klessen}}{{Mac Low} \&
  {Klessen}}{2004}]{MacLowKlessen04}
{Mac Low} M.,  {Klessen} R.~S.,  2004, Reviews of Modern Physics, 76, 125

\bibitem[\protect\citeauthoryear{{Maeder} \& {Behrend}}{{Maeder} \&
  {Behrend}}{2002}]{MB01}
{Maeder} A.,  {Behrend} R.,  2002, in ASP Conf. Ser. 267: Hot Star Workshop
  III: The Earliest Phases of Massive Star Birth {Formation and pre-MS
  Evolution of Massive Stars with Growing Accretion}.
pp 179--+

\bibitem[\protect\citeauthoryear{{Maeder} \& {Meynet}}{{Maeder} \&
  {Meynet}}{2000}]{MM00}
{Maeder} A.,  {Meynet} G.,  2000, \araa, 38, 143

\bibitem[\protect\citeauthoryear{{Ma{\'{\i}}z Apell{\' a}niz}, {{\' U}beda},
  {Walborn} \& {Nelan}}{{Ma{\'{\i}}z Apell{\' a}niz} et~al.}{2005}]{Jesus05}
{Ma{\'{\i}}z Apell{\' a}niz} J.,  {{\' U}beda} L.,  {Walborn} N.~R.,    {Nelan}
  E.~P.,  2005, ArXiv Astrophysics e-prints: stro-ph/0506283

\bibitem[\protect\citeauthoryear{{Ma{\'{\i}}z Apell{\'a}niz}, {Walborn},
  {Morrell}, {Niemela} \& {Nelan}}{{Ma{\'{\i}}z Apell{\'a}niz}
  et~al.}{2006}]{Jesus07}
{Ma{\'{\i}}z Apell{\'a}niz} J.,  {Walborn} N.~R.,  {Morrell} N.~I.,  {Niemela}
  V.~S.,    {Nelan} E.~P.,  2006, ArXiv Astrophysics e-prints

\bibitem[\protect\citeauthoryear{{Malkov} \& {Zinnecker}}{{Malkov} \&
  {Zinnecker}}{2001}]{MZ01}
{Malkov} O.,  {Zinnecker} H.,  2001, \mnras, 321, 149

\bibitem[\protect\citeauthoryear{{Mart{\'{\i}}n}, {Barrado y Navascu{\' e}s},
  {Baraffe}, {Bouy} \& {Dahm}}{{Mart{\'{\i}}n} et~al.}{2003}]{Martin_etal03}
{Mart{\'{\i}}n} E.~L.,  {Barrado y Navascu{\' e}s} D.,  {Baraffe} I.,  {Bouy}
  H.,    {Dahm} S.,  2003, \apj, 594, 525

\bibitem[\protect\citeauthoryear{{Massey}}{{Massey}}{1998}]{M98}
{Massey} P.,  1998, in ASP Conf. Ser. 142: The Stellar Initial Mass Function
  (38th Herstmonceux Conference) {The Initial Mass Function of Massive Stars in
  the Local Group}.
pp 17--+

\bibitem[\protect\citeauthoryear{{Massey}}{{Massey}}{2003}]{Massey03}
{Massey} P.,  2003, \araa, 41, 15

\bibitem[\protect\citeauthoryear{{Massey} \& {Hunter}}{{Massey} \&
  {Hunter}}{1998}]{MH98}
{Massey} P.,  {Hunter} D.~A.,  1998, \apj, 493, 180

\bibitem[\protect\citeauthoryear{{Maxted} \& {Jeffries}}{{Maxted} \&
  {Jeffries}}{2005}]{MF05}
{Maxted} P.~F.~L.,  {Jeffries} R.~D.,  2005, ArXiv Astrophysics e-prints

\bibitem[\protect\citeauthoryear{{Mayor}, {Duquennoy}, {Halbwachs} \&
  {Mermilliod}}{{Mayor} et~al.}{1992}]{Mayor_etal92}
{Mayor} M.,  {Duquennoy} A.,  {Halbwachs} J.-L.,    {Mermilliod} J.-C.,  1992,
  in ASP Conf. Ser. 32: IAU Colloq. 135: Complementary Approaches to Double and
  Multiple Star Research {CORAVEL Surveys to Study Binaries of Different Masses
  and Ages}.
pp 73--+

\bibitem[\protect\citeauthoryear{{Megeath}, {Herter}, {Beichman}, {Gautier},
  {Hester}, {Rayner} \& {Shupe}}{{Megeath} et~al.}{1996}]{Meg}
{Megeath} S.~T.,  {Herter} T.,  {Beichman} C.,  {Gautier} N.,  {Hester} J.~J.,
  {Rayner} J.,    {Shupe} D.,  1996, \aap, 307, 775

\bibitem[\protect\citeauthoryear{{Meyer}, {Adams}, {Hillenbrand}, {Carpenter}
  \& {Larson}}{{Meyer} et~al.}{2000}]{Meyer00}
{Meyer} M.~R.,  {Adams} F.~C.,  {Hillenbrand} L.~A.,  {Carpenter} J.~M.,
  {Larson} R.~B.,  2000, Protostars and Planets IV, pp 121--+

\bibitem[\protect\citeauthoryear{{Meynet} \& {Maeder}}{{Meynet} \&
  {Maeder}}{2003}]{MM03}
{Meynet} G.,  {Maeder} A.,  2003, \aap, 404, 975

\bibitem[\protect\citeauthoryear{{Miller} \& {Scalo}}{{Miller} \&
  {Scalo}}{1979}]{MS79}
{Miller} G.~E.,  {Scalo} J.~M.,  1979, \apjs, 41, 513

\bibitem[\protect\citeauthoryear{{Moraux}, {Bouvier} \& {Stauffer}}{{Moraux}
  et~al.}{2001}]{MBS01}
{Moraux} E.,  {Bouvier} J.,    {Stauffer} J.~R.,  2001, \aap, 367, 211

\bibitem[\protect\citeauthoryear{{Moraux}, {Kroupa} \& {Bouvier}}{{Moraux}
  et~al.}{2004}]{MKB04}
{Moraux} E.,  {Kroupa} P.,    {Bouvier} J.,  2004, \aap, 426, 75

\bibitem[\protect\citeauthoryear{{Motte}, {Andre} \& {Neri}}{{Motte}
  et~al.}{1998}]{Motte}
{Motte} F.,  {Andre} P.,    {Neri} R.,  1998, \aap, 336, 150

\bibitem[\protect\citeauthoryear{{Muench}, {Lada} \& {Lada}}{{Muench}
  et~al.}{2000}]{MLL00}
{Muench} A.~A.,  {Lada} E.~A.,    {Lada} C.~J.,  2000, \apj, 533, 358

\bibitem[\protect\citeauthoryear{{Muench}, {Lada}, {Lada} \& {Alves}}{{Muench}
  et~al.}{2002}]{Muench_etal02}
{Muench} A.~A.,  {Lada} E.~A.,  {Lada} C.~J.,    {Alves} J.,  2002, \apj, 573,
  366

\bibitem[\protect\citeauthoryear{{Najarro}, {Figer}, {Hillier} \&
  {Kudritzki}}{{Najarro} et~al.}{2004}]{Najarro_etal04}
{Najarro} F.,  {Figer} D.~F.,  {Hillier} D.~J.,    {Kudritzki} R.~P.,  2004,
  \apjl, 611, L105

\bibitem[\protect\citeauthoryear{{Najita}, {Tiede} \& {Carr}}{{Najita}
  et~al.}{2000}]{Netal00}
{Najita} J.~R.,  {Tiede} G.~P.,    {Carr} J.~S.,  2000, \apj, 541, 977

\bibitem[\protect\citeauthoryear{{Nakano}}{{Nakano}}{1989}]{Nakano89}
{Nakano} T.,  1989, \apj, 345, 464

\bibitem[\protect\citeauthoryear{{Nordlund} \& {Padoan}}{{Nordlund} \&
  {Padoan}}{2003}]{NordlundPadoan02}
{Nordlund} {\AA}.,  {Padoan} P.,  2003, LNP Vol.~614: Turbulence and Magnetic
  Fields in Astrophysics, 614, 271

\bibitem[\protect\citeauthoryear{{Oey} \& {Clarke}}{{Oey} \&
  {Clarke}}{1998}]{OC98}
{Oey} M.~S.,  {Clarke} C.~J.,  1998, \aj, 115, 1543

\bibitem[\protect\citeauthoryear{{Oey} \& {Clarke}}{{Oey} \&
  {Clarke}}{2005}]{OC05}
{Oey} M.~S.,  {Clarke} C.~J.,  2005, \apjl, 620, L43

\bibitem[\protect\citeauthoryear{{Onishi}, {Mizuno}, {Kawamura}, {Tachihara} \&
  {Fukui}}{{Onishi} et~al.}{2002}]{Onishi_etal02}
{Onishi} T.,  {Mizuno} A.,  {Kawamura} A.,  {Tachihara} K.,    {Fukui} Y.,
  2002, \apj, 575, 950

\bibitem[\protect\citeauthoryear{{Padoan} \& {Nordlund}}{{Padoan} \&
  {Nordlund}}{2002}]{PadoanNordlund02}
{Padoan} P.,  {Nordlund} {\AA}.,  2002, \apj, 576, 870

\bibitem[\protect\citeauthoryear{{Padoan} \& {Nordlund}}{{Padoan} \&
  {Nordlund}}{2004}]{PadoanNordlund04}
{Padoan} P.,  {Nordlund} {\AA}.,  2004, \apj, 617, 559

\bibitem[\protect\citeauthoryear{{Paresce}, {de Marchi} \&
  {Romaniello}}{{Paresce} et~al.}{1995}]{PdeMR95}
{Paresce} F.,  {de Marchi} G.,    {Romaniello} M.,  1995, \apj, 440, 216

\bibitem[\protect\citeauthoryear{{Park}, {Sung}, {Bessell} \& {Kang}}{{Park}
  et~al.}{2000}]{Petal00}
{Park} B.,  {Sung} H.,  {Bessell} M.~S.,    {Kang} Y.~H.,  2000, \aj, 120, 894

\bibitem[\protect\citeauthoryear{{Parker}, {Zaritsky}, {Stecher}, {Harris} \&
  {Massey}}{{Parker} et~al.}{2001}]{Parker_etal01}
{Parker} J.~W.,  {Zaritsky} D.,  {Stecher} T.~P.,  {Harris} J.,    {Massey} P.,
   2001, \aj, 121, 891

\bibitem[\protect\citeauthoryear{{Penny}, {Massey} \& {Vukovich}}{{Penny}
  et~al.}{2001}]{MPV01}
{Penny} L.~R.,  {Massey} P.,    {Vukovich} J.,  2001, Bulletin of the American
  Astronomical Society, 33, 1310

\bibitem[\protect\citeauthoryear{{Pflamm-Altenburg} \&
  {Kroupa}}{{Pflamm-Altenburg} \& {Kroupa}}{2006}]{P-AK06a}
{Pflamm-Altenburg} J.,  {Kroupa} P.,  2006, \mnras, 373, 295

\bibitem[\protect\citeauthoryear{{Phan-Bao}, {Martin}, {Reyle}, {Forveille} \&
  {Lim}}{{Phan-Bao} et~al.}{2005}]{PhanBao_etal05}
{Phan-Bao} N.,  {Martin} E.~L.,  {Reyle} C.,  {Forveille} T.,    {Lim} J.,
  2005, ArXiv Astrophysics e-prints

\bibitem[\protect\citeauthoryear{{Pinfield}, {Dobbie}, {Jameson}, {Steele},
  {Jones} \& {Katsiyannis}}{{Pinfield} et~al.}{2003}]{Pinfield_etal03}
{Pinfield} D.~J.,  {Dobbie} P.~D.,  {Jameson} R.~F.,  {Steele} I.~A.,  {Jones}
  H.~R.~A.,    {Katsiyannis} A.~C.,  2003, \mnras, 342, 1241

\bibitem[\protect\citeauthoryear{{Piotto} \& {Zoccali}}{{Piotto} \&
  {Zoccali}}{1999}]{PZ99}
{Piotto} G.,  {Zoccali} M.,  1999, \aap, 345, 485

\bibitem[\protect\citeauthoryear{{Piskunov}, {Belikov}, {Kharchenko}, {Sagar}
  \& {Subramaniam}}{{Piskunov} et~al.}{2004}]{Piskunov_etal04}
{Piskunov} A.~E.,  {Belikov} A.~N.,  {Kharchenko} N.~V.,  {Sagar} R.,
  {Subramaniam} A.,  2004, \mnras, 349, 1449

\bibitem[\protect\citeauthoryear{{Portegies Zwart}, {Makino}, {McMillan} \&
  {Hut}}{{Portegies Zwart} et~al.}{2002}]{Port_etal02}
{Portegies Zwart} S.~F.,  {Makino} J.,  {McMillan} S.~L.~W.,    {Hut} P.,
  2002, \apj, 565, 265

\bibitem[\protect\citeauthoryear{{Portegies Zwart}, {McMillan}, {Hut} \&
  {Makino}}{{Portegies Zwart} et~al.}{2001}]{Port_etal01}
{Portegies Zwart} S.~F.,  {McMillan} S.~L.~W.,  {Hut} P.,    {Makino} J.,
  2001, \mnras, 321, 199

\bibitem[\protect\citeauthoryear{{Preibisch}, {Balega}, {Hofmann}, {Weigelt} \&
  {Zinnecker}}{{Preibisch} et~al.}{1999}]{Preibisch99}
{Preibisch} T.,  {Balega} Y.,  {Hofmann} K.,  {Weigelt} G.,    {Zinnecker} H.,
  1999, New Astronomy, 4, 531

\bibitem[\protect\citeauthoryear{{Ramspeck}, {Heber} \& {Moehler}}{{Ramspeck}
  et~al.}{2001}]{Ram01}
{Ramspeck} M.,  {Heber} U.,    {Moehler} S.,  2001, \aap, 378, 907

\bibitem[\protect\citeauthoryear{{Reid}, {Cruz}, {Allen}, {Mungall},
  {Kilkenny}, {Liebert}, {Hawley}, {Fraser}, {Covey} \& {Lowrance}}{{Reid}
  et~al.}{2003}]{Retal03b}
{Reid} I.~N.,  {Cruz} K.~L.,  {Allen} P.,  {Mungall} F.,  {Kilkenny} D.,
  {Liebert} J.,  {Hawley} S.~L.,  {Fraser} O.~J.,  {Covey} K.~R.,    {Lowrance}
  P.,  2003, \aj, 126, 3007

\bibitem[\protect\citeauthoryear{{Reid}, {Cruz}, {Laurie}, {Liebert}, {Dahn},
  {Harris}, {Guetter}, {Stone}, {Canzian}, {Luginbuhl}, {Levine}, {Monet} \&
  {Monet}}{{Reid} et~al.}{2003}]{Retal03a}
{Reid} I.~N.,  {Cruz} K.~L.,  {Laurie} S.~P.,  {Liebert} J.,  {Dahn} C.~C.,
  {Harris} H.~C.,  {Guetter} H.~H.,  {Stone} R.~C.,  {Canzian} B.,  {Luginbuhl}
  C.~B.,  {Levine} S.~E.,  {Monet} A.~K.~B.,    {Monet} D.~G.,  2003, \aj, 125,
  354

\bibitem[\protect\citeauthoryear{{Reid} \& {Gizis}}{{Reid} \&
  {Gizis}}{1997}]{RG97}
{Reid} I.~N.,  {Gizis} J.~E.,  1997, \aj, 113, 2246

\bibitem[\protect\citeauthoryear{{Reid}, {Gizis} \& {Hawley}}{{Reid}
  et~al.}{2002}]{RGH02}
{Reid} I.~N.,  {Gizis} J.~E.,    {Hawley} S.~L.,  2002, \aj, 124, 2721

\bibitem[\protect\citeauthoryear{{Reid}, {Kirkpatrick}, {Liebert}, {Burrows},
  {Gizis}, {Burgasser}, {Dahn}, {Monet}, {Cutri}, {Beichman} \&
  {Skrutskie}}{{Reid} et~al.}{1999}]{Retal99}
{Reid} I.~N.,  {Kirkpatrick} J.~D.,  {Liebert} J.,  {Burrows} A.,  {Gizis}
  J.~E.,  {Burgasser} A.,  {Dahn} C.~C.,  {Monet} D.,  {Cutri} R.,  {Beichman}
  C.~A.,    {Skrutskie} M.,  1999, \apj, 521, 613

\bibitem[\protect\citeauthoryear{{Reid} \& {Gilmore}}{{Reid} \&
  {Gilmore}}{1982}]{RG82}
{Reid} N.,  {Gilmore} G.,  1982, \mnras, 201, 73

\bibitem[\protect\citeauthoryear{{Reipurth} \& {Clarke}}{{Reipurth} \&
  {Clarke}}{2001}]{RC01}
{Reipurth} B.,  {Clarke} C.,  2001, \aj, 122, 432

\bibitem[\protect\citeauthoryear{{Reyl{\' e}} \& {Robin}}{{Reyl{\' e}} \&
  {Robin}}{2001}]{RR01}
{Reyl{\' e}} C.,  {Robin} A.~C.,  2001, \aap, 373, 886

\bibitem[\protect\citeauthoryear{{Sagar} \& {Richtler}}{{Sagar} \&
  {Richtler}}{1991}]{SR91}
{Sagar} R.,  {Richtler} T.,  1991, \aap, 250, 324

\bibitem[\protect\citeauthoryear{{Salpeter}}{{Salpeter}}{1955}]{S55}
{Salpeter} E.~E.,  1955, \apj, 121, 161

\bibitem[\protect\citeauthoryear{{Sanner} \& {Geffert}}{{Sanner} \&
  {Geffert}}{2001}]{SG01}
{Sanner} J.,  {Geffert} M.,  2001, \aap, 370, 87

\bibitem[\protect\citeauthoryear{{Santiago}, {Beaulieu}, {Johnson} \&
  {Gilmore}}{{Santiago} et~al.}{2001}]{Santetal01}
{Santiago} B.,  {Beaulieu} S.,  {Johnson} R.,    {Gilmore} G.~F.,  2001, \aap,
  369, 74

\bibitem[\protect\citeauthoryear{{Scalo}}{{Scalo}}{1998}]{Sc98}
{Scalo} J.,  1998, in ASP Conf. Ser. 142: The Stellar Initial Mass Function
  (38th Herstmonceux Conference) {The IMF Revisited: A Case for Variations}.
pp 201--+

\bibitem[\protect\citeauthoryear{{Scalo}}{{Scalo}}{1986}]{Sc86}
{Scalo} J.~M.,  1986, Fundamentals of Cosmic Physics, 11, 1

\bibitem[\protect\citeauthoryear{{Schaller}, {Schaerer}, {Meynet} \&
  {Maeder}}{{Schaller} et~al.}{1992}]{Schaller_etal92}
{Schaller} G.,  {Schaerer} D.,  {Meynet} G.,    {Maeder} A.,  1992, \aaps, 96,
  269

\bibitem[\protect\citeauthoryear{{Schwarzschild} \& {H{\"
  a}rm}}{{Schwarzschild} \& {H{\" a}rm}}{1959}]{SH59}
{Schwarzschild} M.,  {H{\" a}rm} R.,  1959, \apj, 129, 637

\bibitem[\protect\citeauthoryear{{Selman}, {Melnick}, {Bosch} \&
  {Terlevich}}{{Selman} et~al.}{1999}]{Setal99}
{Selman} F.,  {Melnick} J.,  {Bosch} G.,    {Terlevich} R.,  1999, \aap, 347,
  532

\bibitem[\protect\citeauthoryear{{Siess}, {Dufour} \& {Forestini}}{{Siess}
  et~al.}{2000}]{SDF00}
{Siess} L.,  {Dufour} E.,    {Forestini} M.,  2000, \aap, 358, 593

\bibitem[\protect\citeauthoryear{{Sirianni}, {Nota}, {Leitherer}, {De Marchi}
  \& {Clampin}}{{Sirianni} et~al.}{2000}]{Sirianni_etal00}
{Sirianni} M.,  {Nota} A.,  {Leitherer} C.,  {De Marchi} G.,    {Clampin} M.,
  2000, \apj, 533, 203

\bibitem[\protect\citeauthoryear{{Slesnick}, {Hillenbrand} \&
  {Carpenter}}{{Slesnick} et~al.}{2004}]{Slesnick_etal04}
{Slesnick} C.~L.,  {Hillenbrand} L.~A.,    {Carpenter} J.~M.,  2004, \apj, 610,
  1045

\bibitem[\protect\citeauthoryear{{Smith} \& {Gallagher}}{{Smith} \&
  {Gallagher}}{2001}]{SmGa01}
{Smith} L.~J.,  {Gallagher} J.~S.,  2001, \mnras, 326, 1027

\bibitem[\protect\citeauthoryear{{Soubiran}, {Bienaym{\' e}} \&
  {Siebert}}{{Soubiran} et~al.}{2003}]{Soubiran_etal03}
{Soubiran} C.,  {Bienaym{\' e}} O.,    {Siebert} A.,  2003, \aap, 398, 141

\bibitem[\protect\citeauthoryear{{Stahler}, {Palla} \& {Ho}}{{Stahler}
  et~al.}{2000}]{SPH00}
{Stahler} S.~W.,  {Palla} F.,    {Ho} P.~T.~P.,  2000, Protostars and Planets
  IV, pp 327--+

\bibitem[\protect\citeauthoryear{{Stobie}, {Ishida} \& {Peacock}}{{Stobie}
  et~al.}{1989}]{SIP}
{Stobie} R.~S.,  {Ishida} K.,    {Peacock} J.~A.,  1989, \mnras, 238, 709

\bibitem[\protect\citeauthoryear{{Stolte}, {Grebel}, {Brandner} \&
  {Figer}}{{Stolte} et~al.}{2002}]{Stolteetal02}
{Stolte} A.,  {Grebel} E.~K.,  {Brandner} W.,    {Figer} D.~F.,  2002, \aap,
  394, 459

\bibitem[\protect\citeauthoryear{{Stothers}}{{Stothers}}{1992}]{St92}
{Stothers} R.~B.,  1992, \apj, 392, 706

\bibitem[\protect\citeauthoryear{{Sung} \& {Bessell}}{{Sung} \&
  {Bessell}}{2004}]{SB04}
{Sung} H.,  {Bessell} M.~S.,  2004, \aj, 127, 1014

\bibitem[\protect\citeauthoryear{{Testi} \& {Sargent}}{{Testi} \&
  {Sargent}}{1998}]{TestiSargent98}
{Testi} L.,  {Sargent} A.~I.,  1998, \apjl, 508, L91

\bibitem[\protect\citeauthoryear{{Tilley} \& {Pudritz}}{{Tilley} \&
  {Pudritz}}{2004}]{TilleyPudritz04}
{Tilley} D.~A.,  {Pudritz} R.~E.,  2004, \mnras, 353, 769

\bibitem[\protect\citeauthoryear{{Vanbeveren}}{{Vanbeveren}}{1982}]{Vanbev82}
{Vanbeveren} D.,  1982, \aap, 115, 65

\bibitem[\protect\citeauthoryear{{Vesperini} \& {Heggie}}{{Vesperini} \&
  {Heggie}}{1997}]{VH97}
{Vesperini} E.,  {Heggie} D.~C.,  1997, \mnras, 289, 898

\bibitem[\protect\citeauthoryear{{Vine} \& {Bonnell}}{{Vine} \&
  {Bonnell}}{2003}]{VB03}
{Vine} S.~G.,  {Bonnell} I.~A.,  2003, \mnras, 342, 314

\bibitem[\protect\citeauthoryear{{Vogt}, {Butler}, {Marcy}, {Fischer},
  {Pourbaix}, {Apps} \& {Laughlin}}{{Vogt} et~al.}{2002}]{Vogt01}
{Vogt} S.~S.,  {Butler} R.~P.,  {Marcy} G.~W.,  {Fischer} D.~A.,  {Pourbaix}
  D.,  {Apps} K.,    {Laughlin} G.,  2002, \apj, 568, 352

\bibitem[\protect\citeauthoryear{{von Hippel}, {Gilmore}, {Tanvir}, {Robinson}
  \& {Jones}}{{von Hippel} et~al.}{1996}]{vonHippelGil96}
{von Hippel} T.,  {Gilmore} G.,  {Tanvir} N.,  {Robinson} D.,    {Jones}
  D.~H.~P.,  1996, \aj, 112, 192

\bibitem[\protect\citeauthoryear{{Weidemann}}{{Weidemann}}{1990}]{Weide}
{Weidemann} V.,  1990, in NATO ASIC Proc. 305: Baryonic Dark Matter {White
  Dwarfs and the Local Mass Density}.
pp 87--+

\bibitem[\protect\citeauthoryear{{Weidemann}}{{Weidemann}}{2000}]{W2000}
{Weidemann} V.,  2000, \aap, 363, 647

\bibitem[\protect\citeauthoryear{{Weidemann}, {Jordan}, {Iben} \&
  {Casertano}}{{Weidemann} et~al.}{1992}]{Wetal92}
{Weidemann} V.,  {Jordan} S.,  {Iben} I.~J.,    {Casertano} S.,  1992, \aj,
  104, 1876

\bibitem[\protect\citeauthoryear{{Weidner} \& {Kroupa}}{{Weidner} \&
  {Kroupa}}{2004}]{WK04}
{Weidner} C.,  {Kroupa} P.,  2004, \mnras, 348, 187

\bibitem[\protect\citeauthoryear{{Weidner} \& {Kroupa}}{{Weidner} \&
  {Kroupa}}{2005}]{WK05b}
{Weidner} C.,  {Kroupa} P.,  2005, \apj, 625, 754

\bibitem[\protect\citeauthoryear{{Weidner} \& {Kroupa}}{{Weidner} \&
  {Kroupa}}{2006}]{WK05a}
{Weidner} C.,  {Kroupa} P.,  2006, \mnras, 365, 1333

\bibitem[\protect\citeauthoryear{{Weidner}, {Kroupa} \& {Larsen}}{{Weidner}
  et~al.}{2004}]{WKL04}
{Weidner} C.,  {Kroupa} P.,    {Larsen} S.~S.,  2004, \mnras, 350, 1503

\bibitem[\protect\citeauthoryear{{Weigelt} \& {Baier}}{{Weigelt} \&
  {Baier}}{1985}]{WB85}
{Weigelt} G.,  {Baier} G.,  1985, \aap, 150, L18

\bibitem[\protect\citeauthoryear{{White} \& {Basri}}{{White} \&
  {Basri}}{2003}]{WhiteBasri03}
{White} R.~J.,  {Basri} G.,  2003, \apj, 582, 1109

\bibitem[\protect\citeauthoryear{{Whitworth} \& {Zinnecker}}{{Whitworth} \&
  {Zinnecker}}{2004}]{WhitZinn04}
{Whitworth} A.~P.,  {Zinnecker} H.,  2004, \aap, 427, 299

\bibitem[\protect\citeauthoryear{{Wolfire} \& {Cassinelli}}{{Wolfire} \&
  {Cassinelli}}{1986}]{Wolf86}
{Wolfire} M.~G.,  {Cassinelli} J.~P.,  1986, \apj, 310, 207

\bibitem[\protect\citeauthoryear{{Wolfire} \& {Cassinelli}}{{Wolfire} \&
  {Cassinelli}}{1987}]{Wolf87}
{Wolfire} M.~G.,  {Cassinelli} J.~P.,  1987, \apj, 319, 850

\bibitem[\protect\citeauthoryear{{Wuchterl} \& {Klessen}}{{Wuchterl} \&
  {Klessen}}{2001}]{WK01}
{Wuchterl} G.,  {Klessen} R.~S.,  2001, \apjl, 560, L185

\bibitem[\protect\citeauthoryear{{Wuchterl} \& {Tscharnuter}}{{Wuchterl} \&
  {Tscharnuter}}{2003}]{WT03}
{Wuchterl} G.,  {Tscharnuter} W.~M.,  2003, \aap, 398, 1081

\bibitem[\protect\citeauthoryear{{Zheng}, {Flynn}, {Gould}, {Bahcall} \&
  {Salim}}{{Zheng} et~al.}{2001}]{Zheng01}
{Zheng} Z.,  {Flynn} C.,  {Gould} A.,  {Bahcall} J.~N.,    {Salim} S.,  2001,
  \apj, 555, 393

\bibitem[\protect\citeauthoryear{{Zinnecker}}{{Zinnecker}}{1984}]{Zinn84}
{Zinnecker} H.,  1984, \mnras, 210, 43

\bibitem[\protect\citeauthoryear{{Zoccali}, {Cassisi}, {Frogel}, {Gould},
  {Ortolani}, {Renzini}, {Rich} \& {Stephens}}{{Zoccali}
  et~al.}{2000}]{Zetal00}
{Zoccali} M.,  {Cassisi} S.,  {Frogel} J.~A.,  {Gould} A.,  {Ortolani} S.,
  {Renzini} A.,  {Rich} R.~M.,    {Stephens} A.~W.,  2000, \apj, 530, 418

\bibitem[\protect\citeauthoryear{{Zucker} \& {Mazeh}}{{Zucker} \&
  {Mazeh}}{2001}]{ZuMaz01}
{Zucker} S.,  {Mazeh} T.,  2001, \apj, 562, 1038

\end{thebibliography}
%
%
%

\end{document}